\documentclass[fleqn,usenatbib,useAMS]{mnras}

\usepackage{graphicx}	
\usepackage{amsmath}	
\usepackage{amssymb}	
\usepackage{multicol}        
\usepackage{bm}		
\usepackage{pdflscape}	
\usepackage[normalem]{ulem}

\usepackage[T1]{fontenc}
\usepackage{ae,aecompl}

\usepackage{newtxtext,newtxmath}

\title[The DIVING$^\mathrm{3D}$ Survey II]{The DIVING$^\mathrm{3D}$ Survey - Deep IFS View of Nuclei of Galaxies - II. First results: nuclear emission-line properties of the mini-DIVING$^\mathrm{3D}$ sample}

\author[R. B. Menezes et al.]{R. B. Menezes$^{1}$\thanks{ \href{mailto:roberto.menezes@maua.br}{roberto.menezes@maua.br}}, J. E. Steiner$^2$ \thanks{Deceased},  T. V. Ricci$^3$\thanks{\href{mailto:tiago.ricci@uffs.edu.br}{tiago.ricci@uffs.edu.br}} and Patr\'icia da Silva$^2$\thanks{\href{mailto:p.silva2201@gmail.com}{p.silva2201@gmail.com}} \\
\\
$^{1}$Instituto Mau\'a de Tecnologia, Pra\c{c}a Mau\'a 1, 09580-900, S\~ao Caetano do Sul, SP, Brazil\\  
$^{2}$Instituto de Astronomia, Geof\'isica e Ci\^encias Atmosf\'ericas, Departamento de Astronomia, Universidade de S\~ao Paulo, 05508-090, SP, Brazil\\
$^{3}$Universidade Federal da Fronteira Sul, Cerro Largo, 97900000, RS, Brazil}

\date{Accepted 2022 April 24. Received 2022 April 24; in original form 2021 July 28}

\pubyear{2022}

\begin{document}
\label{firstpage}
\pagerange{\pageref{firstpage}--\pageref{lastpage}}
\maketitle

\begin{abstract}
This work presents the first results of the Deep IFS View of Nuclei of Galaxies (DIVING$^\mathrm{3D}$) survey. We analysed the nuclear emission-line spectra of a sub-sample we call mini-DIVING$^\mathrm{3D}$, which includes all Southern galaxies with B < 11.2 and |b| > 15$\degr$. We verified that $23\% \pm 4\%$ of the galaxies show nuclear emission-line properties characteristic of Low Ionization Nuclear Emission-Line Regions (LINERs). Diagnostic diagram analysis reveals an apparent dichotomy, not detected in previous studies, between objects classified as H II regions and as LINERs or Seyferts, with very few galaxies classified as transition objects. A possible explanation for this result is that at least part of the transition objects are composite systems, with a central LINER contaminated by the emission from circumnuclear H II regions. The higher spatial resolution of the DIVING$^\mathrm{3D}$ survey, in comparison with previous studies, allowed us to isolate the nuclear emission from circumnuclear contaminations, reducing the number of transition objects. We also propose an alternative scenario, in which the emission-line spectra of some transition objects are the result of shock heating by central outflows, together with photoionization by young stars. Clear evidence of active galactic nuclei (AGNs), in the optical and X-ray spectral bands, were detected in 69\% of the LINERs in the mini-DIVING$^\mathrm{3D}$ sample. Considering the entire mini-DIVING$^\mathrm{3D}$ sample, evidence of AGNs were detected in 65\% of the objects.

\end{abstract}

\begin{keywords}
techniques: imaging spectroscopy -- galaxies: active -- galaxies: nuclei -- galaxies: Seyfert  
\end{keywords}




\section{Introduction}\label{sec1}
The central regions of galaxies are certainly very important components of these large structures, as they may provide relevant information about the galactic formation and evolution. The nuclei of some galaxies show continuum and emission-line properties in their spectra that cannot be attributed only to stars. These are the so-called Active Galactic Nuclei (AGNs), which, according to their spectroscopic characteristics and luminosities, are divided into different classes, such as Quasars and Seyfert galaxies, for example. It is now well accepted that the energy emitted by an AGN comes from the accretion of matter onto a central supermassive black hole (SMBH; see \citealt{net13} for a detailed discussion).

SMBHs seem to be present at the centre of every massive galaxy \citep{kor95,ric98}. The masses of these SMBHs show a clear correlation with the values of the stellar velocity dispersion in the bulges of the host galaxies (the M - $\sigma$ relation; \citealt{fer00,geb00,gul09}). There are also similar correlations between the masses of the central SMBHs and other parameters of the host galaxies, such as the luminosities and the masses of the bulges (e.g. \citealt{kor95,mag98,hop07,sag16}). Although less common, lower mass black holes have also been detected at the centres of late-type or even dwarf galaxies (e.g. \citealt{fil03,don07}). According to \citet{xia11}, such low-mass galaxies seem to follow the same M - $\sigma$ relation of the high-mass ones, but with a larger scatter. These correlations between the central SMBH and properties of the host galaxy are very significant, as they reveal an apparent connection between the central SMBHs (possibly associated with AGNs) and the formation/evolution of the galaxies. It is worth mentioning, however, that, according to \citet{kor13}, the coevolution scenario between SMBHs and the host galaxies is more complex than the statement above and depends, for example, on the masses of the SMBHs. In addition, \citet{kor13} also argue that it is questionable whether the low-mass galaxies follow an M - $\sigma$ relation with a larger scatter than the massive galaxies or they do not follow any correlation. 

The presence of a SMBH at the centre of a galaxy is still difficult to be confirmed in many situations. Measurements of SMBH masses are usually performed via the analysis of mega-maser emission, gas kinematics or stellar kinematics around the nucleus. Such methods, however, have a number of limitations. The gas kinematics technique can be affected by the presence of non-keplerian motions (inflows or outflows, for example). The stellar kinematics method usually can only be applied to objects without a large amount of dust and with a high surface brightness. Considering the achievable spatial resolutions, it is currently not possible to measure SMBH masses much lower than $\sim 10^6 M_{\sun}$. For more details, see \citet{fer05} and \citet{kor13}. Due to all these limitations, the identification of central SMBHs in galaxies often depend on alternative approaches, such as the detection of AGNs, which are always associated with SMBHs. Considering that AGNs may have a significant influence on the surrounding environment \citep{gra04,spr05,hop06} and that the corresponding SMBHs may show a connection with the host galaxy, the search for AGNs and the determination of the fractions of galaxies, with different morphological types, showing some nuclear activity is certainly a field of great importance for the study of the formation/evolution of these objects.

There are different methods for detecting central AGNs in galaxies. A strong point-like X-ray or radio source at the nucleus is usually indicative of an AGN \citep{ho08}. In the optical, the most used technique, developed by \citet{bal81} and modified by \citet{vei87}, involves the analysis of diagnostic diagrams, which are graphs showing two pairs of emission-line ratios. This method allows one to classify line-emitting nuclei in HII regions, Seyferts, and Low Ionization Nuclear Emission-Line Regions (LINERs; \citealt{kew01,kau03,kew06,sch07}). 

The physical mechanisms responsible for the low-ionization spectra of LINERs are still controversial. At first, the emission-line spectra of LINERs were taken as the result of shock heating \citep{hec80}. This model has been considerably improved in later studies \citep{dop95,dop96} and is still used to explain the LINER emission in some objects. The photoionization by young stars has also been proposed as a LINER mechanism \citep{ter85,fil92,shi92}, although this scenario is usually not taken into account anymore. Models of photoionization by hot low-mass evolved stars (HOLMES; sometimes referred to as post-AGB stars, but including nuclei of planetary nebulae and white dwarfs), on the other hand, seem to be much more promising for explaining the observed emission-line spectra of many LINERs \citep{bin94,sta08,era10,cid11}. The acronym HOLMES was actually proposed by \citet{flo11}. Despite the variety of proposed models, one of the most accepted ideas today is that a large amount of LINERs are powered by accretion of matter onto SMBHs \citep{fer83,hal83}, i.e. many LINERs are authentic AGNs, but with a lower ionization parameter. This last scenario has been confirmed by the detection, in a significant fraction of LINERs, of central point-like X-ray and radio sources and of a broad component of the H$\alpha$ emission line, which is also considered a proof of the presence of an AGN \citep{ho08}.   

Certainly a very effective way of studying the properties of the central regions of galaxies, including the detection of AGNs, involves surveys. One of the most popular surveys of the central regions of galaxies is the PALOMAR survey \citep{fil85,ho08}, which obtained, with a slit of 2 arcsec $\times$ 4 arcsec, optical spectra of the central regions of all northern galaxies brighter than B = 12.5. However, since this survey analysed only long-slit spectra, it did not provide significant information about the spatial morphology of the line-emitting regions in the galaxies. Such information requires surveys performed with instruments capable of providing 3D spectroscopy, such as Integral Field Units (IFUs). Examples of surveys involving this kind of technology are: the Spectroscopic Areal Unit for Research on Optical Nebulae (SAURON; \citealt{bac01}); the ATLAS$^\mathrm{3D}$ \citep{cap11}; the Calar Alto Legacy Integral Field Area (CALIFA; \citealt{san12}); the Sydney-Australian-Astronomical-Observatory Multi-object Integral-Field Spectrograph Galaxy Survey (SAMI; \citealt{bry15}); the Mapping Nearby Galaxies at APO (MaNGA; \citealt{bun15}); the Siding Spring Southern Seyfert Spectroscopic Snapshot Survey (S7; \citealt{dop15}). None of these IFU surveys involved data with a spatial resolution better than $\sim$ 1 arcsec, which made it impossible for such studies to efficiently isolate the nuclear emission from the circumnuclear one.

We are conducting the Deep IFS View of Nuclei of Galaxies (DIVING$^\mathrm{3D}$; \citealt{ste22} - paper I) survey, with the goal of observing, using 3D spectroscopy, the central regions of all galaxies in the Southern hemisphere with B < 12.0 (according to the Revised Shapley-Ames Catalogue of Bright Galaxies, RSA - \citealt{san81}) and with a Galactic latitude |b| > 15$\degr$. From this selection, we excluded 11 Sm/Im objects, since it was not possible to clearly identify their nuclei in the 2MASS images. The complete final sample has a total of 170 objects. Most the observations were taken, in the optical, with the IFU of the Gemini Multi-object Spectrograph (GMOS), on the Gemini South and Gemini North telescopes. Some observations are being taken with the SOAR Integral Field Spectrograph (SIFS), on the SOAR telescope. The data cubes provided by these instruments are seeing-limited, having a combination of spatial and spectral resolutions not matched by previous surveys. It is worth emphasizing that the sub-arcsecond spatial resolution of the DIVING$^\mathrm{3D}$ survey is mainly due to the observing conditions we established for the observations, which resulted in a median seeing of 0.7 arcsec for the sample. However, the data treatment we apply (see Section~\ref{sec2}) improves the spatial resolution even more. In principle, such a treatment procedure could be applied to data of previous surveys, as long as a reliable estimate of the point spread function (PSF) was available. However, considering the typical improvements in the spatial resolution provided by our treatment procedure and, depending on the spatial resolution of the survey, even such a treatment would not be enough to result in a sub-arcsecond spatial resolution, as the one obtained in the DIVING$^\mathrm{3D}$ survey. 

\begin{figure}
    \centering
    \includegraphics[scale=0.53]{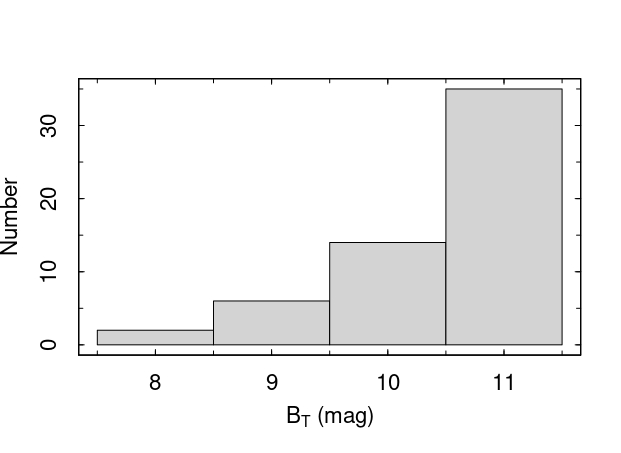}
    \caption{Distribution of the apparent B magnitudes of the mini-DIVING$^\mathrm{3D}$ sample.}
    \label{fig1}
\end{figure}

In this paper, we present the first results of the analysis of the DIVING$^\mathrm{3D}$ sample, focused on the nuclear emission-line properties of all galaxies brighter than B=11.2. We call this sub-sample the mini-DIVING$^\mathrm{3D}$, which has a total of 57 objects. Fig. \ref{fig1} shows the distribution of the apparent B magnitudes of the sample. We also computed the $V/V_{max}$ ratio of each source, where $V$ is the volume of a sphere centred on the observer and with a radius that corresponds to the distance of the object and $V_{max}$ is the maximum volume that the object may have and still be included in the sample. We found $\langle V/V_{MAX} \rangle =$ 0.506$\pm$0.038 for the mini-DIVING$^\mathrm{3D}$ sample. According to \citet{sch68}, a sample is uniformly distributed in space if $\langle V/V_{MAX} \rangle = 0.5$, which is the case for the mini-DIVING$^\mathrm{3D}$ sample. 

The paper is structured as follows. In Section~\ref{sec2}, we describe the observations, the reduction, and also the treatment procedure applied to the data cubes. In Section~\ref{sec3}, we present the data analysis procedure and the results are shown in Section~\ref{sec4}. In Section~\ref{sec5}, we discuss and compare our results with those of previous studies. We draw our conclusions in Section~\ref{sec6}.  

\section{Observations, reduction, and data treatment}\label{sec2}

The GMOS/IFU data cubes were reduced using the Gemini package under the \textsc{iraf} environment. The reduction procedure included the following basic steps: trimming, overscan and bias subtraction, cosmic-ray removal, correction of bad pixels, extraction of the spectra, correction for fibre-to-fibre and for pixel-to-pixel gain variations, wavelength calibration, sky subtraction, flux calibration, and data cube construction. The reduction procedure resulted in data cubes with spatial pixels (spaxels) of $0.05$ arcsec. Each lenslet of GMOS/IFU corresponds to a hexagon with a radius of 0.18 arcsec; therefore reducing the data cubes obtained with this instrument with spaxels of $0.05$ arcsec corresponds to an oversampling of the data. We opted for this procedure because such an oversampling resulted in higher spatial resolutions when the Richardson-Lucy deconvolution, which is part of our data treatment procedure, as explained below, was applied (for more detail, see \citealt{men19}). 

The SIFS data cubes were reduced using scripts developed in Interactive Data Language (IDL). The process is similar to that applied to GMOS/IFU data and resulted in data cubes with spaxels of $0.3$ arcsec. See paper I for further detail about the observations and data reduction. 

After the data reduction, we applied a treatment procedure, developed by our working group, to the obtained data cubes \citep{men19}. This process included: differential atmospheric refraction correction, combination of the data cubes in the form of a median (for the objects with multiple exposures), Butterworth spatial filtering, `instrumental fingerprint' removal, and Richardson-Lucy deconvolution. The treatment procedure was essentially the same for GMOS/IFU and SIFS data cubes, except for the fact that, in the case of SIFS data cubes, no differential atmospheric correction was applied, since SOAR has an Atmospheric Dispersion Corrector installed, and a spatial re-sampling (to obtain spaxels of $0.1$ arcsec) was applied before the Butterworth spatial filtering. For more information about this treatment procedure, see \citet{men14,men15} and paper I.

Table~\ref{tbl1} shows the observational parameters of the galaxies in the mini-DIVING$^\mathrm{3D}$ sample. In particular, the values of the final full-width at half-maximum (FWHM) of the point spread function (PSF), at 5500 \AA, of each data cube, after the data treatment (FWHM(5500)$_{treated}$), are presented. For the objects with a central AGN and with a broad component of the H$\alpha$ emission line, the FWHM of the PSF, at the wavelength of H$\alpha$, was estimated, after the data treatment, by constructing an image of the blue or red wing of the broad component of H$\alpha$. The reason for this procedure is that the emitting region of the broad components of permitted emission lines in AGNs (Broad Line Region - BLR) is very compact and should be point-like in images with the GMOS/IFU or SIFS spatial resolutions. FWHM(5500)$_{treated}$ was then calculated using the following equation, which was deduced from the behaviour of the FWHM with the wavelength ($\lambda$) observed in data cubes of standard stars \citep{men19}:

\begin{equation}
FWHM \left(\lambda\right) = FWHM_{ref} \cdot \left(\frac{\lambda}{\lambda_{ref}}\right)^{\alpha},
\end{equation}
where FWHM$_{ref}$ is the FWHM at $\lambda_{ref}$. For the calculation of FWHM(5500)$_{treated}$, FWHM$_{ref}$ was taken as the FWHM of the point-like structure in the image of the broad component of H$\alpha$ and $\lambda_{ref}$ was taken as the wavelength of such an image. We assumed $\alpha$ = -0.3 for GMOS/IFU data cubes and $\alpha$ = -0.05 for SIFS data cubes, which are average values estimated from different data cubes of standard stars. 

\begin{small}
\begin{table*}
\centering
\caption{Observational parameters of the galaxies in the mini-DIVING$^\mathrm{3D}$ sample. The morphological types were obtained from NASA Extragalactic Database (NED) and the magnitudes at the B band were taken from RSA. The values of the FWHM of the PSF at 5500 \AA~(FWHM(5500)$_{treated}$), in arcsec, were estimated from the data cubes after the treatment procedure and the distances, together with the corresponding references, required to calculate these same values in pc are shown in Table 1 of paper I. \label{tbl1}}
\begin{tabular}{ccccccc}
\hline
Galaxy & Morphological & Magnitude & Programme ID (PI) & Exposure & FWHM(5500)$_{treated}$ & FWHM(5500)$_{treated}$ \\
       & type         & at the B band &            & time (s) & (arcsec) & (pc)
\\ \hline
\multicolumn{7}{c}{GMOS-IFU}                                                                                            \\ \hline
NGC 134 & SAB(s)bc & 10.96 & GS-2014B-Q-30 (J. E. Steiner) & 3 $\times$ 930 & 0.46 & 41 \\
NGC 157 & SAB(rs)bc & 11.04 & GS-2014B-Q-30 (J. E. Steiner) & 3 $\times$ 930 & 0.34 & 21  \\
NGC 247 & SAB(s)d & 9.51 & GS-2014B-Q-30 (J. E. Steiner) & 3 $\times$ 930 & 0.49 & 7.9 \\
NGC 253  & SAB(s)c & 8.13 & GS-2013B-Q-33 (R. B. Menezes) & 3 $\times$ 910 & 0.48 & 7.1 \\
NGC 300 & SA(s)d & 8.70 & GS-2013B-Q-20 (T. V. Ricci) & 1 $\times$ 1800 & 0.49 & 5.2 \\
NGC 613 & SB(rs)bc & 10.75 & GS-2014B-Q-30 (J. E. Steiner) & 3 $\times$ 930 & 0.51 & 53 \\
NGC 720 & E5 & 11.15 & GS-2013B-Q-20 (T. V. Ricci) & 1 $\times$ 1800 & 1.29 & 190 \\
NGC 908 & SA(s)c & 10.87 & GS-2014B-Q-30 (J. E. Steiner) & 3 $\times$ 930 & 0.57 & 51 \\
NGC 936 & SB0 & 11.19 & GS-2014B-Q-30 (J. E. Steiner) & 3 $\times$ 630 & 0.63 & 61 \\
NGC 1068 & (R)SA(rs)b & 9.55 & GS-2010B-Q-81 (J. E. Steiner) & 6 $\times$ 830 & 0.73 & 36 \\
NGC 1097 & SB(s)b & 10.16 & GS-2016B-Q-25 (J. E. Steiner) & 3 $\times$ 565 & 0.46 & 34 \\
NGC 1187 & SB(r)c & 10.93 & GS-2014B-Q-30 (J. E. Steiner) & 3 $\times$ 930 & 0.57 & 59 \\
NGC 1291 & (R)SB0/a(s) & 9.42 & GS-2014B-Q-30 (J. E. Steiner) & 3 $\times$ 630 & 0.50 & 21 \\
NGC 1300 & SB(rs)bc & 11.10 & GS-2013B-Q-20 (T. V. Ricci) & 1 $\times$ 1800 & 0.41 & 29 \\
NGC 1313 & SB(s)d & 9.37 & GS-2012B-Q-52 (R. B. Menezes) & 3 $\times$ 589 & 0.55 & 9.9 \\
NGC 1316 & SAB0 & 9.60 & GS-2013B-Q-20 (T. V. Ricci) & 1 $\times$ 1800 & 0.75 & 76 \\
NGC 1365 & SB(s)b & 10.21 & GS-2014B-Q-30 (J. E. Steiner) & 3 $\times$ 630 & 0.57 & 38 \\
NGC 1380 & SA0 & 11.10 & GS-2008B-Q-21 (J. E. Steiner) & 1 $\times$ 1800 & 1.01 & 100 \\
NGC 1395 & E2 & 11.18 & GS-2013B-Q-20 (T. V. Ricci) & 1 $\times$ 1800 & 0.85 & 100 \\
NGC 1398 & (R')SB(r)ab & 10.60 & GS-2015B-Q-25 (J. E. Steiner) & 3 $\times$ 565 & 0.35 & 49 \\
NGC 1399 & E1 & 10.79 & GS-2008B-Q-21 (J. E. Steiner) & 1 $\times$ 1800 & 0.95 & 97 \\
NGC 1404 & E1 & 11.06 & GS-2008B-Q-21 (J. E. Steiner) & 1 $\times$ 1800 & 0.71 & 65 \\
NGC 1407 & E0 & 10.93 & GS-2013B-Q-20 (T. V. Ricci) & 1 $\times$ 1800 & 0.73 & 89 \\
NGC 1433 & (R')SB(r)ab & 10.68 & GS-2015B-Q-25 (J. E. Steiner) & 3 $\times$ 565 & 0.52 & 21 \\
NGC 1549 & E0-1 & 10.76 & GS-2013B-Q-20 (T. V. Ricci) & 1 $\times$ 1800 & 0.65 & 51 \\
NGC 1553 & SA0 & 10.42 & GS-2014B-Q-30 (J. E. Steiner) & 3 $\times$ 630 & 0.60 & 28 \\
NGC 1559 & SB(s)cd & 10.97 & GS-2013B-Q-74 (R. B. Menezes) & 3 $\times$ 1000 & 0.42 & 24 \\
NGC 1566 & SAB(s)bc & 10.21 & GS-2013B-Q-33 (R. B. Menezes) & 3 $\times$ 910 & 0.72 & 63 \\
NGC 1574 & SA0 & 11.19 & GS-2013B-Q-20 (T. V. Ricci) & 1 $\times$ 1800 & 0.81 & 79 \\
NGC 1672 & SB(s)b & 11.03 & GS-2015B-Q-25 (J. E. Steiner) & 3 $\times$ 565 & 0.40 & 22 \\
NGC 1792 & SA(rs)bc & 10.85 & GS-2014B-Q-30 (J. E. Steiner) & 3 $\times$ 930 & 0.50 & 26 \\
NGC 1808 & (R)SAB(s)a & 10.70 & GS-2015B-Q-25 (J. E. Steiner) & 3 $\times$ 565 & 0.45 & 21 \\
NGC 2442 & SAB(s)bc & 11.16 & GS-2014A-Q-5 (J. E. Steiner) & 3 $\times$ 815 & 0.61 & 59 \\
NGC 2835 & SB(rs)c & 10.95 & GS-2015A-Q-3 (J. E. Steiner) & 3 $\times$ 865 & 0.37 & 16 \\
NGC 2997 & SAB(rs)c & 10.32 & GS-2014B-Q-30 (J. E. Steiner) & 3 $\times$ 930 & 0.52 & 33 \\
NGC 3115 & S0 & 9.98 & GS-2013A-Q-52 (T. V. Ricci) & 1 $\times$ 1800 & 0.67 & 31 \\
NGC 3585 & E6 & 10.93 & GS-2013A-Q-52 (T. V. Ricci) & 1 $\times$ 1800 & 0.58 & 37 \\
NGC 3621 & SA(s)d & 10.03 & GS-2014A-Q-5 (J. E. Steiner) & 3 $\times$ 815 & 0.40 & 12 \\
NGC 3923 & E4-5 & 10.91 & GS-2013A-Q-52 (T. V. Ricci) & 1 $\times$ 1800 & 0.71 & 55 \\
NGC 4030 & SA(s)bc & 11.07 & GN-2014A-Q-3 (J. E. Steiner) & 3 $\times$ 960 & 0.52 & 75 \\
NGC 4594 & SA(s)a & 9.28 & GS-2011A-Q-67 (R. B. Menezes) & 3 $\times$ 595 & 0.73 & 77 \\
NGC 4697 & E6 & 10.11 & GN-2014A-Q-3 (J. E. Steiner) & 3 $\times$ 960 & 0.45 & 27 \\
NGC 4699 & SAB(rs)b & 10.44 & GS-2013A-Q-52 (T. V. Ricci) & 1 $\times$ 1800 & 0.64 & 61 \\
NGC 4753 & I0 & 10.85 & GS-2015A-Q-3 (J. E. Steiner) & 3 $\times$ 565 & 0.47 & 55 \\
NGC 5068 & SAB(rs)cd & 10.53 & GS-2015A-Q-3 (J. E. Steiner) & 3 $\times$ 865 & 0.27 & 8.8 \\
NGC 5102 & SA0 & 10.64 & GS-2015A-Q-3 (J. E. Steiner) & 3 $\times$ 565 & 0.42 & 8.1 \\
NGC 5128 & S0 & 7.89 & GS-2015A-Q-3 (J. E. Steiner) & 3 $\times$ 565 & 0.27 & 4.6 \\
NGC 5236 & SAB(s)c & 8.51 & GS-2014A-Q-5 (J. E. Steiner) & 3 $\times$ 815 & 0.44 & 10 \\ 
NGC 5247 & SA(s)bc & 11.10 & GS-2015A-Q-3 (J. E. Steiner) & 3 $\times$ 865 & 0.39 & 42 \\
NGC 5643 & SAB(rs)c & 10.89 & GS-2014A-Q-5 (J. E. Steiner) & 3 $\times$ 815 & 0.41 & 34 \\
NGC 6744 & SAB(r)bc & 9.24 & GS-2014A-Q-5 (J. E. Steiner) & 3 $\times$ 815 & 0.35 & 13 \\
NGC 7090 & SBc & 11.10 & GS-2015B-Q-25 (J. E. Steiner) & 3 $\times$ 884 & 0.45 & 14 \\
NGC 7213 & SA(s)a & 11.18 & GS-2015A-Q-3 (J. E. Steiner) & 3 $\times$ 565 & 0.70 & 75 \\
NGC 7424 & SAB(rs)cd & 10.99 & GS-2013A-Q-82 (R. B. Menezes) & 3 $\times$ 808 & 0.38 & 21 \\
NGC 7793 & SA(s)d & 9.65 & GS-2016B-Q-25 (J. E. Steiner) & 3 $\times$ 865 & 0.44 & 8.0 \\
\hline
\multicolumn{7}{c}{SIFS}                                                                                            \\ \hline
NGC 1232 & SAB(rs)c & 10.50 & Science verification & 3 $\times$ 1200 & 1.66 & 120 \\
\hline
\end{tabular}
\end{table*}
\end{small}

For the objects without a central AGN but with available Hubble Space Telescope (HST) images, FWHM(5500)$_{treated}$ was estimated using the following procedure: first we assumed an initial value of FWHM(5500) and then constructed a synthetic data cube, using equation (1), containing the PSFs for the entire spectral range of the original data cube. After that, we integrated the resulting data cube along the spectral axis taking into account the response curve of the filter of the HST image available for the object. We convolved the integrated image with the HST image and calculated the $\chi^{2}$ between the result of this convolution and the image of the original data cube integrated along the spectral axis, also taking into account the response curve of the HST filter. Finally we varied the initial value of FWHM(5500) and repeated this procedure in order to minimize the $\chi^{2}$ and determine the best estimate for FWHM(5500)$_{treated}$. 

For the remaining objects without a central AGN and without an available HST image, the following procedure was adopted to determine FWHM(5500)$_{treated}$. First, we estimated the value of the FWHM of the PSF at 6300 \AA~before the data treatment (FWHM(6300)$_{non~treated}$) from the acquisition image. Then, the FWHM of the PSF at 5500 \AA, also before the data treatment (FWHM(5500)$_{non~treated}$), was calculated using equation (1). Finally, the value of FWHM(5500)$_{treated}$ was taken as 79\% of FWHM(5500)$_{non~treated}$ for data cubes in which the Richardson-Lucy deconvolution was applied using 6 iterations and as 68\% of FWHM(5500)$_{non~treated}$ for data cubes in which the Richardson-Lucy deconvolution was applied using 10 iterations. These percentages were estimated based on previous tests performed on data cubes of standard stars. It is worth mentioning that the Richardson-Lucy deconvolution was applied to each frame of the data cube with the width of the PSF given by equation (1). Such a procedure reduced the FWHM values of the PSF at each frame of the data cube, but did not change the dependence of the PSF with the wavelength. For further detail, see \citet{men19}.

\section{Data analysis}\label{sec3}

\subsection{Extraction of the nuclear spectra and subtraction of the stellar continuum}\label{sec31}

The analysis of the nuclear emission of each galaxy in the mini-DIVING$^\mathrm{3D}$ sample, avoiding contaminations from the circumnuclear region, requires the extraction of the nuclear spectrum of each treated data cube. Such a spectrum was extracted from a circular region, centred on the peak of the stellar emission of the galaxy. The radius of this circular region was taken as half of the FWHM of the PSF at the median wavelength of the data cube, which was calculated using equation (1) and the values of the FWHM of the PSF at 5500 \AA~(see Table~\ref{tbl1}). The radius of the extraction region was kept constant along the entire spectral axis of the data cube. The reason is that the alpha exponent in equation (1) is somewhat uncertain. Our tests showed that such an uncertainty usually does not cause problems for the deconvolution procedure, but may introduce significant imprecisions during the nuclear spectra extraction, possibly resulting, in certain cases, in extracting apertures larger than expected and, as a consequence, in contaminations of the nuclear spectra by the circumnuclear emission. 

Based on tests performed using data cubes of standard stars, we verified that, after the Richardson-Lucy deconvolution (the last step of our data treatment procedure), performed assuming a PSF given by a Moffat function, the resulting PSF of the data cube has a nearly Gaussian shape \citep{men19} and the flux of a point-like source extracted from a circular region (centred at the source) with a radius equal to half of the FWHM of the PSF corresponds to $\sim 47 \%$ of the total flux of the source. Considering that, we multiplied all the extracted spectra by a constant, in order to correct the flux values for the inaccuracy introduced by the extraction procedure involving a circular region.

We corrected the extracted spectra for the Galactic extinction, using the $A_V$ from \citet{sch11} and the extinction law of \citet{car89}. 

The subtraction of the stellar continuum from the extracted nuclear spectra, which is necessary for a reliable analysis of the emission-line properties, was performed with the Penalized Pixel Fitting technique (pPXF - \citealt{cap17}). This method corresponds to the fitting of the stellar spectrum of a given object by a combination of template stellar population spectra from a base. Such template spectra are convolved with a Gauss-Hermite expansion, in order to reproduce the widths and profiles of the stellar absorption lines. We used a base of stellar population spectra with simple stellar population (SSP) models obtained with the code described in \citet{vaz10}, based on the Medium-resolution Isaac Newton Telescope Library of Empirical Spectra (MILES; \citealt{san06}). The base was elaborated using a Kroupa universal initial mass function, with a slope of 1.3, and the isochrones of the Bag of Stellar Tracks and Isochrones (BaSTI; \citealt{hid18}). The final base has a range of stellar metallicities ([M/H]) from -0.66 to +0.40 and a range of stellar population ages from $3 \times 10^7$ yr to $1.3 \times 10^{10}$ yr. The pPXF technique provides the values of different parameters related to the stellar kinematics of the observed object, such as the stellar radial velocity, the stellar velocity dispersion, and the $h_3$ and $h_4$ Gauss-Hermite coefficients. Besides that, it also provides a synthetic stellar spectrum corresponding to the fit obtained with the procedure. We applied the pPXF method to the nuclear spectrum extracted from each galaxy of the sample and then subtracted from it the resulting synthetic stellar spectrum, which resulted in spectra containing only emission lines. 

In this work, we included additive Legendre polynomials to the pPXF fits applied to the extracted nuclear spectra, to remove possible low-frequency spectral features remaining from the data reduction and, therefore, to improve the quality of the fits. The order of such polynomials was kept in the range between 4 and 10 and was determined for each object, with the main goal of achieving a precise starlight subtraction, without introducing artefacts to the remaining emission lines. Since only the stellar continuum and absorption lines are fitted by the pPXF method, all the wavelength intervals possibly containing the emission lines H$\beta$, [O \textsc{iii}]$\lambda$5007, [O \textsc{i}]$\lambda$6300, [N \textsc{ii}]$\lambda\lambda$6548,6583, H$\alpha$, and [S \textsc{ii}]$\lambda\lambda$6716,6731 (which are essential for the analyses in the following sections) were masked. The specific boundaries of these intervals were established for each object, at the borders of the emission lines. The typical width of a wavelength interval used to mask an emission line with a given FWHM was $\sim 4 \times FWHM$. More intervals were added to objects showing additional emission lines. As explained in paper I, the GMOS/IFU observations of early-type galaxies (with morphological types E, S0 - Sb) in the DIVING$^\mathrm{3D}$ survey were taken with the B600 grating, which provided a spectral coverage of 4250 - 7000 \AA. On the other hand, the late-type galaxies (with morphological types Sbc - Sd) in the DIVING$^\mathrm{3D}$ sample were observed with the R831 grating, resulting in a wavelength coverage of 4800 - 6890 \AA. The SIFS observations were taken with a grating of 700 lmm$^{-1}$, providing a spectral coverage of 4500 - 7300 \AA. The pPXF fits were applied to the entire spectral range of the extracted nuclear spectra.

We performed tests of running pPXF fits on the extracted nuclear spectra, adding Gaussian random distributions of noise (which were also obtained from the extracted spectra). At the end, we concluded that these different runs resulted in fit residuals without significant differences, as the $\chi^2$ of the fits were very similar and the same residual features (usually associated with stellar absorption lines not properly subtracted) were detected in all runs. This test revealed that the error related to the problem of reproducing a real stellar spectrum with the use of a limited base is more significant than the error associated with the spectral noise. This information is also relevant for the determination of the uncertainties of integrated fluxes of emission lines, as discussed in Section~\ref{sec32}. 

Appendix A shows the nuclear spectra extracted from the data cubes of the mini-DIVING$^\mathrm{3D}$ sample, together with the fits provided by the pPXF technique and the fit residuals.

\subsection{Correction of the interstellar extinction and calculation of the emission-line ratios}\label{sec32}

In order to correct the starlight-subtracted nuclear spectra for the interstellar extinction, at the observed objects, we calculated the integrated fluxes of the H$\alpha$ and H$\beta$ emission lines and also the H$\alpha$/H$\beta$ emission-line ratio (Balmer decrement). For the galaxies without blended emission lines, these fluxes were obtained via direct integration of the emission lines. On the other hand, for the objects with blended emission lines, we adopted the following approach: first we fitted the [S \textsc{ii}]$\lambda\lambda$6716,6731 emission lines with a sum of two sets of Gaussian functions. Each of these sets contained two Gaussians with the same width and radial velocity. The result was that each [S \textsc{ii}] emission line was fitted with a sum of two Gaussian functions, each one with a specific width and radial velocity. After that, we fitted the H$\alpha$+[N \textsc{ii}]$\lambda\lambda$6548,6583 emission lines with a sum of two sets of three Gaussian functions. In this case, each set contained three Gaussians with the same width and radial velocity of the corresponding set used to fit the [S \textsc{ii}] emission lines. In other words, we used the [S \textsc{ii}]$\lambda\lambda$6716,6731 emission lines as an empirical template to fit the H$\alpha$+[N \textsc{ii}]$\lambda\lambda$6548,6583 emission lines. We also took into account, for each set of Gaussians, the theoretical ratio of [N \textsc{ii}]$\lambda$6583/[N \textsc{ii}]$\lambda$6548 = 3.06 \citep{ost06}. 

For the objects in which the sets of narrow Gaussians were not sufficient to reproduce the profile of the H$\alpha$+[N \textsc{ii}]$\lambda\lambda$6548,6583 emission lines, a broad Gaussian was added to the fit, in order to reproduce a broad component of H$\alpha$. This procedure (also adopted by \citealt{ho97}) provided the integrated flux of the narrow component of H$\alpha$ (corresponding to the sum of the integrated fluxes of the two narrow Gaussians used to fit H$\alpha$). A few variations of this procedure, however, had to be applied to specific objects. For NGC 720, NGC 1097, NGC 1365, NGC 1399, NGC 1404, NGC 1574, NGC 3585, and NGC 7213, only one Gaussian was necessary to fit the narrow component of each emission line. In the case of NGC 1404, the [S \textsc{ii}] lines were not detected and, because of that, the Gaussian fits were only applied to the H$\alpha$+[N \textsc{ii}]$\lambda\lambda$6548,6583 emission lines. For NGC 1365, the broad component of H$\alpha$ was fitted by two broad Gaussians. For NGC 1097 (e.g. \citealt{sto93,sto03}) and NGC 7213 (e.g. \citealt{phi79,sch17}), a broad emission from a relativistic disc was detected in H$\alpha$ and H$\beta$. Such an emission was modelled and subtracted using the formalism of \citet{che89}. Finally, due to the fact that the presence of a broad component of H$\alpha$ is uncertain for NGC 720 and NGC 1291, two versions (with and without a broad H$\alpha$) of the Gaussian fits were applied to the H$\alpha$+[N \textsc{ii}]$\lambda\lambda$6548,6583 emission lines in the nuclear spectra of these galaxies. 

The broad component of H$\alpha$ was not taken into account in the correction of the interstellar extinction. The integrated flux of the narrow component of H$\beta$ was obtained via direct integration, except for the objects with a broad component of this line. In such cases (NGC 1068, NGC 1097, NGC 1365, NGC 1566, and NGC 7213), we fitted the H$\beta$ emission line with a sum of one (NGC 1097 and NGC 7213) or two (NGC 1068, NGC 1365, and NGC 1566) narrow and one broad Gaussians, each one with the same widths and radial velocities of the Gaussians used to fit the H$\alpha$+[N \textsc{ii}]$\lambda\lambda$6548,6583 lines. The Balmer decrement was calculated by dividing the integrated flux of the narrow component of H$\alpha$ by the integrated flux of the narrow component of H$\beta$ (see Table~\ref{tbl2}).

The correction of the interstellar extinction was applied to the starlight-subtracted nuclear spectra taking into account the obtained values of the Balmer decrement, the extinction law of \citet{car89}, and assuming an intrinsic Balmer decrement of 3.10 for objects with emission line ratios characteristic of Seyferts or LINERs and of 2.86 for the remaining objects \citep{ost06}. For the galaxies in which the H$\beta$ emission line was not detected, the Balmer decrement was not determined and, as consequence, no interstellar extinction was applied. Appendix B shows all the Gaussian fits applied to the spectra of the objects with blended emission lines, after the correction of the interstellar extinction.

Finally, the emission-line ratios [N \textsc{ii}]$\lambda$6583/H$\alpha$, [S \textsc{ii}]($\lambda$6716 + $\lambda$6731)/H$\alpha$, [O \textsc{i}]$\lambda$6300/H$\alpha$, and [O \textsc{iii}]$\lambda$5007/H$\beta$ were calculated. For the objects without blended emission lines, the required fluxes of the emission lines were determined via direct integration. However, for the objects with blended emission lines, the Gaussian fitting methodology described above was applied. The uncertainty for the integrated flux of each line, determined via direct integration, was estimated using the following procedure: we established three wavelength intervals for the integration of the line: the first one includes the line and a small margin close to its wings, the second is a shorter interval, containing the line but essentially with no margin, and the third is a larger interval, containing the line and also a larger margin. We also constructed a histogram of the flux values within a wavelength interval without emission lines and then fitted such a histogram with a Gaussian function. After that, we created Gaussian distributions of random noise with the same width of the Gaussian fitted to the original histogram. These random distributions were added to the emission line and the resulting ``noisy'' lines were integrated using the three wavelength intervals mentioned above. The uncertainty of the integrated flux of the emission line was taken as the standard deviation of all integrated fluxes obtained with this procedure. 

In the case of objects with blended emission lines, the approach for estimating the uncertainties of the fluxes of the lines was analogous: the Gaussian fitting of the blended lines was repeated after adding Gaussian distributions of random noise to the lines and assuming different wavelength intervals in the process. The approach of adding Gaussian random noise to the data and integrating the line with different wavelength intervals allows us to obtain an uncertainty that takes into account not only the effect of the spectral noise, but also the effect of the residuals from the starlight subtraction (which are more significant when larger wavelength intervals are used for the integration). For the objects with blended emission lines, the obtained uncertainties also take into account degeneracies and instabilities of the line-fitting procedure with Gaussians. We used, for all objects, the same criterion for determining the wavelength interval for the integration of each emission line, which included the line and a small margin close to its wings. The shorter and larger intervals mentioned above were only used for estimating the uncertainties of the integrated fluxes. 

It is worth emphasizing that, as long as all the fluxes of the emission lines are measured with the same process, the effect of using a shorter or a larger wavelength interval for the integration of all of them is not random. Using, for example, a larger wavelength interval for the integration will result in lower fluxes for, at least, certain groups of objects (which is caused by the inclusion, in these larger intervals, of residuals from the starlight subtraction). We also verified that, for most of the objects, the effect of the starlight subtraction residuals (which is not random) on the uncertainties is significantly higher than the effect of the spectral noise (which is random). This is consistent with what was discussed in Section~\ref{sec31}. The uncertainties of the [N \textsc{ii}]$\lambda$6583/H$\alpha$, [S \textsc{ii}]($\lambda$6716 + $\lambda$6731)/H$\alpha$, [O \textsc{i}]$\lambda$6300/H$\alpha$, and [O \textsc{iii}]$\lambda$5007/H$\beta$ emission-line ratios were determined with a simple propagation, using the variance formula, based on the uncertainties obtained for the integrated fluxes of the individual emission lines.

\begin{table*}
\centering
\caption{Column (1): identification. Column (2): Balmer decrement of the starlight-subtracted spectra. Column (3): luminosity of the narrow H$\alpha$ component, calculated, after the interstellar extinction correction, using the distance values given in paper I. Column (4): ratio of the broad (when detected) and narrow H$\alpha$ components. Column (5): ratio [S \textsc{ii}]$\lambda$6716/[S \textsc{ii}]$\lambda$6731, after the interstellar extinction correction. \label{tbl2}}
\begin{tabular}{ccccc}
\hline
Galaxy & H$\alpha$/H$\beta$ & L$_{H\alpha}(Narrow)$ (10$^{38}$) erg s$^{-1}$ & H$\alpha(Broad)$/H$\alpha(Narrow)$ & [S \textsc{ii}]$\lambda$6716/[S \textsc{ii}]$\lambda$6731 \\
    (1) & (2) & (3) & (4) & (5) 
\\ \hline
NGC 134 & $8 \pm 4$ & $4.99 \pm 0.19$ & - & $0.98 \pm 0.10$ \\
NGC 157 & $11 \pm 9$ & $3.55 \pm 0.08$ & - & $1.33 \pm 0.28$ \\
NGC 247 & - & $(2.0 \pm 0.5) \times 10^{-3}$ & - & - \\
NGC 253 & $5.7 \pm 0.3$ & $(5.863 \pm 0.013) \times 10^{-1}$ & - & $1.05 \pm 0.04$ \\
NGC 300 & - & - & - & - \\
NGC 613 & $6.3 \pm 0.4$ & $(5.002 \pm 0.029) \times 10$ & - & $1.18 \pm 0.05$ \\
NGC 720(n)* & - & $4.07 \pm 0.10$ & - & $0.91 \pm 0.23$ \\
NGC 720(b)* & - & $2.54 \pm 0.21$ & $1.8 \pm 0.4$ &$0.91 \pm 0.23$ \\
NGC 908 & $4.19 \pm 0.25$ & $(1.30 \pm 0.04) \times 10$ & - & $1.03 \pm 0.06$ \\
NGC 936 & $4.8 \pm 0.6$ & $(1.59 \pm 0.04) \times 10$ & - & $1.16 \pm 0.21$ \\
NGC 1068 & $3.383 \pm 0.013$ & $(1.264 \pm 0.003) \times 10^2$ & $4.537 \pm 0.011$ & $0.68 \pm 0.03$ \\
NGC 1097** & $5.2 \pm 0.4$ & $3.94 \pm 0.05$ & $2.51 \pm 0.07$ & $1.12 \pm 0.13$ \\
NGC 1187 & $4.277 \pm 0.025$ & $(1.740 \pm 0.008) \times 10^2$ & - & $0.862 \pm 0.023$ \\
NGC 1232 & - & - & - & - \\
NGC 1291(n)* & $5.0 \pm 0.5$ & $(6.06 \pm 0.17) \times 10$ & - & $1.14 \pm 0.05$ \\ 
NGC 1291(b)* & $4.3 \pm 0.4$ & $(3.77 \pm 0.12) \times 10$ & $0.72 \pm 0.07$ & $1.14 \pm 0.05$ \\
NGC 1300 & $6.6 \pm 0.7$ & $4.43 \pm 0.13$ & - & $1.09 \pm 0.09$ \\
NGC 1313 & $3.36 \pm 0.13$ & $(1.84 \pm 0.06) \times 10^{-1}$ & - & $1.44 \pm 0.08$ \\
NGC 1316 & $8.7 \pm 1.2$ & $(1.63 \pm 0.04) \times 10^2$ & - & $1.23 \pm 0.14$ \\
NGC 1365 & $3.10 \pm 0.09$ & $4.697 \pm 0.022$ & $37.4 \pm 0.3$ & $0.989 \pm 0.021$ \\
NGC 1380 & $8 \pm 3$ & $(3.54 \pm 0.06) \times 10$ & - & $1.13 \pm 0.05$ \\
NGC 1395 & - & $2.50 \pm 0.27$ & - & $1.44 \pm 0.07$ \\
NGC 1398 & $3.4 \pm 0.6$ & $(7.67 \pm 0.22) \times 10^{-1}$ & - & $1.34 \pm 0.11$ \\
NGC 1399 & - & $(1.50 \pm 0.17) \times 10$ & - & - \\
NGC 1404 & - & $(7.9 \pm 0.7) \times 10^{-1}$ & - & - \\
NGC 1407 & - & - & - & - \\
NGC 1433 & $5.32 \pm 0.13$ & $4.07 \pm 0.09$ & - & $1.15 \pm 0.07$ \\
NGC 1549 & - & - & - & - \\
NGC 1553 & $14 \pm 7$ & $(3.26 \pm 0.22) \times 10^2$ & - & $1.03 \pm 0.05$ \\
NGC 1559 & $5.1 \pm 1.5$ & $(5.86 \pm 0.26) \times 10^{-1}$ & - & $1.4 \pm 0.5$ \\
NGC 1566 & $3.72 \pm 0.04$ & $(3.822 \pm 0.009) \times 10$ & $3.473 \pm 0.014$ & $0.84 \pm 0.11$ \\
NGC 1574 & $5.4 \pm 1.5$ & $(1.274 \pm 0.027) \times 10$ & $1.19 \pm 0.15$ & $0.97 \pm 0.13$ \\
NGC 1672 & $7.94 \pm 0.27$ & $9.95 \pm 0.18$ & - & $1.02 \pm 0.05$ \\
NGC 1792 & $7.8 \pm 1.1$ & $7.30 \pm 0.04$ & - & $1.03 \pm 0.03$ \\
NGC 1808 & $8.10 \pm 0.25$ & $(1.447 \pm 0.003) \times 10^2$ & - & $0.63 \pm 0.05$ \\
NGC 2442 & $5.06 \pm 0.15$ & $(3.60 \pm 0.06) \times 10$ & - & $0.97 \pm 0.05$ \\ 
NGC 2835 & - & $(6.2 \pm 0.4) \times 10^{-3}$ & - & $1.23 \pm 0.27$ \\
NGC 2997 & $2.8 \pm 1.7$ & $(2.38 \pm 0.11) \times 10^{-1}$ & - & $1.17 \pm 0.16$ \\
NGC 3115 & - & $(4.2 \pm 0.5) \times 10^{-1}$ & $0.98 \pm 0.28$ & $0.84 \pm 0.18$ \\
NGC 3585 & $6 \pm 3$ & $(9.2 \pm 0.4) \times 10$ & - & $1.37 \pm 0.07$ \\
NGC 3621 & - & $(1.03 \pm 0.26) \times 10^{-2}$ & - & - \\
NGC 3923 & - & - & - & - \\
NGC 4030 & $6 \pm 3$ & $4.30 \pm 0.24$ & - & $1.4 \pm 1.1$ \\
NGC 4594 & $4.6 \pm 0.5$ & $(8.78 \pm 0.27) \times 10$ & $2.24 \pm 0.13$ & $0.84 \pm 0.05$ \\
NGC 4697 & $4.8 \pm 2.2$ & $(6.3 \pm 0.5) \times 10^{-1}$ & - & $1.6 \pm 0.4$ \\
NGC 4699 & $8 \pm 4$ & $(4.28 \pm 0.03) \times 10^2$ & - & $1.12 \pm 0.11$ \\
NGC 4753 & $5.6 \pm 1.1$ & $(5.85 \pm 0.17) \times 10$ & - & $1.06 \pm 0.08$ \\
NGC 5068 & $3.11 \pm 0.27$ & $(4.09 \pm 0.15) \times 10^{-2}$ & - & $1.32 \pm 0.10$ \\
NGC 5102 & $3.3 \pm 0.7$ & $(2.59 \pm 0.06) \times 10$ & - & - \\
NGC 5128 & $5.3 \pm 1.7$ & $(1.41 \pm 0.04) \times 10^{-2}$ & - & $1.10 \pm 0.09$ \\
NGC 5236 & $2.70 \pm 0.14$ & $(3.29 \pm 0.04) \times 10^{-1}$ & - & $1.04 \pm 0.06$ \\
NGC 5247 & - & $(1.98 \pm 0.15) \times 10^{-1}$ & - & $1.7 \pm 1.0$ \\
NGC 5643 & $4.76 \pm 0.08$ & $(1.480 \pm 0.004) \times 10^2$ & - & $0.85 \pm 0.04$ \\
NGC 6744 & $4.06 \pm 0.21$ & $(1.857 \pm 0.011) \times 10^{-1}$ & - & $1.27 \pm 0.07$ \\
NGC 7090 & $2.8 \pm 0.8$ & $(6.90 \pm 0.06) \times 10^{-2}$ & - & $1.35 \pm 0.17$ \\
NGC 7213 & $3.11 \pm 0.08$ & $(1.038 \pm 0.003) \times 10^4$ & $11.40 \pm 0.10$ & $0.899 \pm 0.005$ \\
NGC 7424 & $3.4 \pm 0.6$ & $6.9 \pm 0.3$ & - & $1.4 \pm 0.4$ \\
NGC 7793 & $6.4 \pm 2.9$ & $(1.07 \pm 0.07) \times 10^{-1}$ & - & $0.9 \pm 0.3$ \\
IC 1459 & $3.94 \pm 0.14$ & $(1.786 \pm 0.020) \times 10^3$ & $0.45 \pm 0.03$ & $0.951 \pm 0.010$ \\
\hline
\end{tabular}
\\
\begin{tabular}{c}
* For these galaxies, the Gaussian fits were applied to the H$\alpha$+[N \textsc{ii}]$\lambda\lambda$6548,6583 emission lines with (b) and without (n) a broad component of H$\alpha$ \\
** For these galaxies, the contribution of the broad H$\alpha$ component for the values in column (5) does not include the emission from the relativistic disc

\end{tabular}
\end{table*}

\begin{table*}
\centering
\caption{Emission-line ratios (2) [N \textsc{ii}]$\lambda$6583/H$\alpha$, (3) [S \textsc{ii}]($\lambda$6716 + $\lambda$6731)/H$\alpha$, (4) [O \textsc{i}]$\lambda$6300/H$\alpha$, and (5) [O \textsc{iii}]$\lambda$5007/H$\beta$ of the starlight-subtracted nuclear spectra, after the interstellar extinction correction. The identifications of the galaxies are shown in column (1).\label{tbl3}}
\begin{tabular}{ccccc}
\hline
Galaxy & [N \textsc{ii}]$\lambda$6583/H$\alpha$ & [S \textsc{ii}]($\lambda$6716 + $\lambda$6731)/H$\alpha$ & [O \textsc{i}]$\lambda$6300/H$\alpha$ & [O \textsc{iii}]$\lambda$5007/H$\beta$ \\
    (1) & (2) & (3) & (4) & (5) 
\\ \hline
NGC 134 & $1.79 \pm 0.08$ & $0.91 \pm 0.06$ & - & $2.0 \pm 1.0$ \\
NGC 157 & $0.587 \pm 0.024$ & $0.252 \pm 0.027$ & $0.042 \pm 0.027$ & $0.7 \pm 0.5$ \\
NGC 247 & $0.3^{+0.5}_{-0.3}$ & - & - & - \\
NGC 253 & $0.7458 \pm 0.0027$ & $0.398 \pm 0.004$ & $0.052 \pm 0.005$ & $0.273 \pm 0.021$ \\
NGC 300 & - & - & - & - \\
NGC 613 & $1.797 \pm 0.012$ & $0.925 \pm 0.014$ & $0.275 \pm 0.012$ & $1.08 \pm 0.08$ \\
NGC 720(n)* & $0.74 \pm 0.03$ & $0.63 \pm 0.05$ & - & - \\
NGC 720(b)* & $0.65 \pm 0.09$ & $1.01 \pm 0.12$ & - & - \\
NGC 908 & $0.320 \pm 0.013$ & $0.161 \pm 0.006$ & $0.0049 \pm 0.0013$ & $0.024 \pm 0.014$ \\
NGC 936 & $1.19 \pm 0.04$ & $0.759 \pm 0.020$ & $0.133 \pm 0.011$ & $1.01 \pm 0.35$ \\
NGC 1068 & $2.158 \pm 0.005$ & $0.4768 \pm 0.0014$ & $0.403 \pm 0.028$ & $18.61 \pm 0.06$ \\
NGC 1097 & $2.213 \pm 0.021$ & $0.873 \pm 0.021$ & $0.36 \pm 0.04$ & $3.5 \pm 0.3$ \\
NGC 1187 & $0.450 \pm 0.009$ & $0.185 \pm 0.003$ & $0.0066 \pm 0.0010$ & $0.0761 \pm 0.0024$ \\
NGC 1232 & - & - & - & - \\
NGC 1291(n)* & $1.90 \pm 0.06$ & $0.98 \pm 0.03$ & $0.259 \pm 0.020$ & $2.7 \pm 0.3$ \\
NGC 1291(b)* & $2.02 \pm 0.07$ & $1.14 \pm 0.04$ & $0.295 \pm 0.023$ & $2.7 \pm 0.3$ \\
NGC 1300 & $1.29 \pm 0.05$ & $1.04 \pm 0.04$ & $0.281 \pm 0.021$ & $2.8 \pm 0.3$ \\
NGC 1313 & $0.123 \pm 0.007$ & $0.188 \pm 0.008$ & $0.006 \pm 0.004$ & $2.06 \pm 0.07$ \\
NGC 1316 & $1.59 \pm 0.04$ & $0.94 \pm 0.03$ & $0.125 \pm 0.015$ & $3.0 \pm 0.5$ \\
NGC 1365 & $1.049 \pm 0.006$ & $0.276 \pm 0.003$ & $0.06 \pm 0.04$ & $6.5 \pm 0.3$ \\
NGC 1380 & $1.58 \pm 0.03$ & $0.84 \pm 0.04$ & $0.10 \pm 0.05$ & $2.2 \pm 0.9$ \\
NGC 1395 & $1.23 \pm 0.18$ & $0.65 \pm 0.07$ & - & - \\
NGC 1398 & $2.09 \pm 0.13$ & $0.89 \pm 0.05$ & - & $3.9 \pm 1.0$ \\
NGC 1399 & $0.61 \pm 0.12$ & - & - & - \\
NGC 1404 & $0.33 \pm 0.06$ & - & - & - \\
NGC 1407 & - & - & - & - \\
NGC 1433 & $0.96 \pm 0.04$ & $0.632 \pm 0.025$ & $0.060 \pm 0.007$ & $1.96 \pm 0.07$ \\
NGC 1549 & - & - & - & - \\
NGC 1553 & $2.13 \pm 0.16$ & $0.71 \pm 0.07$ & $0.50 \pm 0.15$ & $4.8 \pm 2.6$ \\
NGC 1559 & $0.34 \pm 0.03$ & $0.23 \pm 0.03$ & - & $0.32 \pm 0.20$ \\
NGC 1566 & $1.090 \pm 0.004$ & $0.769 \pm 0.016$ & $0.235 \pm 0.010$ & $5.98 \pm 0.14$ \\
NGC 1574 & $1.68 \pm 0.04$ & $0.40 \pm 0.03$ & - & $2.4 \pm 0.7$ \\
NGC 1672 & $1.207 \pm 0.025$ & $0.515 \pm 0.016$ & $0.094 \pm 0.007$ & $4.51 \pm 0.27$ \\
NGC 1792 & $0.444 \pm 0.019$ & $0.200 \pm 0.003$ & $0.024 \pm 0.007$ & $0.48 \pm 0.13$ \\
NGC 1808 & $0.9273 \pm 0.0020$ & $0.1450 \pm 0.0018$ & $0.0179 \pm 0.0022$ & $0.213 \pm 0.013$ \\
NGC 2442 & $2.23 \pm 0.04$ & $1.703 \pm 0.028$ & $0.494 \pm 0.013$ & $2.17 \pm 0.05$ \\
NGC 2835 & $0.81 \pm 0.07$ & $0.78 \pm 0.09$ & - & - \\
NGC 2997 & $1.69 \pm 0.13$ & $1.17 \pm 0.10$ & $0.17 \pm 0.07$ & $1.1 \pm 0.9$ \\
NGC 3115 & $0.59 \pm 0.10$ & $0.64 \pm 0.10$ & - & - \\
NGC 3585 & $0.80 \pm 0.05$ & $0.44 \pm 0.06$ & - & $1.2 \pm 0.6$ \\
NGC 3621 & $1.8 \pm 0.5$ & - & - & - \\
NGC 3923 & - & - & - & - \\
NGC 4030 & $1.08 \pm 0.06$ & $0.32 \pm 0.04$ & - & $1.6 \pm 0.9$ \\
NGC 4594 & $2.94 \pm 0.10$ & $1.54 \pm 0.06$ & $0.58 \pm 0.03$ & $2.46 \pm 0.25$ \\
NGC 4697 & $1.51 \pm 0.16$ & $0.42 \pm 0.06$ & - & $2.7 \pm 1.2$ \\
NGC 4699 & $2.17 \pm 0.06$ & $0.67 \pm 0.03$ & $0.22 \pm 0.07$ & $4.4 \pm 2.8$ \\
NGC 4753 & $3.12 \pm 0.10$ & $0.78 \pm 0.04$ & - & $5.4 \pm 2.7$ \\
NGC 5068 & $0.332 \pm 0.025$ & $0.58 \pm 0.03$ & $0.059 \pm 0.024$ & $0.72 \pm 0.09$ \\
NGC 5102 & $0.69 \pm 0.05$ & - & - & - \\
NGC 5128 & $1.48 \pm 0.06$ & $1.24 \pm 0.06$ & $0.20 \pm 0.11$ & $4.3 \pm 1.5$ \\
NGC 5236 & $1.165 \pm 0.026$ & $0.858 \pm 0.028$ & $0.158 \pm 0.008$ & $0.93 \pm 0.10$ \\
NGC 5247 & $0.86 \pm 0.10$ & $0.46 \pm 0.13$ & - & - \\
NGC 5643 & $0.989 \pm 0.004$ & $0.5317 \pm 0.0028$ & $0.159 \pm 0.009$ & $12.17 \pm 0.24$ \\
NGC 6744 & $1.068 \pm 0.017$ & $0.849 \pm 0.024$ & $0.08 \pm 0.05$ & $1.94 \pm 0.18$ \\
NGC 7090 & $0.24 \pm 0.05$ & $0.63 \pm 0.10$ & - & $0.20 \pm 0.09$ \\
NGC 7213 & $1.740 \pm 0.006$ & $2.472 \pm 0.010$ & $1.047 \pm 0.009$ & $1.05 \pm 0.04$ \\
NGC 7424 & $0.35 \pm 0.04$ & $0.41 \pm 0.05$ & - & $0.22 \pm 0.11$ \\
NGC 7793 & $0.32 \pm 0.03$ & $0.43 \pm 0.05$ & - & $1.5 \pm 0.5$ \\
IC 1459 & $2.49 \pm 0.03$ & $1.281 \pm 0.015$ &¨$0.427 \pm 0.012$ & $1.96 \pm 0.12$ \\
\hline
\end{tabular}
\\
\begin{tabular}{c}
* For these galaxies, the Gaussian fits were applied to the H$\alpha$+[N \textsc{ii}]$\lambda\lambda$6548,6583 emission lines with (b) and without (n) a broad component of H$\alpha$
\end{tabular}
\end{table*}

\section{Results}\label{sec4}

\begin{figure*}
\begin{center}

   \includegraphics[scale=0.11]{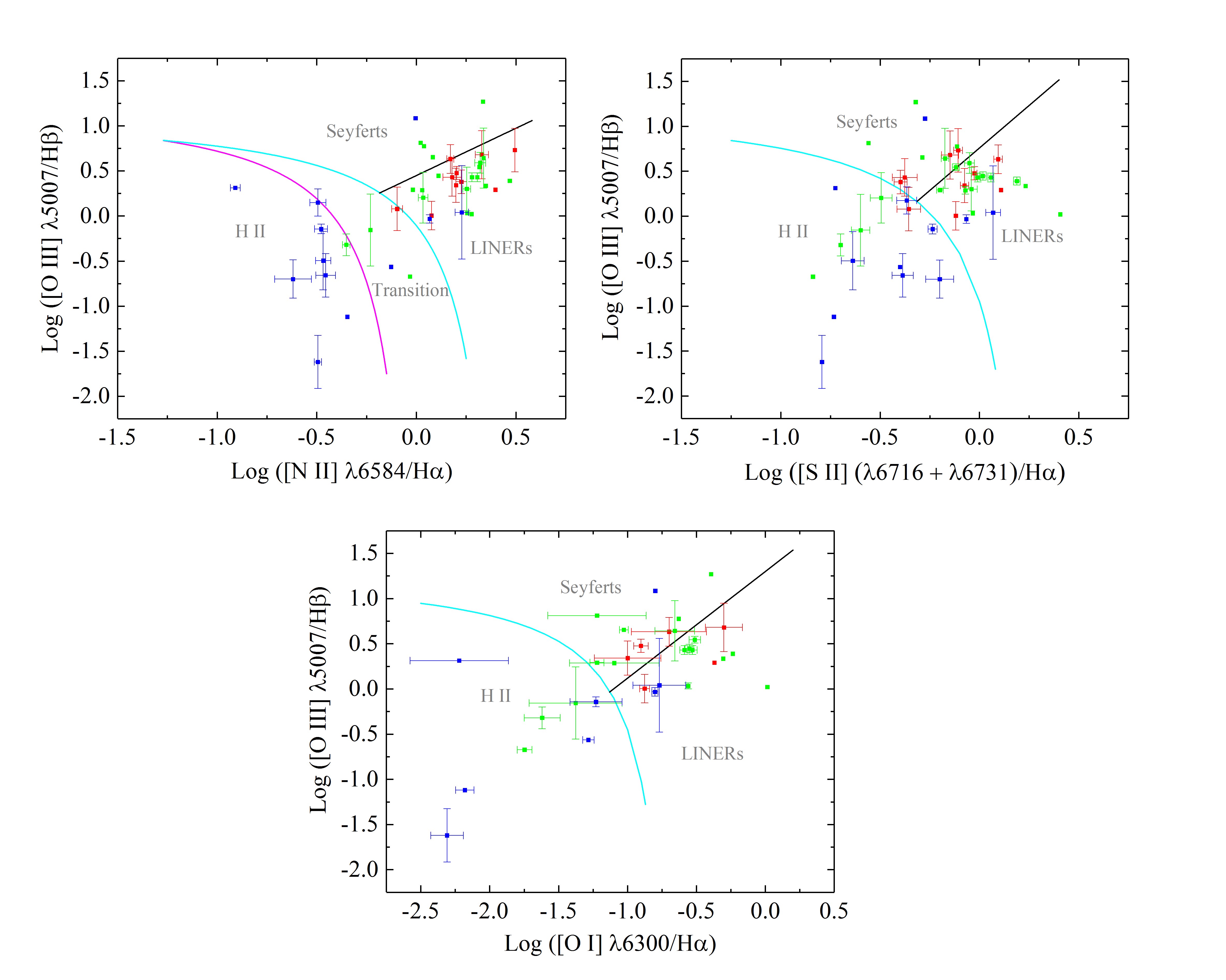} 
  \caption{Diagnostic diagrams of the nuclear spectra of the objects in the mini-DIVING$^\mathrm{3D}$ sample. The red, green, and blue points represent early-type (E+S0), ``early-spiral'' (S0/a + Sa + Sab + Sb + Sbc), and ``late-spiral'' (Sc + Scd + Sd) galaxies, respectively. The magenta curve corresponds to the empirical division between H II regions and AGNs obtained by \citet{kau03}, the cyan curves represent the maximum limit for the ionization by a starburst obtained by \citet{kew01}, the black curves in the diagrams of [O \textsc{iii}]$\lambda$5007/H$\beta$ $\times$ [S \textsc{ii}]($\lambda$6716 + $\lambda$6731)/H$\alpha$ and [O \textsc{iii}]$\lambda$5007/H$\beta$ $\times$ [O \textsc{i}]$\lambda$6300/H$\alpha$ correspond to the division between LINERs and Seyferts determined by \citet{kew06}, and the black curve in the diagram of [O \textsc{iii}]$\lambda$5007/H$\beta$ $\times$ [N \textsc{ii}]$\lambda$6584/H$\alpha$ represents the division between LINERs and Seyferts obtained by \citet{sch07}.}\label{fig2}
\end{center}
\end{figure*}

Table~\ref{tbl2} shows the values of the following quantities calculated from the starlight-subtracted nuclear spectra: the Balmer decrement, before the interstellar extinction correction; the luminosity of the narrow component of the H$\alpha$ emission line, $L_{H\alpha}(Narrow)$, after the interstellar extinction correction (for the cases in which such a correction was applied); the ratio between the integrated fluxes of the broad and narrow components of H$\alpha$, H$\alpha(Broad)$/H$\alpha(Narrow)$, whenever the former was detected; the ratio [S \textsc{ii}]$\lambda$6716/[S \textsc{ii}]$\lambda$6731, determined also after the interstellar extinction correction (again, for the cases in which such a correction was applied). It is worth mentioning that the uncertainties of the $L_{H\alpha}(Narrow)$ values were calculated taking into account only the uncertainties of the integrated fluxes of these H$\alpha$ components, but not the uncertainties of the distances or of the Balmer decrements (used for the correction of the interstellar extinction).

The nuclear spectra of NGC 1316, NGC 1672, NGC 1792, and NGC 1808 showed considerably high (> 7.5) values of the Balmer decrement, indicating significant interstellar extinction. High values of H$\alpha$/H$\beta$ were also detected in the nuclear spectra of NGC 134, NGC 157, NGC 1380, NGC 1553, NGC 3585, NGC 4030, and NGC 4699 but the corresponding uncertainties are too high for these values to be taken as reliable. For these objects, the interstellar extinction correction and the resulting $L_{H\alpha}$ values are also very uncertain, although, as mentioned before, the uncertainties of $L_{H\alpha}$ do not take into account the uncertainties of H$\alpha$/H$\beta$.

Table~\ref{tbl3} shows the values of the main emission-line ratios determined for the starlight-subtracted nuclear spectra (after the correction of the interstellar extinction). Based on these values of emission-line ratios, we constructed the diagnostic diagrams of [O \textsc{iii}]$\lambda$5007/H$\beta$ $\times$ [N \textsc{ii}]$\lambda$6583/H$\alpha$, [O \textsc{iii}]$\lambda$5007/H$\beta$ $\times$ [S \textsc{ii}]($\lambda$6716 + $\lambda$6731)/H$\alpha$, and [O \textsc{iii}]$\lambda$5007/H$\beta$ $\times$ [O \textsc{i}]$\lambda$6300/H$\alpha$ \citep{bal81,vei87}. The results are shown in Fig.~\ref{fig2}. The [O \textsc{iii}]$\lambda$5007/H$\beta$ $\times$ [N \textsc{ii}]$\lambda$6583/H$\alpha$ diagram deserves special attention. A group of objects falls on the branches of LINERs and Seyferts and another group of objects falls on the branch of H II regions. However, only a few galaxies are classified as transition objects (falling on the branch between the maximum limit for the ionization by a starburst, determined by \citealt{kew01}, and the empirical division between H II regions and AGNs, obtained by \citealt{kau03}). In other words, there is an apparent dichotomy in that diagnostic diagram, which was not detected in previous studies. This topic will be discussed in further detail in Section~\ref{sec5}.  

Using the diagnostic diagrams in Fig.~\ref{fig2}, we classified the nuclear emission-line spectra of the galaxies in the mini-DIVING$^\mathrm{3D}$ sample as characteristic of Seyferts, LINERs, H II regions, or transition objects (this last classification can only be obtained from the [O \textsc{iii}]$\lambda$5007/H$\beta$ $\times$ [N \textsc{ii}]$\lambda$6583/H$\alpha$ diagram). However, the diagnostic diagrams in Fig.~\ref{fig2} could provide relatively different classifications for the same object. Therefore, in order to obtain a final classification for each object, we adopted the following approach: first, we determined a classification based on each one of the diagnostic diagrams. Then, the final classification was determined taking into account all classifications provided by the diagrams. Any classification based on an emission-line ratio with a value lower than 3$\sigma$ was not taken into account for the final classification. In the case of a galaxy whose classification provided by the [O \textsc{iii}]$\lambda$5007/H$\beta$ $\times$ [N \textsc{ii}]$\lambda$6583/H$\alpha$ diagram was transition object, then the final classification was also transition object. For a given diagnostic diagram, when the [O \textsc{iii}]$\lambda$5007/H$\beta$ was not available, then a partial classification (L/S for LINER/Seyfert, H II/L for H II region/LINER, H II/T/S for H II region/transition/Seyfert, H II/T/L/S for H II region/transition/LINER/Seyfert, etc) based only on the available ratio was given. Table~\ref{tbl4} shows the classifications obtained for the nuclear emission-line spectra of all galaxies in the sample.  

Using the final classifications shown in Table~\ref{tbl4}, we determined the fractions of nuclear spectra of early-type (E + S0), early-spiral (S0/a + Sa + Sab + Sb + Sbc), and late-spiral (Sc + Scd + Sd) galaxies classified as H II regions, transition objects, LINERs or Seyferts. Such fractions were also determined considering the entire mini-DIVING$^\mathrm{3D}$ sample. The results are shown in Table~\ref{tbl5}. The fractions of galaxies whose nuclear spectra received the partial classifications of LINER/Seyfert or transition/LINER/Seyfert are also shown. 

In order to estimate the uncertainties of the percentages in Table~\ref{tbl5}, we tested two approaches: in the first one, a Monte Carlo simulation, we constructed the group of three diagnostic diagrams in Fig.~\ref{fig2} a hundred times, placing the points in positions given by a Gaussian random distribution, based on the error bars of the original points. Then we classified the nuclear spectra of the galaxies and determined the percentages of objects with different classifications for each obtained group of three diagnostic diagrams. The uncertainty of the percentage, for each classification of the galaxies, was taken as the standard deviation of all the percentages resulting from this procedure. In the second approach, we classified the nuclear spectra of the galaxies in three situations: considering the original values of the emission-line ratios (which gives the classifications in Table~\ref{tbl4}), considering the emission-line ratios plus the corresponding error bars (1$\sigma$) and considering the emission-line ratios minus the corresponding error bars. Then, we determined the fractions of galaxies with nuclear spectra classified as H II regions, transition objects, LINERs, etc in the three situations mentioned above. Finally, the uncertainties of the percentages were taken as the standard deviation of the percentages determined (for each classification) for these three situations. The main problem with the first approach is that it assumes that the uncertainties represented by the error bars in the diagnostic diagrams in Fig.~\ref{fig2} are totally random, which, as explained in Section~\ref{sec32}, is not correct. The first approach provided considerably small uncertainties ($< 1\%$) for the percentages, which does not seem realistic for this case. The second approach, on the other hand, removes the randomness of the uncertainties, which is more consistent with what was discussed in Section~\ref{sec32}, but may not be totally appropriate for all the groups of objects in the sample. The second approach resulted in higher uncertainties for the percentages than the first approach. We opted to use the uncertainties provided by the second approach, shown in Table~\ref{tbl5}, for our analysis. However, considering the previous discussion, such uncertainties must be taken as upper limits for the real uncertainties of the percentages. 

\begin{table*}
\centering
\caption{Classifications of the nuclear emission-line spectra of the galaxies based on the (2) [N \textsc{ii}]$\lambda$6584/H$\alpha$, (3)[S \textsc{ii}]($\lambda$6716 + $\lambda$6731)/H$\alpha$, and (4) [O \textsc{i}]$\lambda$6300/H$\alpha$ diagnostic diagrams. The final classification is shown in column (5). Classifications H II, T, S, and L represent H II region, transition object, Seyfert, and LINER, respectively. The classifications marked with a * are based on emission-line ratios with values lower than 3$\sigma$ and were not taken into account for the final classification.\label{tbl4}}
\begin{tabular}{ccccc}
\hline
Galaxy & [N \textsc{ii}]$\lambda$6583/H$\alpha$ & [S \textsc{ii}]($\lambda$6716 + $\lambda$6731)/H$\alpha$ & [O \textsc{i}]$\lambda$6300/H$\alpha$ & Final Classification \\
    (1)  & (2) & (3) & (4) & (5)  
\\ \hline
NGC 134 & L & L & - & L \\
NGC 157 & T & H II & H II* & T \\
NGC 247 & H II/T/S & - & - & H II/T/S \\
NGC 253 & T & H II & H II & T \\
NGC 300 & - & - & - & - \\
NGC 613 & L & L & L & L \\
NGC 720(n)* & T/L/S & H II/L/S & - & H II/T/L/S \\
NGC 720(b)* & H II/T/S & H II/L/S & - & H II/T/L/S \\
NGC 908 & H II & H II & H II & H II \\
NGC 936 & L & L & L & L \\
NGC 1068 & S & S & S & S \\
NGC 1097 & L & S & L & L/S \\
NGC 1187 & H II & H II & H II & H II \\
NGC 1232 & - & - & - & - \\
NGC 1291(n)* & L & L & L & L \\
NGC 1291(b)* & L & L & L & L \\
NGC 1300 & L & L & L & L \\
NGC 1313 & H II & H II & H II* & H II \\
NGC 1316 & L & L & S & L/S \\
NGC 1365 & S & S & S* & S \\
NGC 1380 & L & L & S* & L \\
NGC 1395 & T/L/S & H II/L/S & - & H II/T/L/S \\
NGC 1398 & L & L & - & L \\
NGC 1399 & H II/T/S & - & - & H II/T/S \\
NGC 1404 & H II/T/S & - & - & H II/T/S \\
NGC 1407 & - & - & - & - \\
NGC 1433 & L & L & S & L/S \\
NGC 1549 & - & - & - & - \\
NGC 1553 & L & S & L & L/S \\
NGC 1559 & H II & H II & - & H II \\
NGC 1566 & S & S & S & S \\
NGC 1574 & L & S & - & L/S \\
NGC 1672 & S & S & S & S \\
NGC 1792 & H II & H II & H II & H II \\
NGC 1808 & T & H II & H II & T \\
NGC 2442 & L & L & L & L \\
NGC 2835 & T/L/S & H II/L/S & - & H II/T/L/S \\
NGC 2997 & L & L & L* & L \\
NGC 3115 & T/S & L/S & - & T/L/S \\
NGC 3585 & T & H II & - & T \\
NGC 3621 & L/S & - & - & L/S \\
NGC 3923 & - & - & - & - \\
NGC 4030 & L & H II & - & H II/L \\
NGC 4594 & L & L & L & L \\
NGC 4697 & L & S & - & L/S \\
NGC 4699 & L & S & S & L/S \\
NGC 4753 & L & S & - & L/S \\
NGC 5068 & H II & H II & H II* & H II \\
NGC 5102 & H II/T/L/S & - & - & H II/T/L/S \\
NGC 5128 & S & L & S* & L/S \\
NGC 5236 & L & L & H II & H II/L \\
NGC 5247 & T/L/S & H II/S & - & H II/T/L/S \\
NGC 5643 & S & S & S & S \\
NGC 6744 & L & L & S* & L \\
NGC 7090 & H II & H II & - & H II \\
NGC 7213 & L & L & L & L \\
NGC 7424 & H II & H II & - & H II \\
NGC 7793 & H II & H II & - & H II \\
IC 1459 & L & L & L & L \\
\hline
\end{tabular}
\\
\begin{tabular}{c}
* For these galaxies, the Gaussian fits were applied to the H$\alpha$+[N \textsc{ii}]$\lambda\lambda$6548,6583 emission lines with (b) and without (n) a broad component of H$\alpha$
\end{tabular}
\end{table*}

By considering the complete mini-DIVING$^\mathrm{3D}$ sample, we can see, from the results in Table~\ref{tbl5}, that the LINER category is the one with the highest fraction of objects, $23\% \pm 4\%$, followed by the LINER/Seyfert category (partial classification), with $18\% \pm 7\%$ of the objects, and by the H II region category, with $15.8\% \pm 2.1\%$ of the objects. On the other hand, considering the groups of early-type, early-spiral, and late-spiral galaxies, the situation changes significantly. Among early-type galaxies, the LINER/Seyfert category is the one with the highest fraction of objects, $32\% \pm 15\%$, followed by the LINER category, with $16\% \pm 9\%$ of the objects. In the case of early spirals, almost half ($42.9\% \pm 2.9\%$) of the galaxies show a nuclear emission-line spectrum characteristic of LINERs, followed by the objects with a partial classification of LINER/Seyfert ($14.3\% \pm 2.9\%$). Finally, nearly half of the late-spiral galaxies ($47\% \pm 4\%$) show nuclear emission-line spectra typical of H II regions. 

\begin{table*}
\centering
\caption{Fraction and number of galaxies in the mini-DIVING$^\mathrm{3D}$ sample, with different morphological types, showing nuclear emission-line spectra characteristic of LINERs, Seyferts, LINERs or Seyferts (LINERs/Seyferts), transition objects, and H II regions. The corresponding fractions and numbers determined from a sub-sample of the PALOMAR survey, selected with the same criteria used for the selection of the mini-DIVING$^\mathrm{3D}$ sample, are also shown.\label{tbl5}}

\begin{tabular}{c}
Complete sample \\
\end{tabular}
\\
\begin{tabular}{ccccc}
\hline
 & mini-DIVING$^\mathrm{3D}$ & mini-DIVING$^\mathrm{3D}$ & PALOMAR with B < 11.2 & PALOMAR with B < 11.2 \\
 & (number of galaxies) & (per cent) & (number of galaxies) & (per cent) \\
H II regions & 9 & $15.8 \pm 2.1$ & 29 & 25 \\
Transition objects & 4 & $7.0 \pm 1.1$ & 14 & 12 \\
LINERs & 13 & $23 \pm 4$ & 33 & 29 \\
Seyferts & 5 & $9 \pm 3$ & 8 & 7 \\
LINERs/Seyferts & 10 & $18 \pm 7$ & 11 & 10 \\
Transition/LINERs/Seyferts & 1 & $1.8 \pm 1.8$ & 0 & 0 \\
Total & 57 & & 115 & \\
\hline
\hline
\end{tabular}
\\
\begin{tabular}{c}
Early-type galaxies \\
\end{tabular}
\\
\begin{tabular}{ccccc}
\hline
 & mini-DIVING$^\mathrm{3D}$ & mini-DIVING$^\mathrm{3D}$ & PALOMAR with B < 11.2 & PALOMAR with B < 11.2 \\
 & (number of galaxies) & (per cent) & (number of galaxies) & (per cent) \\
H II regions & 0 & $0^{+3}_{-0}$ & 2 & 6 \\
Transition objects & 1 & $5 \pm 3$ & 3 & 9 \\
LINERs & 3 & $16 \pm 9$ & 13 & 39 \\
Seyferts & 0 & $0^{+3}_{-0}$ & 0 & 0 \\
LINERs/Seyferts & 6 & $32 \pm 15$ & 3 & 9.1 \\
Transition/LINERs/Seyferts & 1 & $5^{+6}_{-5}$ & 0 & 0 \\
Total & 19 & & 33 & \\
\hline
\hline
\end{tabular}
\\
\begin{tabular}{c}
Early-spiral galaxies \\
\end{tabular}
\\
\begin{tabular}{ccccc}
\hline
& mini-DIVING$^\mathrm{3D}$ & mini-DIVING$^\mathrm{3D}$ & PALOMAR with B < 11.2 & PALOMAR with B < 11.2 \\
 & (number of galaxies) & (per cent) & (number of galaxies) & (per cent) \\
H II regions & 1 & $4.8 \pm 2.9$ & 6 & 12.5 \\
Transition objects & 2 & $9.5 \pm 2.9$ & 5 & 10.4 \\
LINERs & 9 & $42.9 \pm 2.9$ & 20 & 41.7 \\
Seyferts & 4 & $19 \pm 6$ & 7 & 15 \\
LINERs/Seyferts & 3 & $14.3 \pm 2.9$ & 6 & 12.5 \\
Transition/LINERs/Seyferts & 0 & $0^{+3}_{-0}$ & 0 & 0 \\ 
Total & 21 & & 48 & \\
\hline
\hline
\end{tabular}
\\
\begin{tabular}{c}
Late-spiral galaxies \\
\end{tabular}
\\
\begin{tabular}{ccccc}
\hline
& mini-DIVING$^\mathrm{3D}$ & mini-DIVING$^\mathrm{3D}$ & PALOMAR with B < 11.2 & PALOMAR with B < 11.2 \\
 & (number of galaxies) & (per cent) & (number of galaxies) & (per cent) \\
H II regions & 8 & $47 \pm 4$ & 21 & 62 \\
Transition objects & 1 & $6 \pm 4$ & 6 & 18 \\
LINERs & 1 & $6 \pm 4$ & 0 & 0 \\
Seyferts & 1 & $6 \pm 4$ & 1 & 3 \\
LINERs/Seyferts & 1 & $6 \pm 4$ & 2 & 6 \\
Transition/LINERs/Seyferts & 0 & $0^{+4}_{-0}$ & 0 & 0 \\
Total & 17 & & 34 & \\
\hline
\hline
\end{tabular}
\end{table*}

\section{Discussion}\label{sec5}

\subsection{Comparison with the PALOMAR survey and the nature of transition objects}\label{sec51}

\begin{figure*}
\begin{center}
   \includegraphics[scale=0.6]{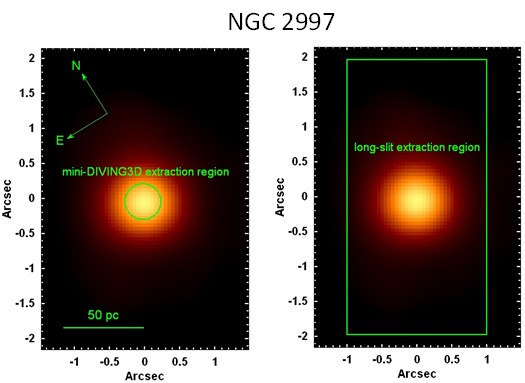} 
  \caption{Left-hand panel: image of the data cube of the central region of NGC 2997, collapsed along its spectral axis, with the circular region used to extract the nuclear spectrum used in the analysis of the mini-DIVING$^\mathrm{3D}$ sample. Right-hand panel: the same as in the image on the left, but with the rectangular region used to extract the ``2 arcsec $\times$ 4 arcsec nuclear spectrum''. The size of this rectangular region is the same of the slit used to obtain the spectra analysed in the PALOMAR survey.\label{fig3}}
\end{center}
\end{figure*}

\begin{figure}
\begin{center}
   \includegraphics[scale=0.24]{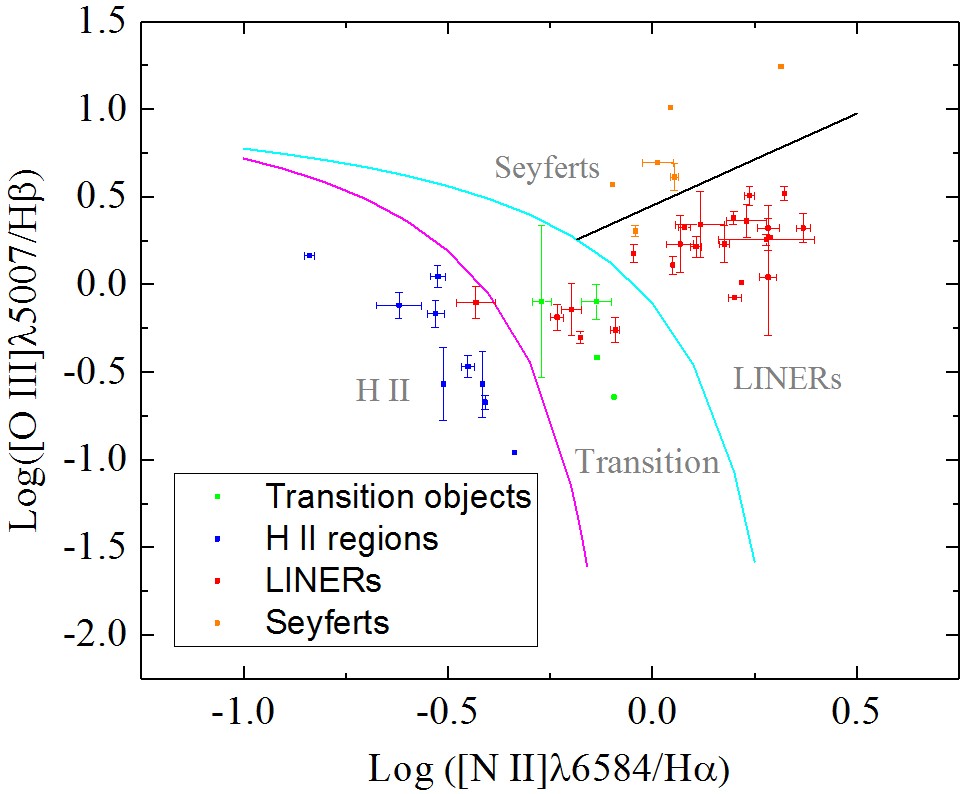} 
  \caption{[O \textsc{iii}]$\lambda$5007/H$\beta$ $\times$ [N \textsc{ii}]$\lambda$6584/H$\alpha$ diagnostic diagram with the points corresponding to the ``2 arcsec $\times$ 4 arcsec nuclear spectra'' extracted from the data cubes in the mini-DIVING$^\mathrm{3D}$ sample. The magenta curve corresponds to the empirical division between H II regions and AGNs obtained by \citet{kau03}, the cyan curve represents the maximum limit for the ionization by a starburst obtained by \citet{kew01}, and the black curve corresponds to the division between LINERs and Seyferts obtained by \citet{sch07}. The blue, green, red, and orange points correspond to galaxies whose classifications, based only on the O \textsc{iii}]$\lambda$5007/H$\beta$ $\times$ [N \textsc{ii}]$\lambda$6584/H$\alpha$ diagram in Fig.~\ref{fig2}, were H II regions, transition objects, LINERs, and Seyferts, respectively.\label{fig4}}
\end{center}
\end{figure}

Since one of the most popular surveys of the central regions of galaxies was the PALOMAR survey, it is certainly convenient to establish a comparison between its results, obtained using slit spectra, and the results of the mini-DIVING$^\mathrm{3D}$ sample. In order to perform an adequate comparison, first of all, we selected a sub-sample of the PALOMAR survey (which we call mini-PALOMAR) with the same parameters adopted for the mini-DIVING$^\mathrm{3D}$ sample: B < 11.2 and |b| > 15$\degr$. The mini-PALOMAR sample has a total of 115 objects, only three of them (NGC 4594, NGC 3115, and NGC 1068) also being part of the mini-DIVING$^\mathrm{3D}$ sample. The reason for this low number of galaxies in the mini-PALOMAR shared with the mini-DIVING$^\mathrm{3D}$ is that the former is mainly focused on objects in the Northern hemisphere, while the latter only includes galaxies in the Southern hemisphere. We classified the mini-PALOMAR objects with the same criteria used to obtain the classifications in Table~\ref{tbl4}. Finally, we calculated the fractions of galaxies, considering the complete sample and also the sub-samples of early types, early spirals, and late spirals, classified as H II regions, transition objects, LINERs, Seyferts, and with partial classifications of LINER/Seyfert and transition/LINER/Seyfert. The results are shown in Table~\ref{tbl5}. Since no error bars were provided by the publications of the PALOMAR survey, we did not estimate uncertainties for these fractions in Table~\ref{tbl5}. 

Considering the complete sample, Table~\ref{tbl5} shows that the fractions of objects in the mini-DIVING$^\mathrm{3D}$ with nuclear spectra characteristic of LINERs, Seyferts, or with partial classifications of LINER/Seyfert and transition/LINER/Seyfert are all compatible with the corresponding fractions obtained for the mini-PALOMAR, at the 1$\sigma$ or 2$\sigma$ levels. On the other hand, the fractions of galaxies in the mini-DIVING$^\mathrm{3D}$ and in the mini-PALOMAR samples with nuclear spectra classified as H II regions or transition objects are not compatible, even at the 3$\sigma$ level. For these two categories, the fractions obtained for the mini-DIVING$^\mathrm{3D}$ are lower than those determined for the mini-PALOMAR. This result is consistent with the apparent dichotomy in the [O \textsc{iii}]$\lambda$5007/H$\beta$ $\times$ [N \textsc{ii}]$\lambda$6584/H$\alpha$ diagnostic diagram in Fig.~\ref{fig2}, which revealed a low number of transition objects in the mini-DIVING$^\mathrm{3D}$ sample. 

For the subsamples of early-type and early-spiral galaxies, all the fractions obtained for the mini-DIVING$^\mathrm{3D}$ and for the mini-PALOMAR are compatible, at the 1$\sigma$, 2$\sigma$, or 3$\sigma$ levels. For the sub-sample of late-spirals, the fractions of galaxies in the mini-DIVING$^\mathrm{3D}$ sample with nuclear spectra characteristic of LINERs, Seyferts, LINERs/Seyferts, and transition/LINERs/Seyferts are compatible with the corresponding fractions of the mini-PALOMAR sample, at the 1$\sigma$ or 2$\sigma$ levels. The fraction of late-spiral galaxies in the mini-DIVING$^\mathrm{3D}$ with nuclear spectra classified as H II regions ($47\% \pm 4\%$) is lower than the corresponding fraction in the mini-PALOMAR (62\%), these two values being only nearly compatible, at the 3$\sigma$ level. The same happens to the fraction of late spirals in the mini-DIVING$^\mathrm{3D}$ with nuclear spectra classified as transition objects ($6\% \pm 4\%$), which is nearly compatible, at the 3$\sigma$ level, with the higher value obtained for the mini-PALOMAR (18\%). Based on these results, we conclude that the discrepancies between the fractions of galaxies in the mini-DIVING$^\mathrm{3D}$ and in the mini-PALOMAR, considering the corresponding complete samples, with nuclear spectra characteristic of H II regions and transition objects are mostly due to the discrepancies between the classifications of the late spirals in these samples.

 The topic of transition objects certainly deserves special attention. A very common interpretation for the transition objects in the literature assumes that they are composite systems, containing a LINER contaminated by an H II region component \citep{ho93}. Such a contamination may come from circumnuclear H II regions or even H II regions along the line of sight. However, as discussed by \citet{ho08}, this scenario for transition objects has its ``problems''. If transition objects were indeed the result of a central LINER contaminated by H II regions, it would be expected that observations with sufficiently high spatial resolution were able to isolate the nuclear emission from circumnuclear contaminations. Therefore, transition objects observed with high spatial resolution would probably be classified as LINERs. However, \citet{shi07} obtained high spatial resolution STIS observations of 23 of the galaxies in the sample of the PALOMAR survey and found only partial support for this scenario. The detection of AGN signatures, such as compact X-Ray or radio emitting cores, in transition objects could be an additional evidence for the composite H II/LINER scenario. \citet{fil00,fil02} analysed 8.4 GHz VLA observations of all transition objects in the PALOMAR survey, but found evidence for the presence of AGNs in only $\sim 25\%$ of the objects. Considering these ``problems'', \citet{ho08} proposed an alternative scenario, assuming that transition objects are indeed accretion powered, but with a very low accretion rate. Considering the group of LINERs and Seyferts, Seyferts 1 show the highest accretion rates, followed by Seyferts 2, LINERs 1, LINERs 2 and, according to the proposed scenario, ending with transition objects. \citet{ho08} also proposed that the anomalously strong H$\alpha$ emission in transition objects may be the result of photoionization by off-nuclear X-ray binaries and/or cosmic ray heating by the central radio core. In addition, \citet{shi07} suggested that evolved hot stars and/or turbulent mixing layers in the interstellar medium may also play a significant role for the strong H$\alpha$ emission in transition objects.

A few natural questions at this point are: what could be the reason for the lower fraction of transition objects detected in the mini-DIVING$^\mathrm{3D}$ (and the resulting dichotomy in the [O \textsc{iii}]$\lambda$5007/H$\beta$ $\times$ [N \textsc{ii}]$\lambda$6584/H$\alpha$ diagnostic diagram in Fig.~\ref{fig2}), in comparison to the mini-PALOMAR? What does this result suggest about the nature of transition objects? The main difference between the DIVING$^\mathrm{3D}$ and the PALOMAR surveys is that the former has a significantly higher spatial resolution than the later. In other words, with the data cubes of the DIVING$^\mathrm{3D}$ survey, it is possible to isolate the nuclear emission and avoid contaminations from the circumnuclear regions of the galaxies with a much higher efficiency than in the case of the slit spectra analysed by the PALOMAR survey. Considering that, we can say that the result obtained with the mini-DIVING$^\mathrm{3D}$ sample certainly suggests that, despite the findings of \citet{shi07}, at least part of the transition objects may be composite systems, with a central LINER whose emission is contaminated by the emission from circumnuclear H II regions. 

In order to try to validate the scenario above of transition objects being composite systems, we performed the following test. First we extracted another spectrum of each data cube of the mini-DIVING$^\mathrm{3D}$ sample from a rectangular region of 2 arcsec $\times$ 4 arcsec, centred on the peak of the stellar emission of the galaxy. From now on we will refer to these new extracted spectra as ``2 arcsec $\times$ 4 arcsec nuclear spectra''. The size of this extracting region is the same of the slit used to obtain the spectra analysed in the PALOMAR survey. Fig.~\ref{fig3} shows, as an example, the rectangular extracting region of the 2 arcsec $\times$ 4 arcsec nuclear spectrum of the data cube of NGC 2997, together with the circular region (also centred on the peak of the stellar emission of the galaxy) from which the original nuclear spectrum (shown in Fig. A12) was extracted. After the extraction, we applied to each extracted 2 arcsec $\times$ 4 arcsec nuclear spectrum the same procedure described in Sections~\ref{sec31} and \ref{sec32}: we subtracted the stellar continuum using the pPXF technique and applied a correction of the interstellar extinction, using the obtained value of the Balmer decrement and the extinction law of \citet{car89}. Finally, we calculated the emission-line ratios [N \textsc{ii}]$\lambda$6583/H$\alpha$ and [O \textsc{iii}]$\lambda$5007/H$\beta$.

Fig.~\ref{fig4} shows the diagnostic diagram of [O \textsc{iii}]$\lambda$5007/H$\beta$ $\times$ [N \textsc{ii}]$\lambda$6583/H$\alpha$ with the points corresponding to the 2 arcsec $\times$ 4 arcsec nuclear spectra of the objects in the mini-DIVING$^\mathrm{3D}$ sample. It is easy to see that the fraction of galaxies in Fig.~\ref{fig4} classified as transition objects ($\sim$ 14\%) is twice the fraction obtained from the corresponding diagnostic diagram in Fig.~\ref{fig2} ($\sim$ 7\%). Actually, this new fraction of transition objects is consistent with the value obtained for the complete mini-PALOMAR sample ($\sim$ 12\%), as can be seen in Table~\ref{tbl5}. Fig.~\ref{fig4} also shows a reduction in the number of LINERs, compared to the [O \textsc{iii}]$\lambda$5007/H$\beta$ $\times$ [N \textsc{ii}]$\lambda$6583/H$\alpha$ diagram in Fig.~\ref{fig2}. Therefore, the result of this test certainly suggests that at least part of the observed transition objects may indeed be the result of the emission from LINERs contaminated by the emission from surrounding H II regions (possibly due to poor spatial resolution observations).

It is worth mentioning that the definition we use for transition objects is not the same as the one adopted, for example, by \citet{ho97b}, which emphasizes the [O \textsc{iii}]$\lambda$5007/H$\beta$ $\times$ [O \textsc{i}]$\lambda$6300/H$\alpha$ diagram and establishes that transition objects show [O \textsc{iii}]$\lambda$5007/H$\beta$ < 3 and  0.08 $\leq$ [O \textsc{i}]$\lambda$6300/H$\alpha$ < 0.17. The reason why we opted to use a definition based mainly on the [N \textsc{ii}]$\lambda$6584/H$\alpha$ ratio, instead of [O \textsc{i}]$\lambda$6300/H$\alpha$, is that the [O \textsc{i}]$\lambda$6300 line was not detected in many objects of our sample or its integrated flux was determined with a high uncertainty. Therefore, a statistical analysis based on classifications obtained mainly with the [O \textsc{i}]$\lambda$6300 line would probably be less reliable. However, the difference between these definitions of transition objects did not affect the comparison with the mini-PALOMAR sample because, as explained above, we classified the objects in the mini-PALOMAR sample with the same criteria used for the classification of the galaxies in the mini-DIVING$^\mathrm{3D}$. If we try to identify transition objects in the mini-DIVING$^\mathrm{3D}$ sample using only the [O \textsc{i}]$\lambda$6300 criterion of \citet{ho97b}, the result would be different classifications for a few objects. NGC 157, NGC 253, and NGC 1808 (all transition, in Table~\ref{tbl4}) would be classified as H II regions; NGC 936 and NGC 1380 (LINERs, in Table~\ref{tbl4}) would be classified as transition; NGC 5236 (H II region/LINER, in Table~\ref{tbl4}) would be classified as transition. It is also worth emphasizing that these two definitions for transition objects, although quantitatively different, are based on the same idea that these objects show intermediate emission-line ratios between those of H II regions and LINERs; therefore the idea of transition objects being the result of a central LINER contaminated by the emission from surrounding H II regions \citep{ho93} is consistent with both definitions.    

Although the hypothesis of composite LINER/H II regions systems described above may explain the detection of some transition objects, there are certain members of this category that remain with this classification, even when observed with high spatial resolution. In the case of the mini-DIVING$^\mathrm{3D}$ sample, the nuclear emission-line spectra of four objects were classified as characteristic of transition objects: NGC 157, NGC 253, NGC 1808, and NGC 3585. 

\begin{figure*}
\begin{center}
   \includegraphics[scale=0.50]{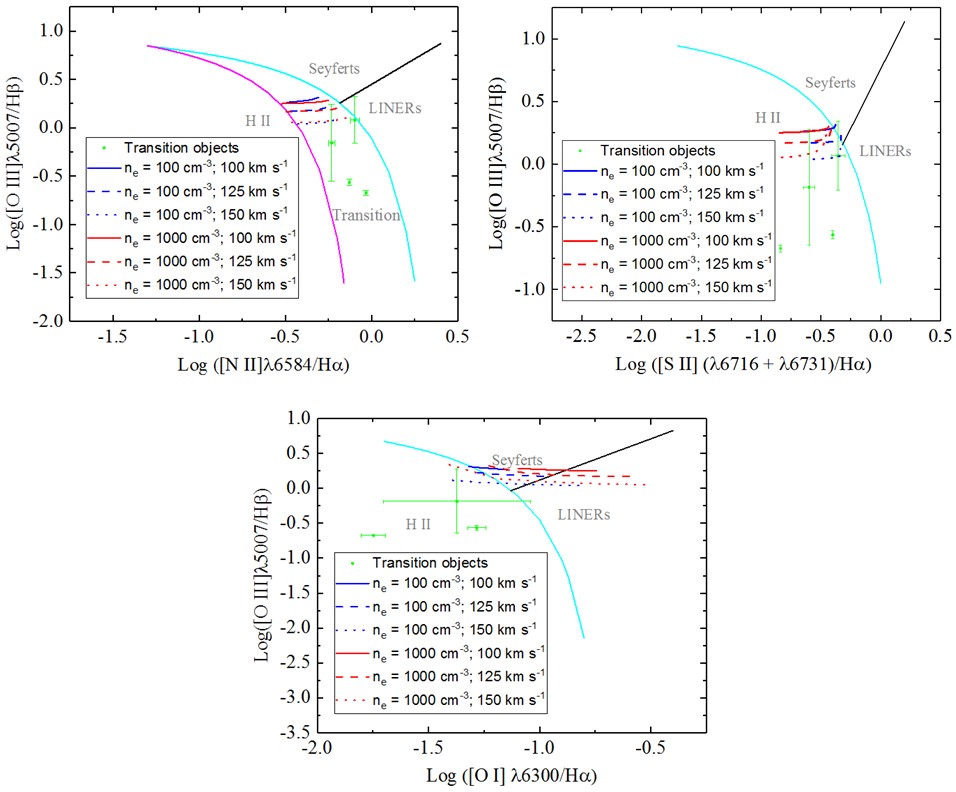} 
  \caption{The same as in Fig.~\ref{fig2}, but only with the points (in green) corresponding to galaxies classified as transition objects, according to the diagrams in Fig.~\ref{fig2}. The red and blue curves represent simulated emission-line ratios taken from the Mappings III library, obtained with shock-heating models, assuming electronic densities and shock velocities consistent with the observed parameters in the spectra.\label{fig5}}
\end{center}
\end{figure*}

NGC 157 is an SAB(rs)bc galaxy, at a distance of 16.1 Mpc. There is not much information in the literature about the central region of this galaxy, but, so far, no evidence of an AGN was detected (e.g. \citealt{hug05}). NGC 253 is an SAB(s)c galaxy at a distance of 3.2 Mpc with a nuclear starburst associated with an outflow, which is detected due to its optical line, X-ray and radio emission (e.g. \citealt{str00,sha10,wes11,bol13,wal17}). The Chandra X-ray emission from the central region of NGC 253 was interpreted as being due to a highly obscured AGN \citep{wea02,mul10}, which is not the dominant source \citep{for00,wea02}. NGC 1808 is an (R)SAB(s)a galaxy at a distance of 9.3 Mpc. Similarly to NGC 253, NGC 1808 also shows a nuclear starburst, with an outflow coming from the central kpc. Such an outflow was detected via its [N\textsc{ii}], H$\alpha$ and Na I lines \citep{phi93}, and its CO molecular emission \citep{sal16}. Some authors claimed that the nuclear starburst in this galaxy may co-exist with a weak AGN (e.g. \citealt{ver85,jun95,awa96,jim05}), while others disagree (e.g. \citealt{for92,phi93,kra01}). Finally, NGC 3585 is an E6 galaxy, at a distance of 17.3 Mpc. As in the case of NGC 157, there is not much information in the literature about the central region of this object. Using Chandra data, \citet{she17} found evidence of a low-luminosity AGN (LLAGN) in this galaxy. 

If NGC 157, NGC 253, NGC 1808, and NGC 3585 harbour LLAGNs, then the fact that their classification as transition objects remains even when observed with high spatial resolutions could be explained by the scenario of composite LINER/H II regions systems if we assume that H II regions may be present along the line of sight of their nuclei. In fact, the disc inclinations of NGC 157, NGC 253, NGC 1808, and NGC 3585, according to Hyperleda\footnote{http://leda.univ-lyon1.fr/} \citep{mak14}, are very high and equal to 61.8\degr, 90.0\degr, 82.7\degr, and 90.0\degr, respectively. This certainly increases the probability of the presence of H II regions along the line of sight towards the nuclei of these nearly edge-on galaxies. Such a scenario, however, may not be the most likely one, specially in the case of NGC 157, due to the absence of clear evidence of an AGN in this object. Therefore, alternative models should be considered to explain the nuclear emission-line spectra of the four transition objects in the mini-DIVING$^\mathrm{3D}$ sample. 

The low accretion rate scenario proposed by \citet{ho08} to explain the emission-line spectra of transition objects is applicable to the four galaxies with this classification detected in the mini-DIVING$^\mathrm{3D}$ sample. In addition, one interesting point to be discussed is the fact that two of these objects, NGC 253 and NGC 1808, show significant outflows associated with nuclear starbursts. Based on that, we would like to propose a different scenario, which assumes that the transition-like nuclear emission-line spectra of these two galaxies are the result of a combination of shock heating from the outflows generated by the nuclear starbursts with photoionization by the young stellar populations in these starbursts. 

A detailed modelling of the shock-heating/photoionization mechanism proposed above is beyond the scope of this paper. However, just to have an idea of the plausibility of such a scenario, Fig.~\ref{fig5} shows the same diagnostic diagrams of Fig.~\ref{fig2} with the curves representing the values of the emission-line ratios resulting from pure shock-heating models. Such curves were obtained from the Mappings III library \citep{all08}. The four points in the diagrams in Fig.~\ref{fig5} correspond to the galaxies classified as transition objects and only the curves corresponding to shock-heating models with electronic densities and velocities consistent with the observed parameters in the spectra of the four transition objects were included. One can see that only the emission-line ratios of two (NGC 157 and NGC 3585) of the transition objects detected in this work are nearly reproduced by the models. On the other hand, such models, specially in the [O \textsc{iii}]$\lambda$5007/H$\beta$ $\times$ [N \textsc{ii}]$\lambda$6583/H$\alpha$ and [O \textsc{iii}]$\lambda$5007/H$\beta$ $\times$ [S \textsc{ii}]($\lambda$6716 + $\lambda$6731)/H$\alpha$ diagrams indeed reproduce part of the region corresponding to transition objects. Therefore, we conclude that it is plausible that these shock-heating models, combined with photoionization by young stars, under specific conditions, reproduce the observed emission-line ratios of the four transition objects detected in this work.

The uncertainties of the [O \textsc{iii}]$\lambda$5007/H$\beta$ ratios of NGC 157 and NGC 3585 are considerably high (see Fig.~\ref{fig2} and Table~\ref{tbl3}). At the 1$\sigma$ or 2$\sigma$ levels, the emission line ratios of NGC 3585 are also compatible with those of LINERs and Seyferts. Similarly, at the 2$\sigma$ level, the emission line ratios of NGC 157 are compatible with those of Seyferts and H II regions. Therefore, the classification of the nuclear emission-line spectra of NGC 157 and NGC 3585 is uncertain and the previous discussion about the nature of transition objects may actually not apply to these two galaxies. The emission-line ratios of NGC 1808 and NGC 253, on the other hand, are much more precise and, as consequence, there is no doubt about the classification of the nuclear emission-line spectra of these two objects. 

Considering all the findings and the discussion above, we believe that transition objects may be a heterogeneous class. Some of them are potentially the result of the emission from LINERs contaminated by the emission from surrounding H II regions (specially when observed with poor spatial resolution). Others, as proposed by \citet{ho08}, may be accretion powered, but with a low accretion rate. Finally, we propose that a scenario involving shock heating from outflows, together with photoionization by young stars, may explain the transition-like nuclear spectra of certain galaxies.

\begin{table*}
\centering
\caption{Evidence of AGNs in the objects of the mini-DIVING$^\mathrm{3D}$ sample. Columns (2) and (3) indicate the presence or not of broad components in the H$\alpha$ and H$\beta$ emission lines, respectively. Column (4) reveals the existence or not of evidence of AGNs based on X-ray data. The information in column (4) was taken from \citet{she17}, unless otherwise specified. It is worth mentioning that, in certain cases, proper X-ray data were not available.\label{tbl6}}
\begin{tabular}{cccc}
\hline
Galaxy & Broad H$\alpha$ & Broad H$\beta$ & X-ray \\
    (1)  & (2) & (3) & (4)  
\\ \hline
NGC 134 & No & No & No** \\
NGC 157 & No & No & No** \\
NGC 247 & No & No & No \\
NGC 253 & No & No & Yes \\
NGC 300 & No & No & No \\
NGC 613 & No & No & Yes \\
NGC 720 & Yes* & No & Yes \\
NGC 908 & No & No & - \\
NGC 936 & No & No & - \\
NGC 1068 & Yes & Yes & Yes \\
NGC 1097 & Yes & Yes & Yes \\
NGC 1187 & No & No & Yes \\
NGC 1232 & No & No & Yes \\
NGC 1291 & Yes* & No & Yes \\
NGC 1300 & No & No & Yes \\
NGC 1313 & No & No & No \\
NGC 1316 & No & No & Yes \\
NGC 1365 & Yes & Yes & Yes \\
NGC 1380 & No & No & Yes \\
NGC 1395 & No & No & Yes \\
NGC 1398 & No & No & Yes \\
NGC 1399 & No & No & Yes \\
NGC 1404 & No & No & Yes \\
NGC 1407 & No & No & Yes \\
NGC 1433 & No & No & No** \\
NGC 1549 & No & No & Yes** \\
NGC 1553 & No & No & Yes \\
NGC 1559 & No & No & - \\
NGC 1566 & Yes & Yes & Yes** \\
NGC 1574 & Yes & No & - \\
NGC 1672 & No & No & Yes \\
NGC 1792 & No & No & No** \\
NGC 1808 & No & No & Yes \\
NGC 2442 & No & No & No** \\
NGC 2835 & No & No & No \\
NGC 2997 & No & No & Yes \\
NGC 3115 & Yes & No & Yes \\
NGC 3585 & No & No & Yes \\
NGC 3621 & No & No & Yes \\
NGC 3923 & No & No & Yes \\
NGC 4030 & No & No & No \\
NGC 4594 & Yes & No & Yes \\
NGC 4697 & No & No & Yes \\
NGC 4699 & No & No & - \\
NGC 4753 & No & No & No** \\
NGC 5068 & No & No & No \\
NGC 5102 & No & No & Yes \\
NGC 5128 & No & No & Yes \\
NGC 5236 & No & No & Yes \\
NGC 5247 & No & No & No \\
NGC 5643 & No & No & Yes \\
NGC 6744 & No & No & No \\
NGC 7090 & No & No & No \\
NGC 7213 & Yes & Yes & Yes \\
NGC 7424 & No & No & No \\
NGC 7793 & No & No & Yes \\
IC 1459 & Yes & No & Yes \\
\hline
\end{tabular}
\\
\begin{tabular}{c}
* For these galaxies, the presence of a broad H$\alpha$ component is uncertain \\
** For these galaxies, not discussed by \citet{she17}, the information was obtained directly from Chandra data 
\end{tabular}
\end{table*}

\begin{figure}
\begin{center}
   \includegraphics[scale=0.43]{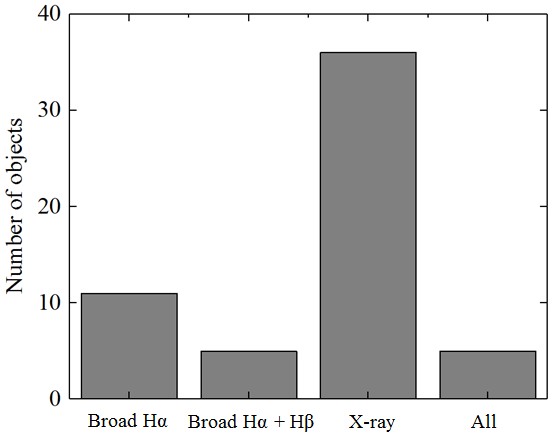} 
  \caption{Histogram showing the number of galaxies in Table~\ref{tbl6} with a broad H$\alpha$ component, with a broad H$\alpha$ and also a broad H$\beta$ component, with a point-like X-ray source, and with all previous features.\label{fig6}}
\end{center}
\end{figure}

One additional topic that deserves some attention is the fraction of early-type galaxies in the mini-DIVING$^\mathrm{3D}$ sample with the partial classification of LINER/Seyfert. Table~\ref{tbl5} shows that the fraction of early-type galaxies with this classification ($32\% \pm 15\%$) is compatible with the corresponding fraction of the mini-PALOMAR (9.1\%), at the 2$\sigma$ level. However, these values are only compatible due to the high uncertainty of the fraction determined for the mini-DIVING$^\mathrm{3D}$ sample (which is higher than the uncertainties of all other fractions in Table~\ref{tbl5}). We believe that a larger sample will probably result in a lower uncertainty for this fraction and may reveal that the fraction of early-type galaxies in the DIVING$^\mathrm{3D}$ survey classified as LINERs/Seyferts is actually higher than the corresponding fraction in the PALOMAR survey. If that is the case, the most likely explanation is also related to the higher spatial resolution of the DIVING$^\mathrm{3D}$ survey, together with the treatment and analysis techniques used in this work. As explained in \citet{men19}, the benefits provided by our data cube treatment methodology (such as high spatial-frequency noise removal and improvement of the spatial resolution) may allow the detection of faint line-emitting regions that otherwise would probably not be detected. This, combined with the high spatial resolution of the DIVING$^\mathrm{3D}$ data cubes, which allows the nuclear emission to be efficiently isolated from the circumnuclear emission, may result in the detection of previously unknown LINERs, Seyferts or, in this case, objects with a partial classification of LINER/Seyfert. An example of the detection of a very faint AGN in an early-type galaxy (NGC 3115), which would not have been detected without the use of our treatment techniques, can be found in \citet{men14b}.

The previous discussions and comparisons with the mini-PALOMAR sample were made based on the uncertainties estimated in this work for the percentages of objects, in the mini-DIVING$^\mathrm{3D}$ sample, with different nuclear spectra classifications. Such uncertainties take into account only our precision to include the points representing these galaxies in the diagnostic diagrams. However, there are also statistical uncertainties, related to the limited sizes of the mini-DIVING$^\mathrm{3D}$ and mini-PALOMAR samples, that should be taken into account. These uncertainties can be obtained using a statistical inference and basically establish how precise are the parameters obtained for the objects in the local Universe based on limited samples. Using Bayes' theorem, with a constant prior and a likelihood given by a binomial distribution (which depends on the total number of objects in the sample, on the number of objects with a given nuclear spectrum classification and on the probability of finding an object with such a classification), we obtain a posterior given by a beta function. Using this resulting beta function, we estimated the uncertainty of the percentage of objects with each classification by determining the range of percentages that included 2/3 of the objects with that classification. This approach resulted in statistical uncertainties of $5\%$, $4\%$, $6\%$, $4\%$, $5\%$, and $2\%$ for the percentages of H II regions, transition objects, LINERs, Seyferts, LINERs/Seyferts, and transition/LINERs/Seyferts, respectively, in the complete mini-DIVING$^\mathrm{3D}$ sample. The final uncertainties for the percentages of these classifications, obtained by combining the uncertainties in Table~\ref{tbl5} with the statistical uncertainties above, are $5\%$, $4\%$, $7\%$, $5\%$, $9\%$, and $3\%$, respectively. 

The procedure described above for estimating the statistical uncertainties also resulted in uncertainties of $5\%$, $4\%$, $6\%$, $4\%$, $5\%$, and $2\%$ for the percentages of H II regions, transition objects, LINERs, Seyferts, LINERs/Seyferts, and transition/LINERs/Seyferts, respectively, in the complete mini-PALOMAR sample. If we consider these uncertainties in the analysis, we conclude that all the fractions of objects, with different classifications, obtained with the mini-DIVING$^\mathrm{3D}$ sample are compatible, at the 1$\sigma$ level, with the ones obtained with the mini-PALOMAR sample. That is actually expected, as both samples analysed here are relatively small and, as a consequence, the statistical uncertainties are significant. Although the statistical uncertainties make the percentages of objects with different classifications in the mini-DIVING$^\mathrm{3D}$ and in the mini-PALOMAR samples compatible, the most relevant result from this work, which should be taken as preliminary, is that the highest discrepancies between these two samples were detected among the objects with nuclear spectra classified as H II regions and transition objects. The larger number of objects in the complete DIVING$^\mathrm{3D}$ sample will reduce the statistical uncertainties by nearly half of the current values and allow a more precise comparison with the results of other surveys.

\subsection{The AGNs in the mini-DIVING$^\mathrm{3D}$ sample}\label{sec52}

One final parameter of this work that should be determined is the number of objects in the mini-DIVING$^\mathrm{3D}$ sample with clear evidence for the presence of AGNs. In order to evaluate that, we focused on the optical and X-ray spectral bands. As mentioned in Section~\ref{sec1}, in the optical, broad components in permitted emission lines (such as H$\alpha$ and H$\beta$) are a clear indication of an AGN. A point-like hard X-ray source is also indicative of an AGN, although a spectral analysis is usually necessary to confirm the nature of the source. We looked for this evidence in the objects of the mini-DIVING$^\mathrm{3D}$ sample, taking into account the literature (X-ray information was taken from \citealt{she17}, unless otherwise specified) and the mini-DIVING$^\mathrm{3D}$ data (to evaluate the presence or not of broad components in the H$\alpha$ and H$\beta$ emission lines). Table~\ref{tbl6} shows the results. In addition, Fig.~\ref{fig6} shows a histogram with the number of galaxies in the mini-DIVING$^\mathrm{3D}$ with a broad H$\alpha$ component, with a broad H$\alpha$ and also a broad H$\beta$ component, with a point-like X-ray source, and with all previous features.

Seyfert galaxies are usually interpreted as authentic AGNs, i.e. they are powered by the accretion of matter onto a central SMBH. Therefore, it is actually expected that all objects classified as Seyfert galaxies in the mini-DIVING$^\mathrm{3D}$ sample show evidence of AGNs. Table~\ref{tbl6} reveals that this is actually the case for the five Seyfert galaxies detected in this work, three of them (60\%) showing broad components in permitted optical emission lines. 

Throughout this paper, we classified as LINERs the objects whose nuclear emission-line ratios fall on the branches of LINERs in the three diagnostic diagrams in Fig.~\ref{fig2}. However, this classification does not imply a specific excitation mechanism. As explained in Section~\ref{sec1}, unlike Seyfert galaxies, the excitation mechanisms of LINERs are somewhat controversial. Models involving shock heating, photoionization by young stars, photoionization by HOLMES, and accretion of matter onto a central SMBH have been proposed to explain the observed low ionization emission-line spectra of LINERs. Today, the most promising models for LINERs assume accretion onto a central SMBH (i.e. LINERs are authentic AGNs, but with a low ionization parameter) or photoionization by HOLMES, although the shock-heating scenario have also been suggested in certain cases. For a detailed review of this topic, see \citet{ho08} and references therein. In the mini-DIVING$^\mathrm{3D}$ sample, 13 objects were classified as LINERs. Table~\ref{tbl6} indicates that nine of them ($\sim 69\%$) show evidence of AGNs and broad components in permitted optical emission lines are only visible in three or four ($\sim 23\%$ or $\sim 31\%$), as a broad H$\alpha$ component in NGC 1291 is uncertain. This result does not rule out the hypothesis of the presence of (weak or obscured) AGNs in the rest of the objects classified as LINERs; however, it certainly suggests that these other objects (without evidence of AGNs) may be good candidates for LINERs powered, at least partially, by different mechanisms, such as photoionization by HOLMES. A detailed modelling, which is beyond the scope of this paper, is required to evaluate the applicability of the different excitation mechanisms to explain the nuclear line emission detected in the objects classified as LINERs in this work. 

Nine objects from the sample were classified as H II regions and, surprisingly, two of them ($\sim 22\%$) show evidence of AGNs, due to the X-ray emission. One possible explanation is that these two AGNs are faint and/or highly obscured, being obfuscated by H II regions in the same area and, as a consequence, not being detected in the optical. In the case of transition objects, as already discussed in Section~\ref{sec51}, four galaxies in the mini-DIVING$^\mathrm{3D}$ sample received this classification, none with broad components of permitted optical emission lines, and three of them ($75\%$) show evidence of AGNs.

Regarding the objects that received partial classifications, 10 were classified as LINERs/Seyferts, seven of them ($70\%$) showing evidence of AGNs, two ($20\%$) with broad components of permitted optical emission lines. Two galaxies (NGC 4030 and NGC 5236) were classified as HII regions/LINERs, none with broad components of permitted optical emission lines, and only one of them (NGC 5236) shows X-ray emission characteristic of an AGN. This indicates that NGC 5236 indeed hosts an AGN and this partial classification possibly resulted from contaminations of the nuclear LINER spectrum by the emission from H II regions in the area, which were sufficient to result in this partial classification, but not to change the classification to transition. Three objects received a more uncertain classification of H II region/transition/Seyfert, none with broad components of permitted optical emission lines, and two of them ($\sim 67\%$) show evidence of AGNs in the X-ray emission. Only one galaxy was classified as transition/LINER/Seyfert (NGC 3115), which shows X-ray emission characteristic of an AGN and also a broad H$\alpha$ component \citep{men14b}. 

Finally, five galaxies received the classification of H II region/transition/LINER/Seyfert, which essentially means only that these objects show nuclear emission lines, but a more precise classification of the nuclear emission-line spectrum is not possible (due to the fact that the [O \textsc{iii}]$\lambda$5007/H$\beta$ ratio could not be determined for any of them). Only one of these objects (NGC 720) may show a broad H$\alpha$ component (although the presence of such a component is uncertain) and three of the objects (60\%) with this uncertain classification show evidence of AGNs in the X-ray emission (including NGC 720). Considering the entire sample, 37 galaxies ($\sim 65\%$) show evidence of AGNs. It is worth mentioning that the Chandra data were not available for all objects in the mini-DIVING$^\mathrm{3D}$ sample, which introduced a certain degree of incompleteness in this analysis. Therefore, all the percentages determined in this section should be taken as lower limits. 

\section{Conclusions}\label{sec6}

We are conducting the DIVING$^\mathrm{3D}$ survey, which has the purpose of observing, using optical 3D spectroscopy, the central regions of all galaxies in the Southern hemisphere with B < 12.0 and |b| > 15$\degr$. In this paper, we showed the first results of the analysis focused on the nuclear emission-line properties of all galaxies brighter than B = 11.2, which correspond to a sub-sample we called mini-DIVING$^\mathrm{3D}$. Bellow we list our main findings.

\begin{itemize}

\item The category of LINERs is the one with the highest fraction of objects ($23\% \pm 4\%$) in the mini-DIVING$^\mathrm{3D}$ sample

\item The [O \textsc{iii}]$\lambda$5007/H$\beta$ $\times$ [N \textsc{ii}]$\lambda$6583/H$\alpha$ diagnostic diagram obtained for the mini-DIVING$^\mathrm{3D}$ sample reveals an apparent dichotomy, with a group of objects falling on the branches of LINERs and Seyferts and another group of objects falling on the branch of H II regions, but with only a few galaxies being classified as transition objects

\item By comparing our results with those obtained from a subsample of the PALOMAR survey (mini-PALOMAR), selected with the same criteria used for the mini-DIVING$^\mathrm{3D}$, and taking into account only the uncertainties due to our precision to perform the diagnostic diagram analysis, we verified that the fractions of objects in these two subsamples classified as LINERs, Seyferts or with partial classifications of LINER/Seyfert and transition/LINER/Seyfert are compatible, at the 1$\sigma$ or 2$\sigma$ levels.

\item The fractions of objects in the mini-DIVING$^\mathrm{3D}$ sample classified as H II regions and transition objects are lower than the corresponding fractions in the mini-PALOMAR sample, not being compatible, even at the 3$\sigma$ level (again, only taking into account the uncertainties associated with the precision of the diagnostic diagram analysis)

\item Considering that the PALOMAR slit spectra have a significantly lower spatial resolution than the DIVING$^\mathrm{3D}$ data cubes, the result obtained for transition objects in the mini-DIVING$^\mathrm{3D}$ sample, in comparison to the one obtained from the mini-PALOMAR sample, suggests that part of the transition objects may be composite systems, with a central LINER whose emission is contaminated by the emission from circumnuclear H II regions

\item The median FWHM of the PSFs of the treated data cubes in the mini-DIVING$^\mathrm{3D}$ sample is $\sim$ 35 pc. Such a spatial resolution allowed a more accurate analysis of the nuclear emission of the galaxies, without circumnuclear contaminations (resulting in a lower number of transition objects). This information will be useful for future surveys of objects at higher redshifts, whose goals also involve a detailed study of the nuclear emission 

\item Despite the fact that the highest discrepancies between the results obtained with the mini-DIVING$^\mathrm{3D}$ and the mini-PALOMAR samples were detected among galaxies with nuclear spectra classified as H II regions and transition objects, one should note that, if we take into account the statistical uncertainties, all the fractions of objects, with different classifications, obtained with these two samples become compatible, at the 1$\sigma$ level. That is a consequence of the relatively small number of objects in the samples. Once we have the complete DIVING$^\mathrm{3D}$ sample, these statistical uncertainties will be reduced by nearly half, which will allow a more precise comparison with the results of other surveys 

\item Two of the four transition objects detected in this work showed significant outflows powered by nuclear starbursts. This led us to propose an alternative scenario, in which the emission-line spectra of some transition objects are the result of shock heating by the central outflow, together with photoionization by young stars in the nuclear starburst. There is also an additional model for transition objects, proposed by previous studies, which is applicable to the four cases detected in this work. Such a model assumes that transition objects are accretion powered, but with a very low accretion rate 

\item 69\% of the LINERs detected in the mini-DIVING$^\mathrm{3D}$ sample show evidence of AGNs. Although this result does exclude the possibility that the rest of the LINERs in the sample are also (weak or obscured) AGNs, it certainly suggests that these other objects may be candidates for LINERs powered, at least partially, by different mechanisms, such as photoionization by HOLMES

\item Considering the entire mini-DIVING$^\mathrm{3D}$ sample, evidence of AGNs were detected in 65\% of the objects

\end{itemize}

\section*{Acknowledgements}

Based on observations obtained at the Gemini Observatory (processed using the Gemini \textsc{IRAF} package), which is operated by the Association of Universities for Research in Astronomy, Inc., under a cooperative agreement with the NSF on behalf of the Gemini partnership: the National Science Foundation (United States), the National Research Council (Canada), CONICYT (Chile), the Australian Research Council (Australia), Minist\'{e}rio da Ci\^{e}ncia, Tecnologia e Inova\c{c}\~{a}o (Brazil) and Ministerio de Ciencia, Tecnolog\'{i}a e Innovaci\'{o}n Productiva (Argentina). This research has made use of the NASA/IPAC Extragalactic Database (NED), which is operated by the Jet Propulsion Laboratory, California Institute of Technology, under contract with the National Aeronautics and Space Administration. We also acknowledge the usage of the HyperLeda data base (http://leda.univ-lyon1.fr). We thank Conselho Nacional de Desenvolvimento Cient\'ifico e Tecnol\'ogico (CNPq) for support under grants 306063/2019-0 (RBM), 306790/2019-0 (TVR) and 141766/2016-6 (PS), and Funda\c{c}\~ao de Amparo \`a Pesquisa do Estado de S\~ao Paulo (FAPESP) - for support under grants 2011/51680-6 and 2020/13315-3 (PS). We also thank an anonymous for valuable comments about the paper.

\addcontentsline{toc}{section}{Acknowledgements}

\section*{Data Availability}

Further detail about the DIVING$^\mathrm{3D}$ survey can be found at https://diving3d.maua.br. The raw GMOS/IFU data are available at the Gemini Science Archive (https://archive.gemini.edu/searchform). The treated GMOS/IFU and SIFS data cubes can be requested at diving3d@gmail.com.





\bibliographystyle{mnras}
\bibliography{references} 

\begin{thebibliography}{}
\makeatletter
\relax
\def\mn@urlcharsother{\let\do\@makeother \do\$\do\&\do\#\do\^\do\_\do\%\do\~}
\def\mn@doi{\begingroup\mn@urlcharsother \@ifnextchar [ {\mn@doi@}
  {\mn@doi@[]}}
\def\mn@doi@[#1]#2{\def\@tempa{#1}\ifx\@tempa\@empty \href
  {http://dx.doi.org/#2} {doi:#2}\else \href {http://dx.doi.org/#2} {#1}\fi
  \endgroup}
\def\mn@eprint#1#2{\mn@eprint@#1:#2::\@nil}
\def\mn@eprint@arXiv#1{\href {http://arxiv.org/abs/#1} {{\tt arXiv:#1}}}
\def\mn@eprint@dblp#1{\href {http://dblp.uni-trier.de/rec/bibtex/#1.xml}
  {dblp:#1}}
\def\mn@eprint@#1:#2:#3:#4\@nil{\def\@tempa {#1}\def\@tempb {#2}\def\@tempc
  {#3}\ifx \@tempc \@empty \let \@tempc \@tempb \let \@tempb \@tempa \fi \ifx
  \@tempb \@empty \def\@tempb {arXiv}\fi \@ifundefined
  {mn@eprint@\@tempb}{\@tempb:\@tempc}{\expandafter \expandafter \csname
  mn@eprint@\@tempb\endcsname \expandafter{\@tempc}}}

\bibitem[\protect\citeauthoryear{{Allen}, {Groves}, {Dopita}, {Sutherland}  \&
  {Kewley}}{{Allen} et~al.}{2008}]{all08}
{Allen} M.~G.,  {Groves} B.~A.,  {Dopita} M.~A.,  {Sutherland} R.~S.,
  {Kewley} L.~J.,  2008, \mn@doi [\apjs] {10.1086/589652}, \href
  {https://ui.adsabs.harvard.edu/abs/2008ApJS..178...20A} {178, 20}

\bibitem[\protect\citeauthoryear{{Awaki}, {Ueno}, {Koyama}, {Tsuru}  \&
  {Iwasawa}}{{Awaki} et~al.}{1996}]{awa96}
{Awaki} H.,  {Ueno} S.,  {Koyama} K.,  {Tsuru} T.,   {Iwasawa} K.,  1996,
  \mn@doi [\pasj] {10.1093/pasj/48.3.409}, \href
  {https://ui.adsabs.harvard.edu/abs/1996PASJ...48..409A} {48, 409}

\bibitem[\protect\citeauthoryear{{Bacon} et~al.,}{{Bacon} et~al.}{2001}]{bac01}
{Bacon} R.,  et~al., 2001, \mn@doi [\mnras] {10.1046/j.1365-8711.2001.04612.x},
  \href {https://ui.adsabs.harvard.edu/abs/2001MNRAS.326...23B} {326, 23}

\bibitem[\protect\citeauthoryear{{Baldwin}, {Phillips}  \&
  {Terlevich}}{{Baldwin} et~al.}{1981}]{bal81}
{Baldwin} J.~A.,  {Phillips} M.~M.,   {Terlevich} R.,  1981, \mn@doi [\pasp]
  {10.1086/130766}, \href
  {https://ui.adsabs.harvard.edu/abs/1981PASP...93....5B} {93, 5}

\bibitem[\protect\citeauthoryear{{Binette}, {Magris}, {Stasi{\'n}ska}  \&
  {Bruzual}}{{Binette} et~al.}{1994}]{bin94}
{Binette} L.,  {Magris} C.~G.,  {Stasi{\'n}ska} G.,   {Bruzual} A.~G.,  1994,
  \aap, \href {https://ui.adsabs.harvard.edu/abs/1994A&A...292...13B} {292, 13}

\bibitem[\protect\citeauthoryear{{Bolatto} et~al.,}{{Bolatto}
  et~al.}{2013}]{bol13}
{Bolatto} A.~D.,  et~al., 2013, \mn@doi [\nat] {10.1038/nature12351}, \href
  {https://ui.adsabs.harvard.edu/abs/2013Natur.499..450B} {499, 450}

\bibitem[\protect\citeauthoryear{{Bryant} et~al.,}{{Bryant}
  et~al.}{2015}]{bry15}
{Bryant} J.~J.,  et~al., 2015, \mn@doi [\mnras] {10.1093/mnras/stu2635}, \href
  {https://ui.adsabs.harvard.edu/abs/2015MNRAS.447.2857B} {447, 2857}

\bibitem[\protect\citeauthoryear{{Bundy} et~al.,}{{Bundy} et~al.}{2015}]{bun15}
{Bundy} K.,  et~al., 2015, \mn@doi [\apj] {10.1088/0004-637X/798/1/7}, \href
  {https://ui.adsabs.harvard.edu/abs/2015ApJ...798....7B} {798, 7}

\bibitem[\protect\citeauthoryear{{Cappellari}}{{Cappellari}}{2017}]{cap17}
{Cappellari} M.,  2017, \mn@doi [\mnras] {10.1093/mnras/stw3020}, \href
  {https://ui.adsabs.harvard.edu/abs/2017MNRAS.466..798C} {466, 798}

\bibitem[\protect\citeauthoryear{{Cappellari} et~al.,}{{Cappellari}
  et~al.}{2011}]{cap11}
{Cappellari} M.,  et~al., 2011, \mn@doi [\mnras]
  {10.1111/j.1365-2966.2010.18174.x}, \href
  {https://ui.adsabs.harvard.edu/abs/2011MNRAS.413..813C} {413, 813}

\bibitem[\protect\citeauthoryear{{Cardelli}, {Clayton}  \& {Mathis}}{{Cardelli}
  et~al.}{1989}]{car89}
{Cardelli} J.~A.,  {Clayton} G.~C.,   {Mathis} J.~S.,  1989, \mn@doi [\apj]
  {10.1086/167900}, \href
  {https://ui.adsabs.harvard.edu/abs/1989ApJ...345..245C} {345, 245}

\bibitem[\protect\citeauthoryear{{Chen} \& {Halpern}}{{Chen} \&
  {Halpern}}{1989}]{che89}
{Chen} K.,  {Halpern} J.~P.,  1989, \mn@doi [\apj] {10.1086/167782}, \href
  {https://ui.adsabs.harvard.edu/abs/1989ApJ...344..115C} {344, 115}

\bibitem[\protect\citeauthoryear{{Cid Fernandes}, {Stasi{\'n}ska}, {Mateus}  \&
  {Vale Asari}}{{Cid Fernandes} et~al.}{2011}]{cid11}
{Cid Fernandes} R.,  {Stasi{\'n}ska} G.,  {Mateus} A.,   {Vale Asari} N.,
  2011, \mn@doi [\mnras] {10.1111/j.1365-2966.2011.18244.x}, \href
  {https://ui.adsabs.harvard.edu/abs/2011MNRAS.413.1687C} {413, 1687}

\bibitem[\protect\citeauthoryear{{Dong} et~al.,}{{Dong} et~al.}{2007}]{don07}
{Dong} X.,  et~al., 2007, \mn@doi [\apj] {10.1086/510899}, \href
  {https://ui.adsabs.harvard.edu/abs/2007ApJ...657..700D} {657, 700}

\bibitem[\protect\citeauthoryear{{Dopita} \& {Sutherland}}{{Dopita} \&
  {Sutherland}}{1995}]{dop95}
{Dopita} M.~A.,  {Sutherland} R.~S.,  1995, \mn@doi [\apj] {10.1086/176596},
  \href {https://ui.adsabs.harvard.edu/abs/1995ApJ...455..468D} {455, 468}

\bibitem[\protect\citeauthoryear{{Dopita} \& {Sutherland}}{{Dopita} \&
  {Sutherland}}{1996}]{dop96}
{Dopita} M.~A.,  {Sutherland} R.~S.,  1996, \mn@doi [\apjs] {10.1086/192255},
  \href {https://ui.adsabs.harvard.edu/abs/1996ApJS..102..161D} {102, 161}

\bibitem[\protect\citeauthoryear{{Dopita} et~al.,}{{Dopita}
  et~al.}{2015}]{dop15}
{Dopita} M.~A.,  et~al., 2015, \mn@doi [\apjs] {10.1088/0067-0049/217/1/12},
  \href {https://ui.adsabs.harvard.edu/abs/2015ApJS..217...12D} {217, 12}

\bibitem[\protect\citeauthoryear{{Eracleous}, {Hwang}  \& {Flohic}}{{Eracleous}
  et~al.}{2010}]{era10}
{Eracleous} M.,  {Hwang} J.~A.,   {Flohic} H. M.~L.~G.,  2010, \mn@doi [\apj]
  {10.1088/0004-637X/711/2/796}, \href
  {https://ui.adsabs.harvard.edu/abs/2010ApJ...711..796E} {711, 796}

\bibitem[\protect\citeauthoryear{{Ferland} \& {Netzer}}{{Ferland} \&
  {Netzer}}{1983}]{fer83}
{Ferland} G.~J.,  {Netzer} H.,  1983, \mn@doi [\apj] {10.1086/160577}, \href
  {https://ui.adsabs.harvard.edu/abs/1983ApJ...264..105F} {264, 105}

\bibitem[\protect\citeauthoryear{{Ferrarese} \& {Ford}}{{Ferrarese} \&
  {Ford}}{2005}]{fer05}
{Ferrarese} L.,  {Ford} H.,  2005, \mn@doi [\ssr] {10.1007/s11214-005-3947-6},
  \href {https://ui.adsabs.harvard.edu/abs/2005SSRv..116..523F} {116, 523}

\bibitem[\protect\citeauthoryear{{Ferrarese} \& {Merritt}}{{Ferrarese} \&
  {Merritt}}{2000}]{fer00}
{Ferrarese} L.,  {Merritt} D.,  2000, \mn@doi [\apjl] {10.1086/312838}, \href
  {https://ui.adsabs.harvard.edu/abs/2000ApJ...539L...9F} {539, L9}

\bibitem[\protect\citeauthoryear{{Filho}, {Barthel}  \& {Ho}}{{Filho}
  et~al.}{2000}]{fil00}
{Filho} M.~E.,  {Barthel} P.~D.,   {Ho} L.~C.,  2000, \mn@doi [\apjs]
  {10.1086/313412}, \href
  {https://ui.adsabs.harvard.edu/abs/2000ApJS..129...93F} {129, 93}

\bibitem[\protect\citeauthoryear{{Filho}, {Barthel}  \& {Ho}}{{Filho}
  et~al.}{2002}]{fil02}
{Filho} M.~E.,  {Barthel} P.~D.,   {Ho} L.~C.,  2002, \mn@doi [\apjs]
  {10.1086/341786}, \href
  {https://ui.adsabs.harvard.edu/abs/2002ApJS..142..223F} {142, 223}

\bibitem[\protect\citeauthoryear{{Filippenko} \& {Ho}}{{Filippenko} \&
  {Ho}}{2003}]{fil03}
{Filippenko} A.~V.,  {Ho} L.~C.,  2003, \mn@doi [\apjl] {10.1086/375361}, \href
  {https://ui.adsabs.harvard.edu/abs/2003ApJ...588L..13F} {588, L13}

\bibitem[\protect\citeauthoryear{{Filippenko} \& {Sargent}}{{Filippenko} \&
  {Sargent}}{1985}]{fil85}
{Filippenko} A.~V.,  {Sargent} W.~L.~W.,  1985, \mn@doi [\apjs]
  {10.1086/191012}, \href
  {https://ui.adsabs.harvard.edu/abs/1985ApJS...57..503F} {57, 503}

\bibitem[\protect\citeauthoryear{{Filippenko} \& {Terlevich}}{{Filippenko} \&
  {Terlevich}}{1992}]{fil92}
{Filippenko} A.~V.,  {Terlevich} R.,  1992, \mn@doi [\apjl] {10.1086/186549},
  \href {https://ui.adsabs.harvard.edu/abs/1992ApJ...397L..79F} {397, L79}

\bibitem[\protect\citeauthoryear{{Flores-Fajardo}, {Morisset}, {Stasi{\'n}ska}
  \& {Binette}}{{Flores-Fajardo} et~al.}{2011}]{flo11}
{Flores-Fajardo} N.,  {Morisset} C.,  {Stasi{\'n}ska} G.,   {Binette} L.,
  2011, \mn@doi [\mnras] {10.1111/j.1365-2966.2011.18848.x}, \href
  {https://ui.adsabs.harvard.edu/abs/2011MNRAS.415.2182F} {415, 2182}

\bibitem[\protect\citeauthoryear{{Forbes}, {Boisson}  \& {Ward}}{{Forbes}
  et~al.}{1992}]{for92}
{Forbes} D.~A.,  {Boisson} C.,   {Ward} M.~J.,  1992, \mn@doi [\mnras]
  {10.1093/mnras/259.2.293}, \href
  {https://ui.adsabs.harvard.edu/abs/1992MNRAS.259..293F} {259, 293}

\bibitem[\protect\citeauthoryear{{Forbes}, {Polehampton}, {Stevens}, {Brodie}
  \& {Ward}}{{Forbes} et~al.}{2000}]{for00}
{Forbes} D.~A.,  {Polehampton} E.,  {Stevens} I.~R.,  {Brodie} J.~P.,   {Ward}
  M.~J.,  2000, \mn@doi [\mnras] {10.1046/j.1365-8711.2000.03120.x}, \href
  {https://ui.adsabs.harvard.edu/abs/2000MNRAS.312..689F} {312, 689}

\bibitem[\protect\citeauthoryear{{Gebhardt} et~al.,}{{Gebhardt}
  et~al.}{2000}]{geb00}
{Gebhardt} K.,  et~al., 2000, \mn@doi [\apjl] {10.1086/312840}, \href
  {https://ui.adsabs.harvard.edu/abs/2000ApJ...539L..13G} {539, L13}

\bibitem[\protect\citeauthoryear{{Granato}, {De Zotti}, {Silva}, {Bressan}  \&
  {Danese}}{{Granato} et~al.}{2004}]{gra04}
{Granato} G.~L.,  {De Zotti} G.,  {Silva} L.,  {Bressan} A.,   {Danese} L.,
  2004, \mn@doi [\apj] {10.1086/379875}, \href
  {https://ui.adsabs.harvard.edu/abs/2004ApJ...600..580G} {600, 580}

\bibitem[\protect\citeauthoryear{{G{\"u}ltekin} et~al.,}{{G{\"u}ltekin}
  et~al.}{2009}]{gul09}
{G{\"u}ltekin} K.,  et~al., 2009, \mn@doi [\apj] {10.1088/0004-637X/698/1/198},
  \href {https://ui.adsabs.harvard.edu/abs/2009ApJ...698..198G} {698, 198}

\bibitem[\protect\citeauthoryear{{Halpern} \& {Steiner}}{{Halpern} \&
  {Steiner}}{1983}]{hal83}
{Halpern} J.~P.,  {Steiner} J.~E.,  1983, \mn@doi [\apjl] {10.1086/184051},
  \href {https://ui.adsabs.harvard.edu/abs/1983ApJ...269L..37H} {269, L37}

\bibitem[\protect\citeauthoryear{{Heckman}}{{Heckman}}{1980}]{hec80}
{Heckman} T.~M.,  1980, \aap, \href
  {https://ui.adsabs.harvard.edu/abs/1980A&A....87..152H} {500, 187}

\bibitem[\protect\citeauthoryear{{Hidalgo} et~al.,}{{Hidalgo}
  et~al.}{2018}]{hid18}
{Hidalgo} S.~L.,  et~al., 2018, \mn@doi [\apj] {10.3847/1538-4357/aab158},
  \href {https://ui.adsabs.harvard.edu/abs/2018ApJ...856..125H} {856, 125}

\bibitem[\protect\citeauthoryear{{Ho}}{{Ho}}{2008}]{ho08}
{Ho} L.~C.,  2008, \mn@doi [\araa] {10.1146/annurev.astro.45.051806.110546},
  \href {https://ui.adsabs.harvard.edu/abs/2008ARA&A..46..475H} {46, 475}

\bibitem[\protect\citeauthoryear{{Ho}, {Filippenko}  \& {Sargent}}{{Ho}
  et~al.}{1993}]{ho93}
{Ho} L.~C.,  {Filippenko} A.~V.,   {Sargent} W. L.~W.,  1993, \mn@doi [\apj]
  {10.1086/173291}, \href
  {https://ui.adsabs.harvard.edu/abs/1993ApJ...417...63H} {417, 63}

\bibitem[\protect\citeauthoryear{{Ho}, {Filippenko}  \& {Sargent}}{{Ho}
  et~al.}{1997a}]{ho97b}
{Ho} L.~C.,  {Filippenko} A.~V.,   {Sargent} W. L.~W.,  1997a, \mn@doi [\apjs]
  {10.1086/313041}, \href
  {https://ui.adsabs.harvard.edu/abs/1997ApJS..112..315H} {112, 315}

\bibitem[\protect\citeauthoryear{{Ho}, {Filippenko}, {Sargent}  \& {Peng}}{{Ho}
  et~al.}{1997b}]{ho97}
{Ho} L.~C.,  {Filippenko} A.~V.,  {Sargent} W. L.~W.,   {Peng} C.~Y.,  1997b,
  \mn@doi [\apjs] {10.1086/313042}, \href
  {https://ui.adsabs.harvard.edu/abs/1997ApJS..112..391H} {112, 391}

\bibitem[\protect\citeauthoryear{{Hopkins} \& {Hernquist}}{{Hopkins} \&
  {Hernquist}}{2006}]{hop06}
{Hopkins} P.~F.,  {Hernquist} L.,  2006, \mn@doi [\apjs] {10.1086/505753},
  \href {https://ui.adsabs.harvard.edu/abs/2006ApJS..166....1H} {166, 1}

\bibitem[\protect\citeauthoryear{{Hopkins}, {Hernquist}, {Cox}, {Robertson}  \&
  {Krause}}{{Hopkins} et~al.}{2007}]{hop07}
{Hopkins} P.~F.,  {Hernquist} L.,  {Cox} T.~J.,  {Robertson} B.,   {Krause} E.,
   2007, \mn@doi [\apj] {10.1086/521601}, \href
  {https://ui.adsabs.harvard.edu/abs/2007ApJ...669...67H} {669, 67}

\bibitem[\protect\citeauthoryear{{Hughes} et~al.,}{{Hughes}
  et~al.}{2005}]{hug05}
{Hughes} M.~A.,  et~al., 2005, \mn@doi [\aj] {10.1086/430531}, \href
  {https://ui.adsabs.harvard.edu/abs/2005AJ....130...73H} {130, 73}

\bibitem[\protect\citeauthoryear{{Jim{\'e}nez-Bail{\'o}n}, {Santos-Lle{\'o}},
  {Dahlem}, {Ehle}, {Mas-Hesse}, {Guainazzi}, {Heckman}  \&
  {Weaver}}{{Jim{\'e}nez-Bail{\'o}n} et~al.}{2005}]{jim05}
{Jim{\'e}nez-Bail{\'o}n} E.,  {Santos-Lle{\'o}} M.,  {Dahlem} M.,  {Ehle} M.,
  {Mas-Hesse} J.~M.,  {Guainazzi} M.,  {Heckman} T.~M.,   {Weaver} K.~A.,
  2005, \mn@doi [\aap] {10.1051/0004-6361:20053022}, \href
  {https://ui.adsabs.harvard.edu/abs/2005A&A...442..861J} {442, 861}

\bibitem[\protect\citeauthoryear{{Junkes}, {Zinnecker}, {Hensler}, {Dahlem}  \&
  {Pietsch}}{{Junkes} et~al.}{1995}]{jun95}
{Junkes} N.,  {Zinnecker} H.,  {Hensler} G.,  {Dahlem} M.,   {Pietsch} W.,
  1995, \aap, \href {https://ui.adsabs.harvard.edu/abs/1995A&A...294....8J}
  {294, 8}

\bibitem[\protect\citeauthoryear{{Kauffmann} et~al.,}{{Kauffmann}
  et~al.}{2003}]{kau03}
{Kauffmann} G.,  et~al., 2003, \mn@doi [\mnras]
  {10.1111/j.1365-2966.2003.07154.x}, \href
  {https://ui.adsabs.harvard.edu/abs/2003MNRAS.346.1055K} {346, 1055}

\bibitem[\protect\citeauthoryear{{Kewley}, {Dopita}, {Sutherland}, {Heisler}
  \& {Trevena}}{{Kewley} et~al.}{2001}]{kew01}
{Kewley} L.~J.,  {Dopita} M.~A.,  {Sutherland} R.~S.,  {Heisler} C.~A.,
  {Trevena} J.,  2001, \mn@doi [\apj] {10.1086/321545}, \href
  {https://ui.adsabs.harvard.edu/abs/2001ApJ...556..121K} {556, 121}

\bibitem[\protect\citeauthoryear{{Kewley}, {Groves}, {Kauffmann}  \&
  {Heckman}}{{Kewley} et~al.}{2006}]{kew06}
{Kewley} L.~J.,  {Groves} B.,  {Kauffmann} G.,   {Heckman} T.,  2006, \mn@doi
  [\mnras] {10.1111/j.1365-2966.2006.10859.x}, \href
  {https://ui.adsabs.harvard.edu/abs/2006MNRAS.372..961K} {372, 961}

\bibitem[\protect\citeauthoryear{{Kormendy} \& {Ho}}{{Kormendy} \&
  {Ho}}{2013}]{kor13}
{Kormendy} J.,  {Ho} L.~C.,  2013, \mn@doi [\araa]
  {10.1146/annurev-astro-082708-101811}, \href
  {https://ui.adsabs.harvard.edu/abs/2013ARA&A..51..511K} {51, 511}

\bibitem[\protect\citeauthoryear{{Kormendy} \& {Richstone}}{{Kormendy} \&
  {Richstone}}{1995}]{kor95}
{Kormendy} J.,  {Richstone} D.,  1995, \mn@doi [\araa]
  {10.1146/annurev.aa.33.090195.003053}, \href
  {https://ui.adsabs.harvard.edu/abs/1995ARA&A..33..581K} {33, 581}

\bibitem[\protect\citeauthoryear{{Krabbe}, {B{\"o}ker}  \& {Maiolino}}{{Krabbe}
  et~al.}{2001}]{kra01}
{Krabbe} A.,  {B{\"o}ker} T.,   {Maiolino} R.,  2001, \mn@doi [\apj]
  {10.1086/321679}, \href
  {https://ui.adsabs.harvard.edu/abs/2001ApJ...557..626K} {557, 626}

\bibitem[\protect\citeauthoryear{{Magorrian} et~al.,}{{Magorrian}
  et~al.}{1998}]{mag98}
{Magorrian} J.,  et~al., 1998, \mn@doi [\aj] {10.1086/300353}, \href
  {https://ui.adsabs.harvard.edu/abs/1998AJ....115.2285M} {115, 2285}

\bibitem[\protect\citeauthoryear{{Makarov}, {Prugniel}, {Terekhova}, {Courtois}
   \& {Vauglin}}{{Makarov} et~al.}{2014}]{mak14}
{Makarov} D.,  {Prugniel} P.,  {Terekhova} N.,  {Courtois} H.,   {Vauglin} I.,
  2014, \mn@doi [\aap] {10.1051/0004-6361/201423496}, \href
  {https://ui.adsabs.harvard.edu/abs/2014A&A...570A..13M} {570, A13}

\bibitem[\protect\citeauthoryear{{Menezes}, {Steiner}  \& {Ricci}}{{Menezes}
  et~al.}{2014a}]{men14}
{Menezes} R.~B.,  {Steiner} J.~E.,   {Ricci} T.~V.,  2014a, \mn@doi [\mnras]
  {10.1093/mnras/stt2381}, \href
  {https://ui.adsabs.harvard.edu/abs/2014MNRAS.438.2597M} {438, 2597}

\bibitem[\protect\citeauthoryear{{Menezes}, {Steiner}  \& {Ricci}}{{Menezes}
  et~al.}{2014b}]{men14b}
{Menezes} R.~B.,  {Steiner} J.~E.,   {Ricci} T.~V.,  2014b, \mn@doi [\apjl]
  {10.1088/2041-8205/796/1/L13}, \href
  {https://ui.adsabs.harvard.edu/abs/2014ApJ...796L..13M} {796, L13}

\bibitem[\protect\citeauthoryear{{Menezes}, {da Silva}, {Ricci}, {Steiner},
  {May}  \& {Borges}}{{Menezes} et~al.}{2015}]{men15}
{Menezes} R.~B.,  {da Silva} P.,  {Ricci} T.~V.,  {Steiner} J.~E.,  {May} D.,
  {Borges} B.~W.,  2015, \mn@doi [\mnras] {10.1093/mnras/stv629}, \href
  {https://ui.adsabs.harvard.edu/abs/2015MNRAS.450..369M} {450, 369}

\bibitem[\protect\citeauthoryear{{Menezes}, {Ricci}, {Steiner}, {da Silva},
  {Ferrari}  \& {Borges}}{{Menezes} et~al.}{2019}]{men19}
{Menezes} R.~B.,  {Ricci} T.~V.,  {Steiner} J.~E.,  {da Silva} P.,  {Ferrari}
  F.,   {Borges} B.~W.,  2019, \mn@doi [\mnras] {10.1093/mnras/sty3337}, \href
  {https://ui.adsabs.harvard.edu/abs/2019MNRAS.483.3700M} {483, 3700}

\bibitem[\protect\citeauthoryear{{M{\"u}ller-S{\'a}nchez},
  {Gonz{\'a}lez-Mart{\'\i}n}, {Fern{\'a}ndez-Ontiveros}, {Acosta-Pulido}  \&
  {Prieto}}{{M{\"u}ller-S{\'a}nchez} et~al.}{2010}]{mul10}
{M{\"u}ller-S{\'a}nchez} F.,  {Gonz{\'a}lez-Mart{\'\i}n} O.,
  {Fern{\'a}ndez-Ontiveros} J.~A.,  {Acosta-Pulido} J.~A.,   {Prieto} M.~A.,
  2010, \mn@doi [\apj] {10.1088/0004-637X/716/2/1166}, \href
  {https://ui.adsabs.harvard.edu/abs/2010ApJ...716.1166M} {716, 1166}

\bibitem[\protect\citeauthoryear{{Netzer}}{{Netzer}}{2013}]{net13}
{Netzer} H.,  2013, {The Physics and Evolution of Active Galactic Nuclei}

\bibitem[\protect\citeauthoryear{{Osterbrock} \& {Ferland}}{{Osterbrock} \&
  {Ferland}}{2006}]{ost06}
{Osterbrock} D.~E.,  {Ferland} G.~J.,  2006, {Astrophysics of gaseous nebulae
  and active galactic nuclei}

\bibitem[\protect\citeauthoryear{{Phillips}}{{Phillips}}{1979}]{phi79}
{Phillips} M.~M.,  1979, \mn@doi [\apjl] {10.1086/182881}, \href
  {https://ui.adsabs.harvard.edu/abs/1979ApJ...227L.121P} {227, L121}

\bibitem[\protect\citeauthoryear{{Phillips}}{{Phillips}}{1993}]{phi93}
{Phillips} A.~C.,  1993, \mn@doi [\aj] {10.1086/116447}, \href
  {https://ui.adsabs.harvard.edu/abs/1993AJ....105..486P} {105, 486}

\bibitem[\protect\citeauthoryear{{Richstone} et~al.,}{{Richstone}
  et~al.}{1998}]{ric98}
{Richstone} D.,  et~al., 1998, \nat, \href
  {https://ui.adsabs.harvard.edu/abs/1998Natur.395A..14R} {385, A14}

\bibitem[\protect\citeauthoryear{{Saglia} et~al.,}{{Saglia}
  et~al.}{2016}]{sag16}
{Saglia} R.~P.,  et~al., 2016, \mn@doi [\apj] {10.3847/0004-637X/818/1/47},
  \href {https://ui.adsabs.harvard.edu/abs/2016ApJ...818...47S} {818, 47}

\bibitem[\protect\citeauthoryear{{Salak}, {Nakai}, {Hatakeyama}  \&
  {Miyamoto}}{{Salak} et~al.}{2016}]{sal16}
{Salak} D.,  {Nakai} N.,  {Hatakeyama} T.,   {Miyamoto} Y.,  2016, \mn@doi
  [\apj] {10.3847/0004-637X/823/1/68}, \href
  {https://ui.adsabs.harvard.edu/abs/2016ApJ...823...68S} {823, 68}

\bibitem[\protect\citeauthoryear{{S{\'a}nchez-Bl{\'a}zquez}
  et~al.,}{{S{\'a}nchez-Bl{\'a}zquez} et~al.}{2006}]{san06}
{S{\'a}nchez-Bl{\'a}zquez} P.,  et~al., 2006, \mn@doi [\mnras]
  {10.1111/j.1365-2966.2006.10699.x}, \href
  {https://ui.adsabs.harvard.edu/abs/2006MNRAS.371..703S} {371, 703}

\bibitem[\protect\citeauthoryear{{S{\'a}nchez} et~al.,}{{S{\'a}nchez}
  et~al.}{2012}]{san12}
{S{\'a}nchez} S.~F.,  et~al., 2012, \mn@doi [\aap]
  {10.1051/0004-6361/201117353}, \href
  {https://ui.adsabs.harvard.edu/abs/2012A&A...538A...8S} {538, A8}

\bibitem[\protect\citeauthoryear{{Sandage} \& {Tammann}}{{Sandage} \&
  {Tammann}}{1981}]{san81}
{Sandage} A.,  {Tammann} G.~A.,  1981, {A Revised Shapley-Ames Catalog of
  Bright Galaxies}

\bibitem[\protect\citeauthoryear{{Schawinski}, {Thomas}, {Sarzi}, {Maraston},
  {Kaviraj}, {Joo}, {Yi}  \& {Silk}}{{Schawinski} et~al.}{2007}]{sch07}
{Schawinski} K.,  {Thomas} D.,  {Sarzi} M.,  {Maraston} C.,  {Kaviraj} S.,
  {Joo} S.-J.,  {Yi} S.~K.,   {Silk} J.,  2007, \mn@doi [\mnras]
  {10.1111/j.1365-2966.2007.12487.x}, \href
  {https://ui.adsabs.harvard.edu/abs/2007MNRAS.382.1415S} {382, 1415}

\bibitem[\protect\citeauthoryear{{Schimoia}, {Storchi-Bergmann}, {Winge},
  {Nemmen}  \& {Eracleous}}{{Schimoia} et~al.}{2017}]{sch17}
{Schimoia} J.~S.,  {Storchi-Bergmann} T.,  {Winge} C.,  {Nemmen} R.~S.,
  {Eracleous} M.,  2017, \mn@doi [\mnras] {10.1093/mnras/stx2107}, \href
  {https://ui.adsabs.harvard.edu/abs/2017MNRAS.472.2170S} {472, 2170}

\bibitem[\protect\citeauthoryear{{Schlafly} \& {Finkbeiner}}{{Schlafly} \&
  {Finkbeiner}}{2011}]{sch11}
{Schlafly} E.~F.,  {Finkbeiner} D.~P.,  2011, \mn@doi [\apj]
  {10.1088/0004-637X/737/2/103}, \href
  {https://ui.adsabs.harvard.edu/abs/2011ApJ...737..103S} {737, 103}

\bibitem[\protect\citeauthoryear{{Schmidt}}{{Schmidt}}{1968}]{sch68}
{Schmidt} M.,  1968, \mn@doi [\apj] {10.1086/149446}, \href
  {https://ui.adsabs.harvard.edu/abs/1968ApJ...151..393S} {151, 393}

\bibitem[\protect\citeauthoryear{{Sharp} \& {Bland-Hawthorn}}{{Sharp} \&
  {Bland-Hawthorn}}{2010}]{sha10}
{Sharp} R.~G.,  {Bland-Hawthorn} J.,  2010, \mn@doi [\apj]
  {10.1088/0004-637X/711/2/818}, \href
  {https://ui.adsabs.harvard.edu/abs/2010ApJ...711..818S} {711, 818}

\bibitem[\protect\citeauthoryear{{She}, {Ho}  \& {Feng}}{{She}
  et~al.}{2017}]{she17}
{She} R.,  {Ho} L.~C.,   {Feng} H.,  2017, VizieR Online Data Catalog, \href
  {https://ui.adsabs.harvard.edu/abs/2017yCat..18350223S} {p. J/ApJ/835/223}

\bibitem[\protect\citeauthoryear{{Shields}}{{Shields}}{1992}]{shi92}
{Shields} J.~C.,  1992, \mn@doi [\apjl] {10.1086/186598}, \href
  {https://ui.adsabs.harvard.edu/abs/1992ApJ...399L..27S} {399, L27}

\bibitem[\protect\citeauthoryear{{Shields} et~al.,}{{Shields}
  et~al.}{2007}]{shi07}
{Shields} J.~C.,  et~al., 2007, \mn@doi [\apj] {10.1086/509059}, \href
  {https://ui.adsabs.harvard.edu/abs/2007ApJ...654..125S} {654, 125}

\bibitem[\protect\citeauthoryear{{Springel}, {Di Matteo}  \&
  {Hernquist}}{{Springel} et~al.}{2005}]{spr05}
{Springel} V.,  {Di Matteo} T.,   {Hernquist} L.,  2005, \mn@doi [\mnras]
  {10.1111/j.1365-2966.2005.09238.x}, \href
  {https://ui.adsabs.harvard.edu/abs/2005MNRAS.361..776S} {361, 776}

\bibitem[\protect\citeauthoryear{{Stasi{\'n}ska} et~al.,}{{Stasi{\'n}ska}
  et~al.}{2008}]{sta08}
{Stasi{\'n}ska} G.,  et~al., 2008, \mn@doi [\mnras]
  {10.1111/j.1745-3933.2008.00550.x}, \href
  {https://ui.adsabs.harvard.edu/abs/2008MNRAS.391L..29S} {391, L29}

\bibitem[\protect\citeauthoryear{{Steiner} et~al.,}{{Steiner}
  et~al.}{2022}]{ste22}
{Steiner} J.~E.,  et~al., 2022, \mn@doi [\mnras] {10.1093/mnras/stac034}, \href
  {https://ui.adsabs.harvard.edu/abs/2022MNRAS.510.5780S} {510, 5780}

\bibitem[\protect\citeauthoryear{{Storchi-Bergmann}, {Baldwin}  \&
  {Wilson}}{{Storchi-Bergmann} et~al.}{1993}]{sto93}
{Storchi-Bergmann} T.,  {Baldwin} J.~A.,   {Wilson} A.~S.,  1993, \mn@doi
  [\apjl] {10.1086/186867}, \href
  {https://ui.adsabs.harvard.edu/abs/1993ApJ...410L..11S} {410, L11}

\bibitem[\protect\citeauthoryear{{Storchi-Bergmann} et~al.,}{{Storchi-Bergmann}
  et~al.}{2003}]{sto03}
{Storchi-Bergmann} T.,  et~al., 2003, \mn@doi [\apj] {10.1086/378938}, \href
  {https://ui.adsabs.harvard.edu/abs/2003ApJ...598..956S} {598, 956}

\bibitem[\protect\citeauthoryear{{Strickland}, {Heckman}, {Weaver}  \&
  {Dahlem}}{{Strickland} et~al.}{2000}]{str00}
{Strickland} D.~K.,  {Heckman} T.~M.,  {Weaver} K.~A.,   {Dahlem} M.,  2000,
  \mn@doi [\aj] {10.1086/316846}, \href
  {https://ui.adsabs.harvard.edu/abs/2000AJ....120.2965S} {120, 2965}

\bibitem[\protect\citeauthoryear{{Terlevich} \& {Melnick}}{{Terlevich} \&
  {Melnick}}{1985}]{ter85}
{Terlevich} R.,  {Melnick} J.,  1985, \mn@doi [\mnras]
  {10.1093/mnras/213.4.841}, \href
  {https://ui.adsabs.harvard.edu/abs/1985MNRAS.213..841T} {213, 841}

\bibitem[\protect\citeauthoryear{{Vazdekis}, {S{\'a}nchez-Bl{\'a}zquez},
  {Falc{\'o}n-Barroso}, {Cenarro}, {Beasley}, {Cardiel}, {Gorgas}  \&
  {Peletier}}{{Vazdekis} et~al.}{2010}]{vaz10}
{Vazdekis} A.,  {S{\'a}nchez-Bl{\'a}zquez} P.,  {Falc{\'o}n-Barroso} J.,
  {Cenarro} A.~J.,  {Beasley} M.~A.,  {Cardiel} N.,  {Gorgas} J.,   {Peletier}
  R.~F.,  2010, \mn@doi [\mnras] {10.1111/j.1365-2966.2010.16407.x}, \href
  {https://ui.adsabs.harvard.edu/abs/2010MNRAS.404.1639V} {404, 1639}

\bibitem[\protect\citeauthoryear{{Veilleux} \& {Osterbrock}}{{Veilleux} \&
  {Osterbrock}}{1987}]{vei87}
{Veilleux} S.,  {Osterbrock} D.~E.,  1987, \mn@doi [\apjs] {10.1086/191166},
  \href {https://ui.adsabs.harvard.edu/abs/1987ApJS...63..295V} {63, 295}

\bibitem[\protect\citeauthoryear{{Veron-Cetty} \& {Veron}}{{Veron-Cetty} \&
  {Veron}}{1985}]{ver85}
{Veron-Cetty} M.~P.,  {Veron} P.,  1985, \aap, \href
  {https://ui.adsabs.harvard.edu/abs/1985A&A...145..425V} {145, 425}

\bibitem[\protect\citeauthoryear{{Walter} et~al.,}{{Walter}
  et~al.}{2017}]{wal17}
{Walter} F.,  et~al., 2017, \mn@doi [\apj] {10.3847/1538-4357/835/2/265}, \href
  {https://ui.adsabs.harvard.edu/abs/2017ApJ...835..265W} {835, 265}

\bibitem[\protect\citeauthoryear{{Weaver}, {Heckman}, {Strickland}  \&
  {Dahlem}}{{Weaver} et~al.}{2002}]{wea02}
{Weaver} K.~A.,  {Heckman} T.~M.,  {Strickland} D.~K.,   {Dahlem} M.,  2002,
  \mn@doi [\apjl] {10.1086/342977}, \href
  {https://ui.adsabs.harvard.edu/abs/2002ApJ...576L..19W} {576, L19}

\bibitem[\protect\citeauthoryear{{Westmoquette}, {Smith}  \&
  {Gallagher}}{{Westmoquette} et~al.}{2011}]{wes11}
{Westmoquette} M.~S.,  {Smith} L.~J.,   {Gallagher} J.~S. I.,  2011, \mn@doi
  [\mnras] {10.1111/j.1365-2966.2011.18675.x}, \href
  {https://ui.adsabs.harvard.edu/abs/2011MNRAS.414.3719W} {414, 3719}

\bibitem[\protect\citeauthoryear{{Xiao}, {Barth}, {Greene}, {Ho}, {Bentz},
  {Ludwig}  \& {Jiang}}{{Xiao} et~al.}{2011}]{xia11}
{Xiao} T.,  {Barth} A.~J.,  {Greene} J.~E.,  {Ho} L.~C.,  {Bentz} M.~C.,
  {Ludwig} R.~R.,   {Jiang} Y.,  2011, \mn@doi [\apj]
  {10.1088/0004-637X/739/1/28}, \href
  {https://ui.adsabs.harvard.edu/abs/2011ApJ...739...28X} {739, 28}

\makeatother
\end{thebibliography}

\appendix

\section[]{Nuclear spectra of the sample}\label{appA}

All the spectra extracted from circular regions, centred on the peak of stellar emission in the data cubes of the galaxies in the mini-DIVING$^\mathrm{3D}$ sample, together with the fits provided by the pPXF technique and the fit residuals (corresponding to the emission-line spectra), are shown in the following figures.

\begin{figure*}
\begin{center}
  \includegraphics[scale=0.6]{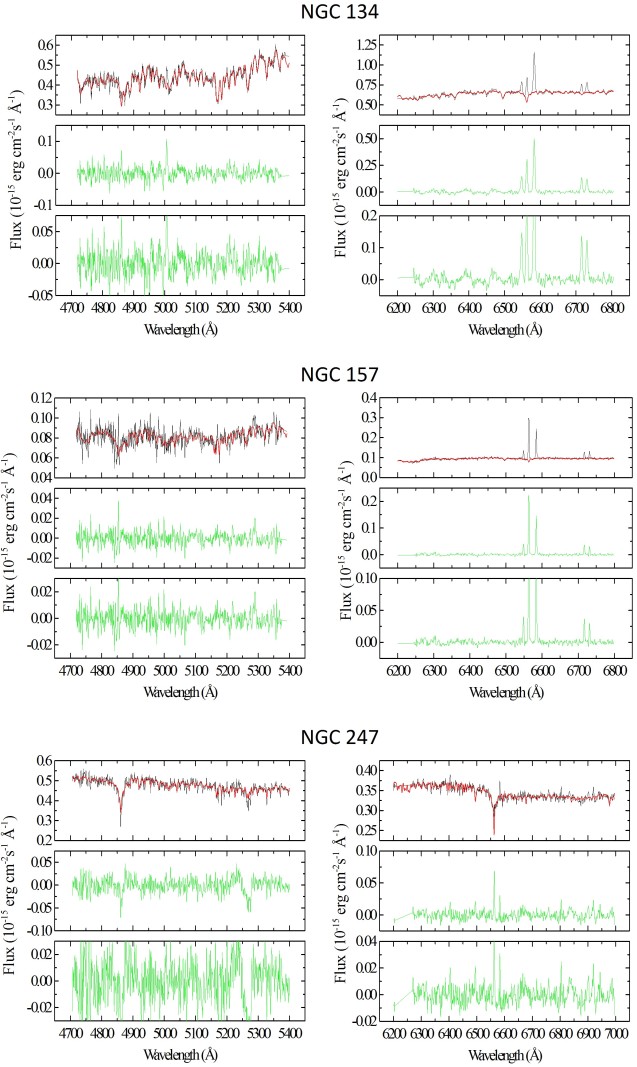}
  \caption{Nuclear spectra extracted from the data cubes of NGC 134, NGC 157, and NGC 247. The fits provided by the pPXF procedure are shown in red and the fit residuals are shown in green. Magnifications of the fit residuals are also shown.\label{fig01A}}
\end{center}
\end{figure*}

\begin{figure*}
\begin{center}
  \includegraphics[scale=0.6]{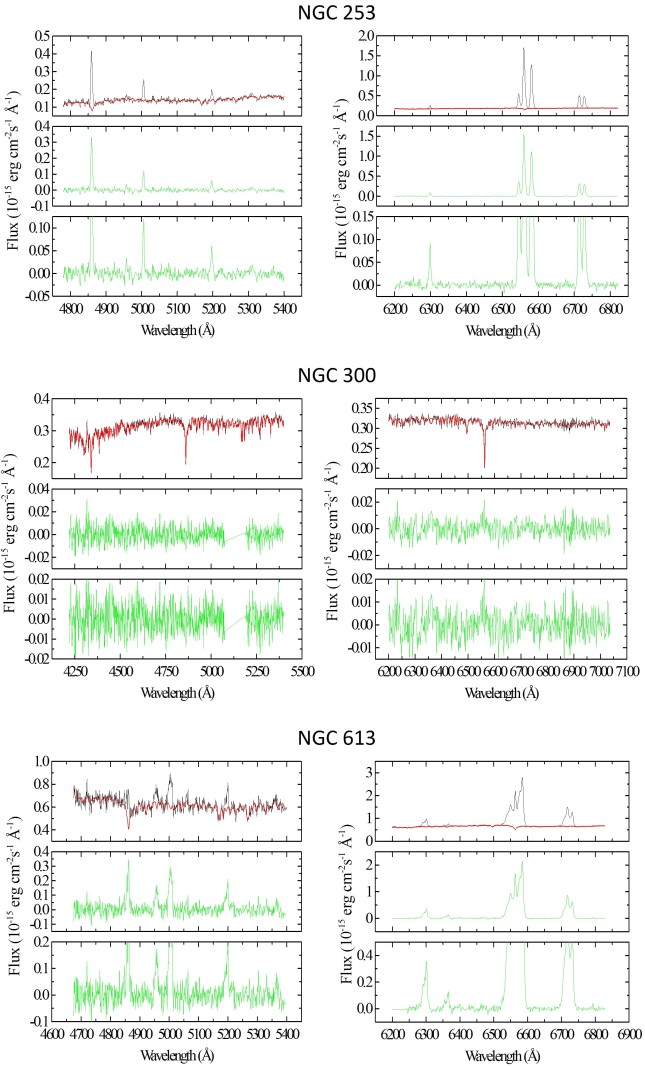}
  \caption{The same as in Fig.~\ref{fig01A}, but for the data cubes of NGC 253, NGC 300, and NGC 613.\label{fig02A}}
\end{center}
\end{figure*}

\begin{figure*}
\begin{center}
  \includegraphics[scale=0.6]{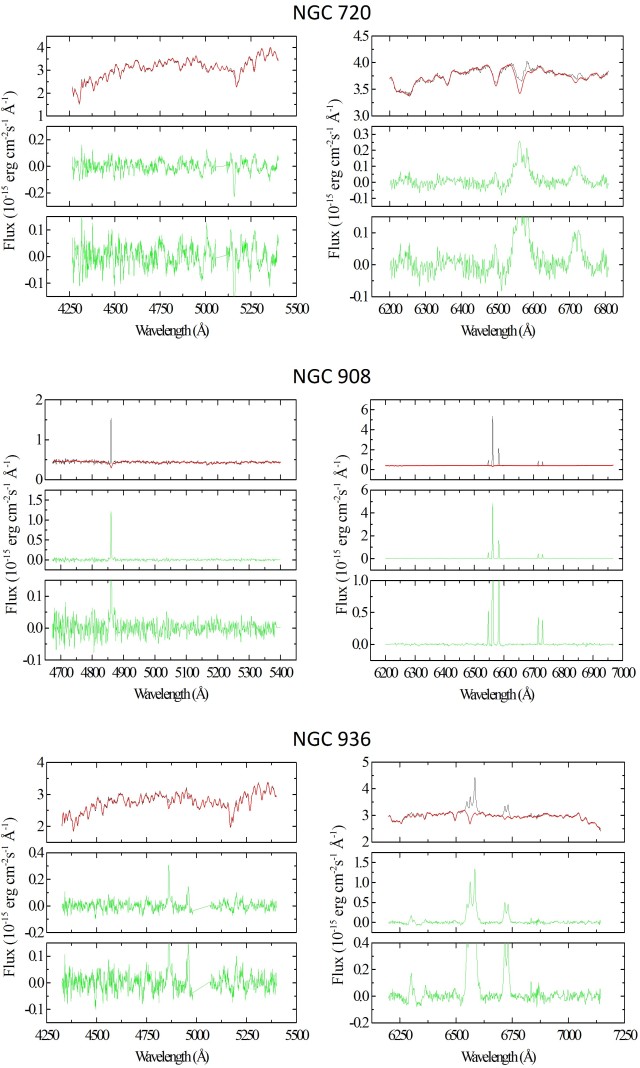}
  \caption{The same as in Fig.~\ref{fig01A}, but for the data cubes of NGC 720, NGC 908, and NGC 936.\label{fig03A}}
\end{center}
\end{figure*}

\begin{figure*}
\begin{center}
  \includegraphics[scale=0.6]{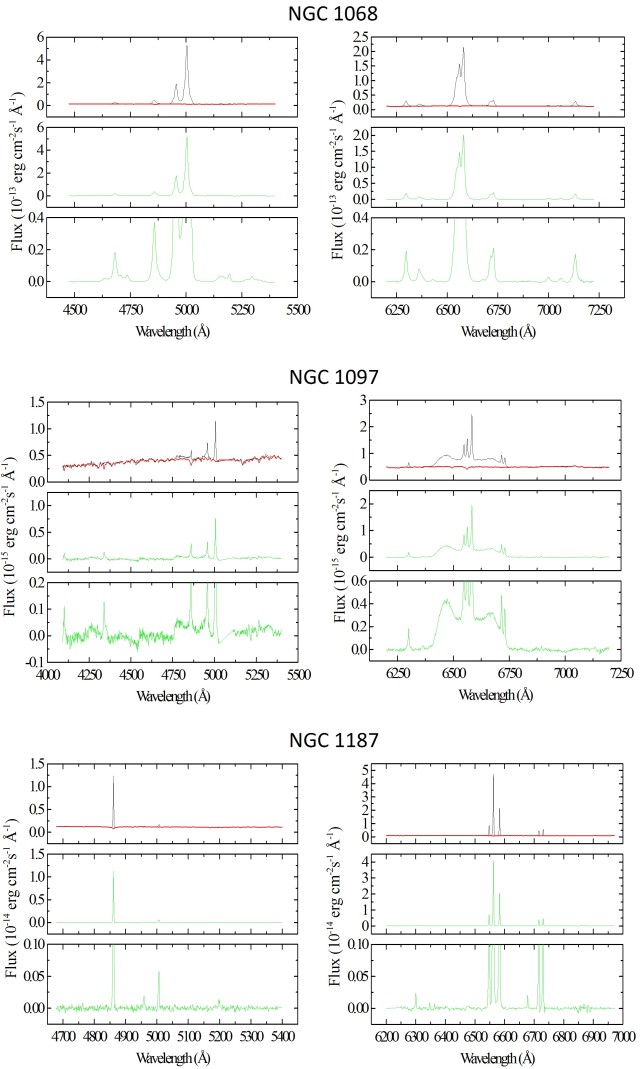}
  \caption{The same as in Fig.~\ref{fig01A}, but for the data cubes of NGC 1068, NGC 1097, and NGC 1187.\label{fig04A}}
\end{center}
\end{figure*}

\begin{figure*}
\begin{center}
  \includegraphics[scale=0.6]{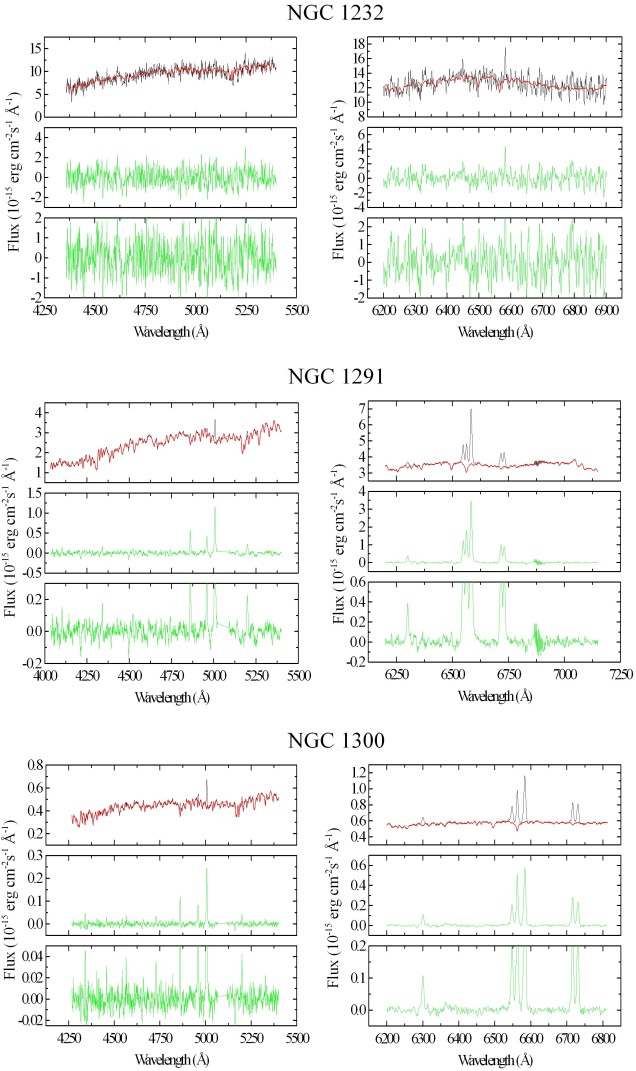}
  \caption{The same as in Fig.~\ref{fig01A}, but for the data cubes of NGC 1232, NGC 1291, and NGC 1300.\label{fig05A}}
\end{center}
\end{figure*}

\begin{figure*}
\begin{center}
  \includegraphics[scale=0.6]{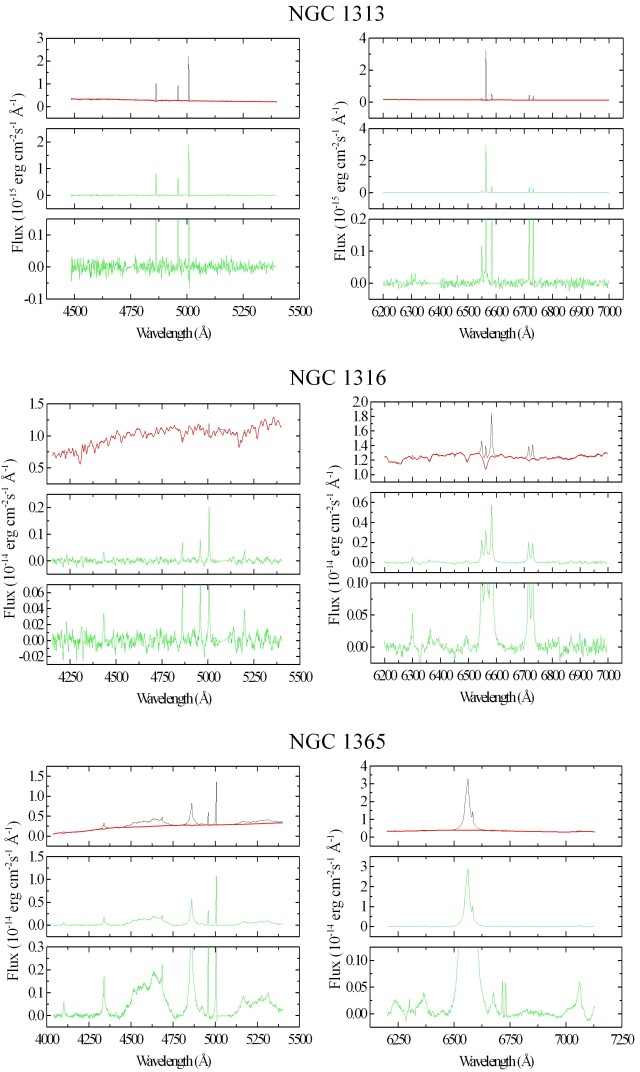}
  \caption{The same as in Fig.~\ref{fig01A}, but for the data cubes of NGC 1313, NGC 1316, and NGC 1365.\label{fig06A}}
\end{center}
\end{figure*}

\begin{figure*}
\begin{center}
  \includegraphics[scale=0.6]{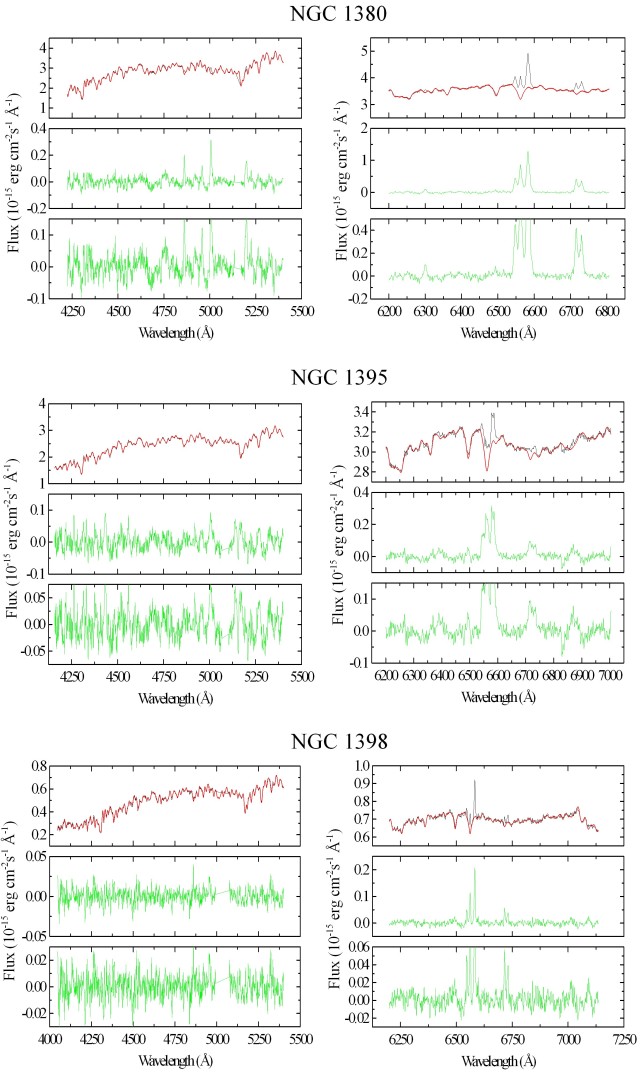}
  \caption{The same as in Fig.~\ref{fig01A}, but for the data cubes of NGC 1380, NGC 1395, and NGC 1398.\label{fig07A}}
\end{center}
\end{figure*}

\begin{figure*}
\begin{center}
  \includegraphics[scale=0.6]{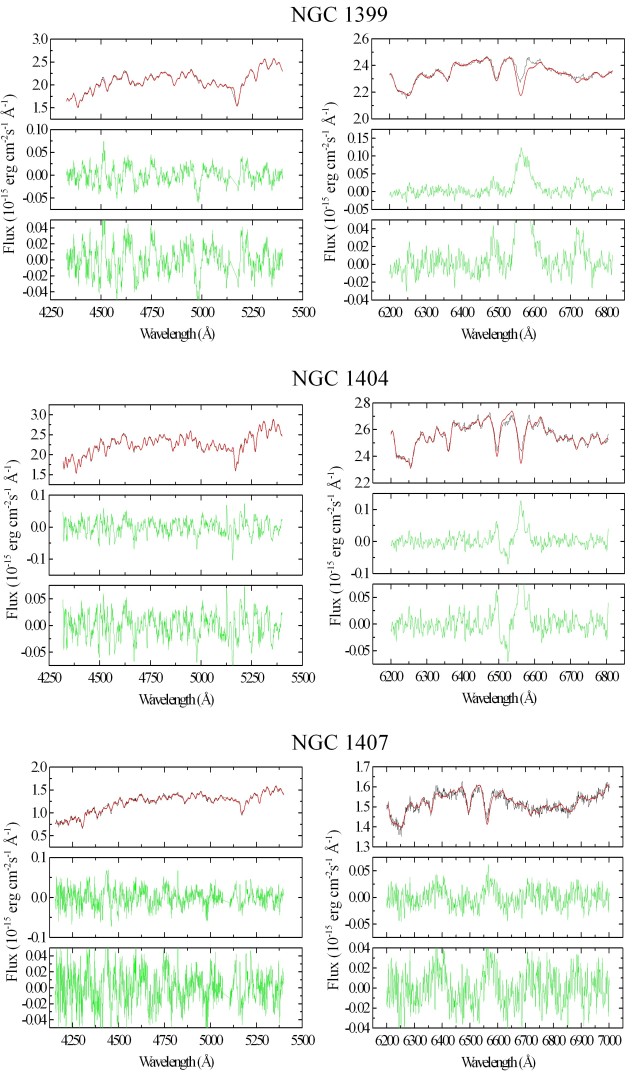}
  \caption{The same as in Fig.~\ref{fig01A}, but for the data cubes of NGC 1399, NGC 1404, and NGC 1407.\label{fig08A}}
\end{center}
\end{figure*}

\begin{figure*}
\begin{center}
  \includegraphics[scale=0.6]{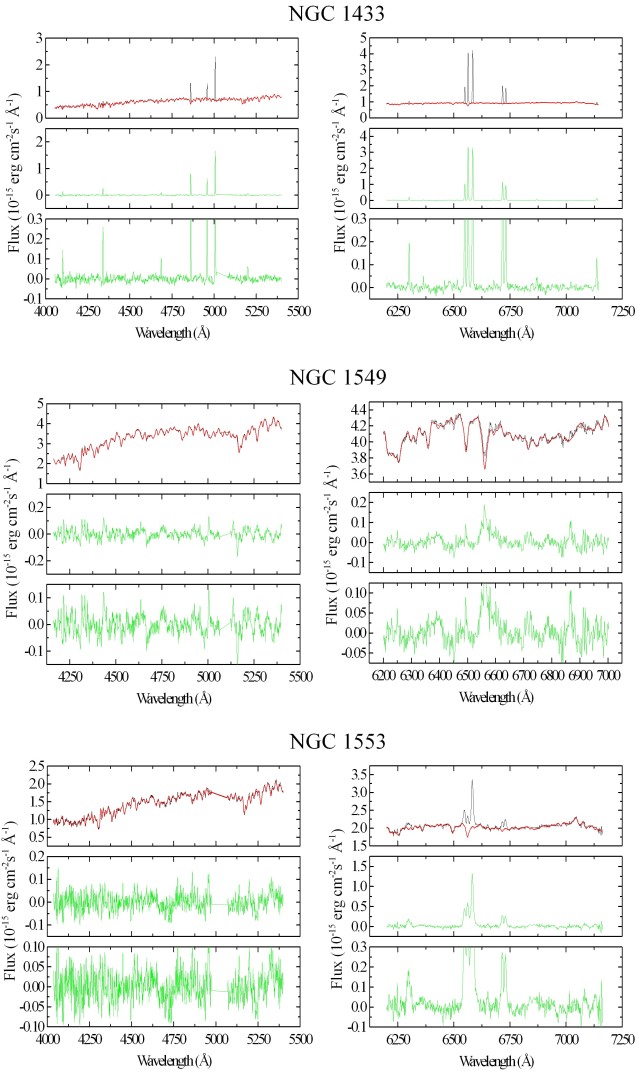}
  \caption{The same as in Fig.~\ref{fig01A}, but for the data cubes of NGC 1433, NGC 1549, and NGC 1553.\label{fig09A}}
\end{center}
\end{figure*}

\begin{figure*}
\begin{center}
  \includegraphics[scale=0.6]{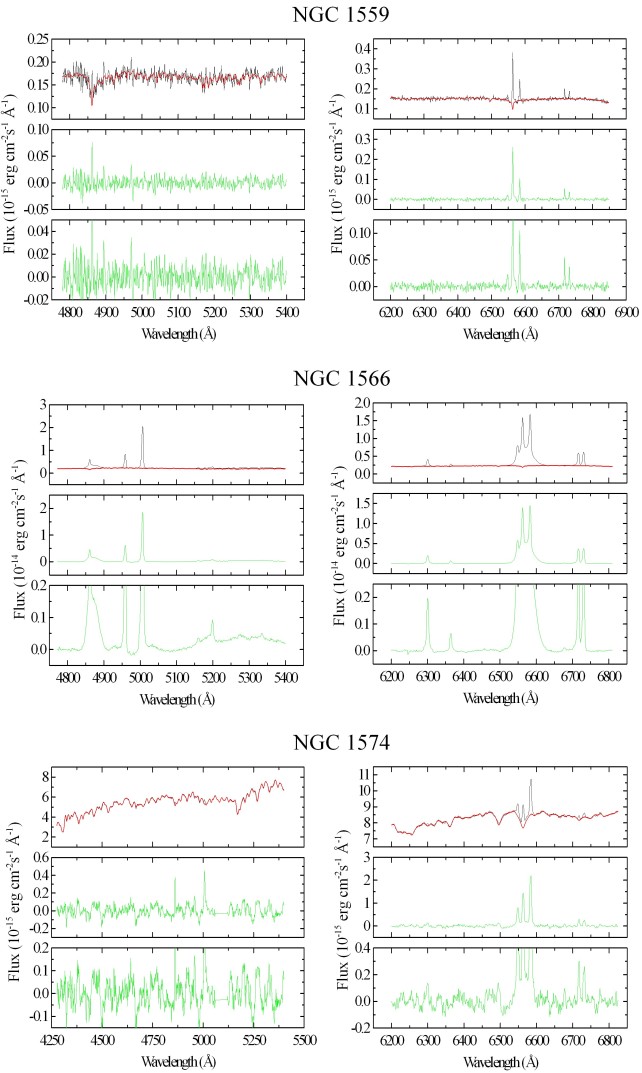}
  \caption{The same as in Fig.~\ref{fig01A}, but for the data cubes of NGC 1559, NGC 1566, and NGC 1574.\label{fig10A}}
\end{center}
\end{figure*}

\begin{figure*}
\begin{center}
  \includegraphics[scale=0.6]{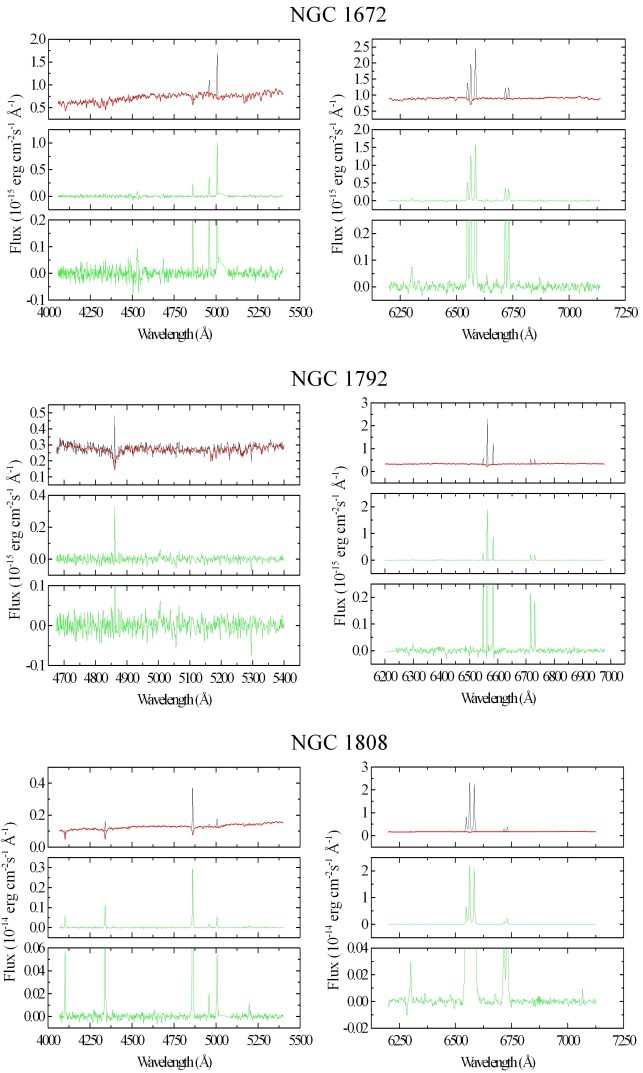}
  \caption{The same as in Fig.~\ref{fig01A}, but for the data cubes of NGC 1672, NGC 1792, and NGC 1808.\label{fig11A}}
\end{center}
\end{figure*}

\begin{figure*}
\begin{center}
  \includegraphics[scale=0.6]{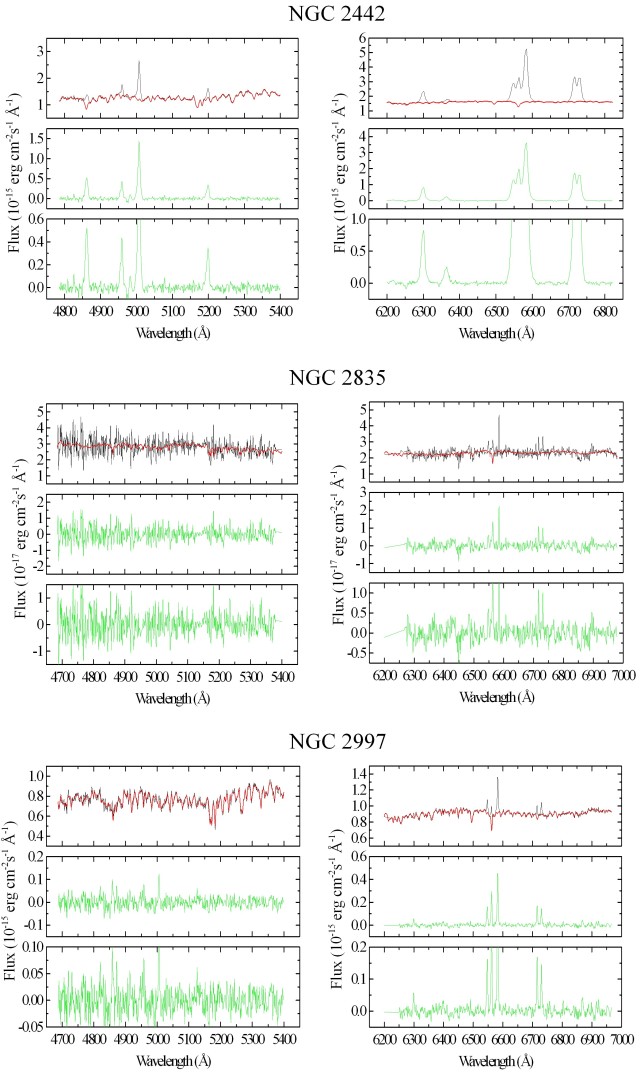}
  \caption{The same as in Fig.~\ref{fig01A}, but for the data cubes of NGC 2442, NGC 2835, and NGC 2997.\label{fig12A}}
\end{center}
\end{figure*}

\begin{figure*}
\begin{center}
  \includegraphics[scale=0.6]{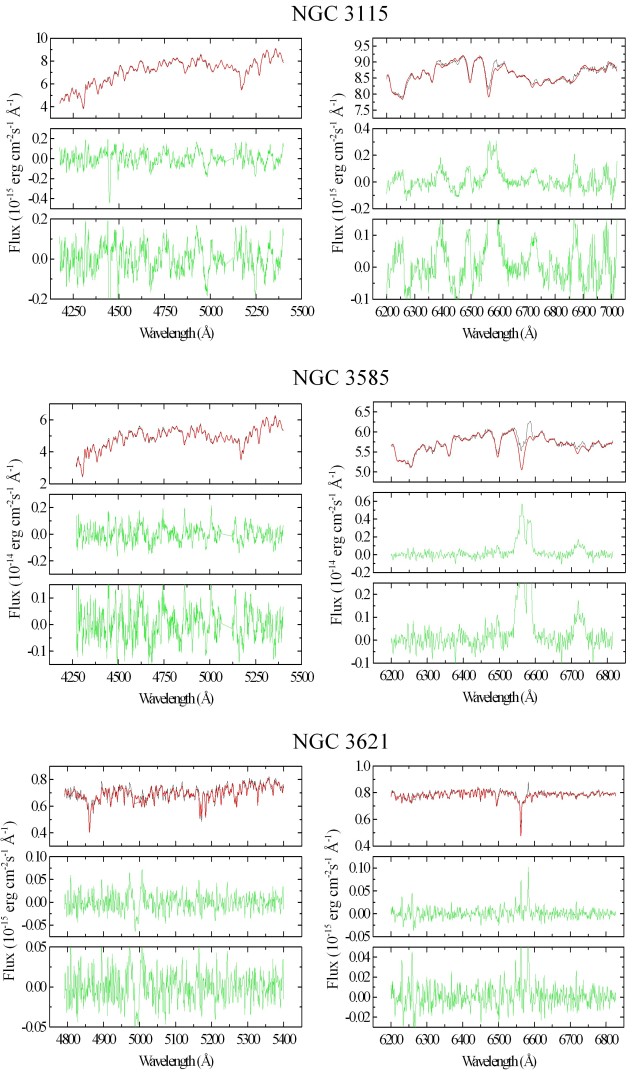}
  \caption{The same as in Fig.~\ref{fig01A}, but for the data cubes of NGC 3115, NGC 3585, and NGC 3621.\label{fig13A}}
\end{center}
\end{figure*}

\begin{figure*}
\begin{center}
  \includegraphics[scale=0.6]{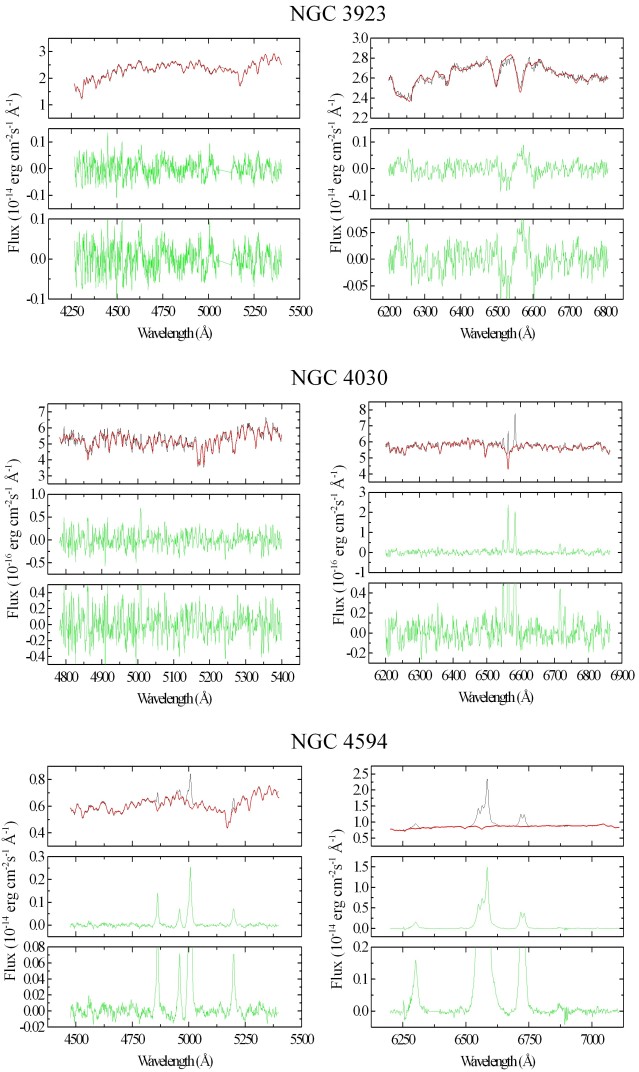}
  \caption{The same as in Fig.~\ref{fig01A}, but for the data cubes of NGC 3923, NGC 4030, and NGC 4594.\label{fig14A}}
\end{center}
\end{figure*}

\begin{figure*}
\begin{center}
  \includegraphics[scale=0.6]{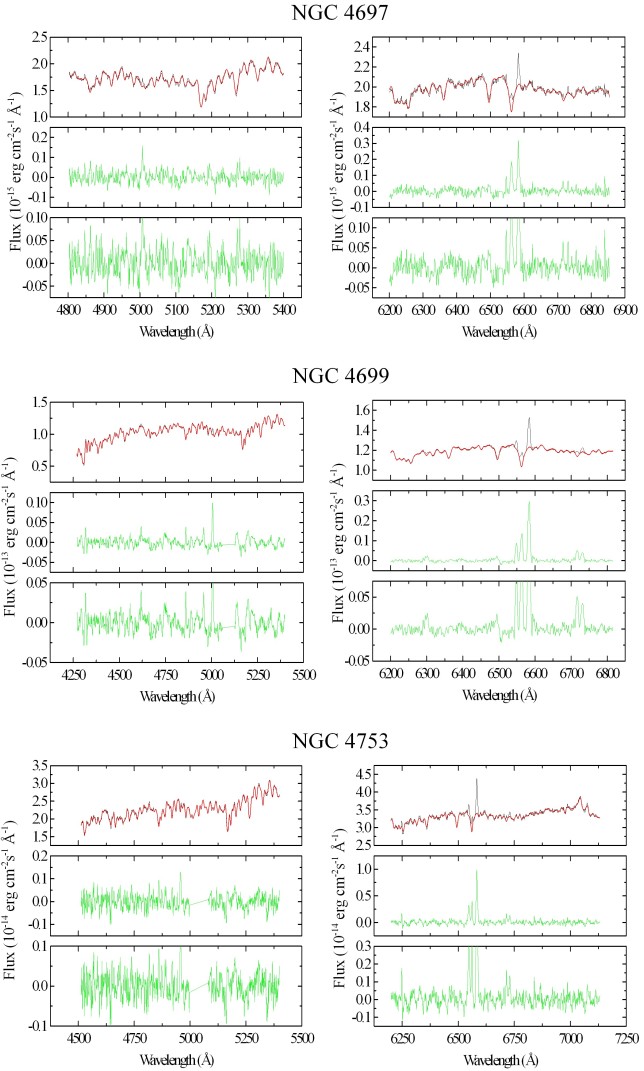}
  \caption{The same as in Fig.~\ref{fig01A}, but for the data cubes of NGC 4697, NGC 4699, and NGC 4753.\label{fig15A}}
\end{center}
\end{figure*}

\begin{figure*}
\begin{center}
  \includegraphics[scale=0.6]{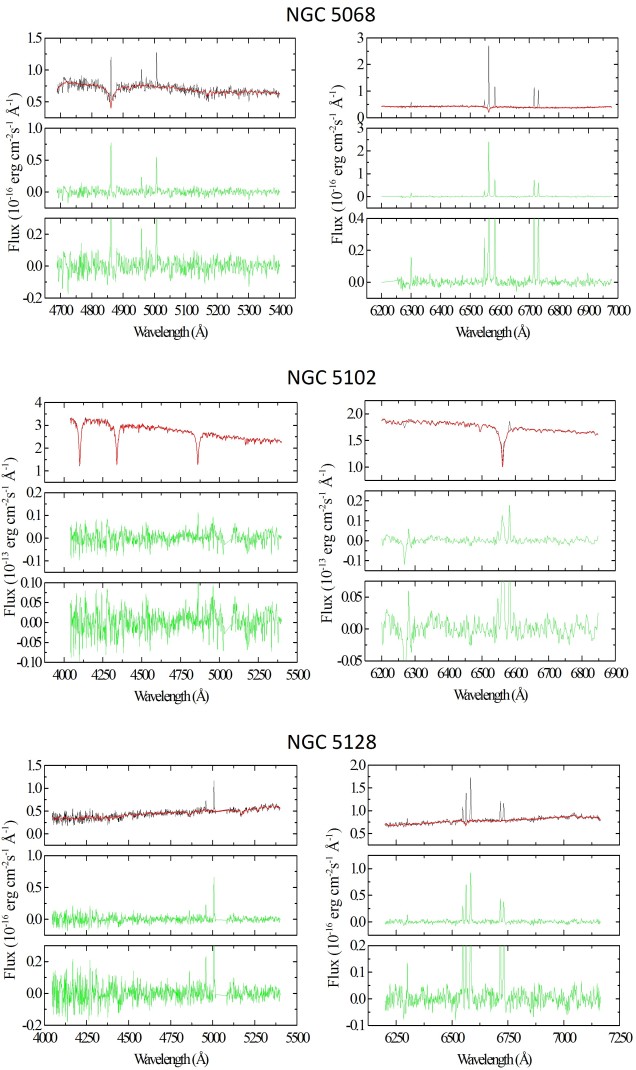}
  \caption{The same as in Fig.~\ref{fig01A}, but for the data cubes of NGC 5068, NGC 5102, and NGC 5128.\label{fig16A}}
\end{center}
\end{figure*}

\begin{figure*}
\begin{center}
  \includegraphics[scale=0.6]{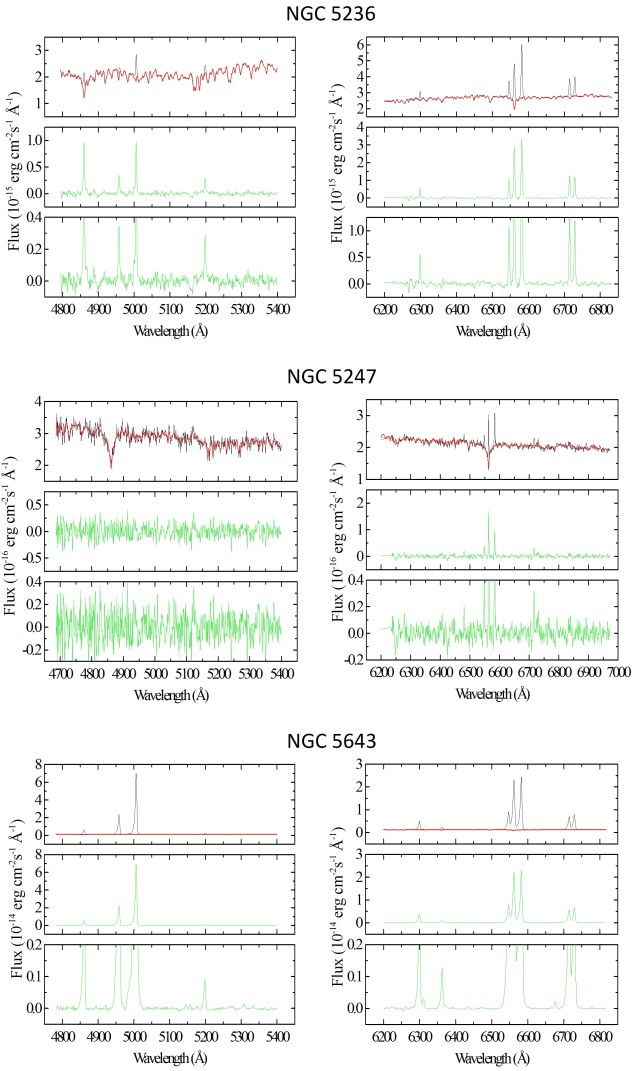}
  \caption{The same as in Fig.~\ref{fig01A}, but for the data cubes of NGC 5236, NGC 5247, and NGC 5643.\label{fig17A}}
\end{center}
\end{figure*}

\begin{figure*}
\begin{center}
  \includegraphics[scale=0.6]{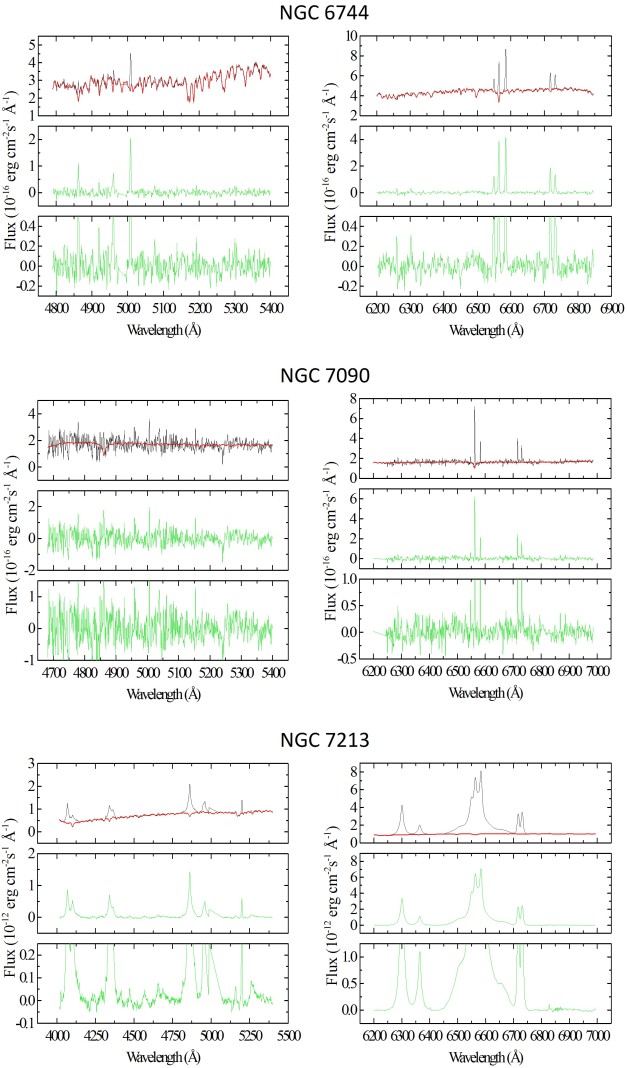}
  \caption{The same as in Fig.~\ref{fig01A}, but for the data cubes of NGC 6744, NGC 7090, and NGC 7213.\label{fig18A}}
\end{center}
\end{figure*}

\begin{figure*}
\begin{center}
  \includegraphics[scale=0.6]{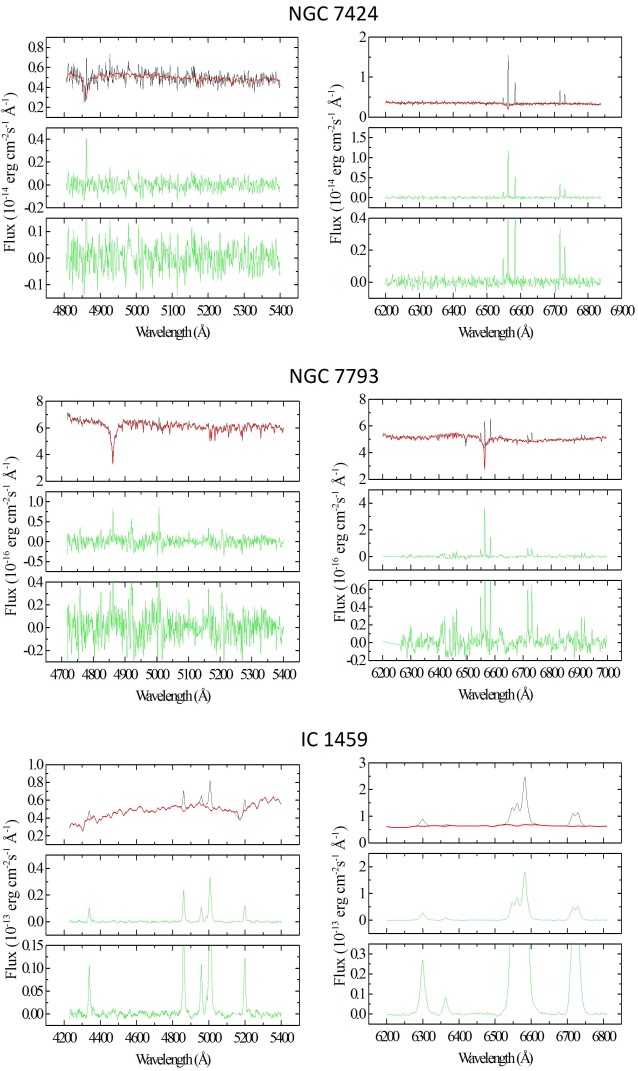}
  \caption{The same as in Fig.~\ref{fig01A}, but for the data cubes of NGC 7424, NGC 7793, and IC 1459.\label{fig19A}}
\end{center}
\end{figure*}

\section[]{Gaussian fits of the blended emission lines}\label{appB}

All the Gaussian fits applied to the blended emission lines in the nuclear spectra of the galaxies in the mini-DIVING$^\mathrm{3D}$ sample are shown in the following figures.

\begin{figure*}
\begin{center}
  \includegraphics[scale=0.6]{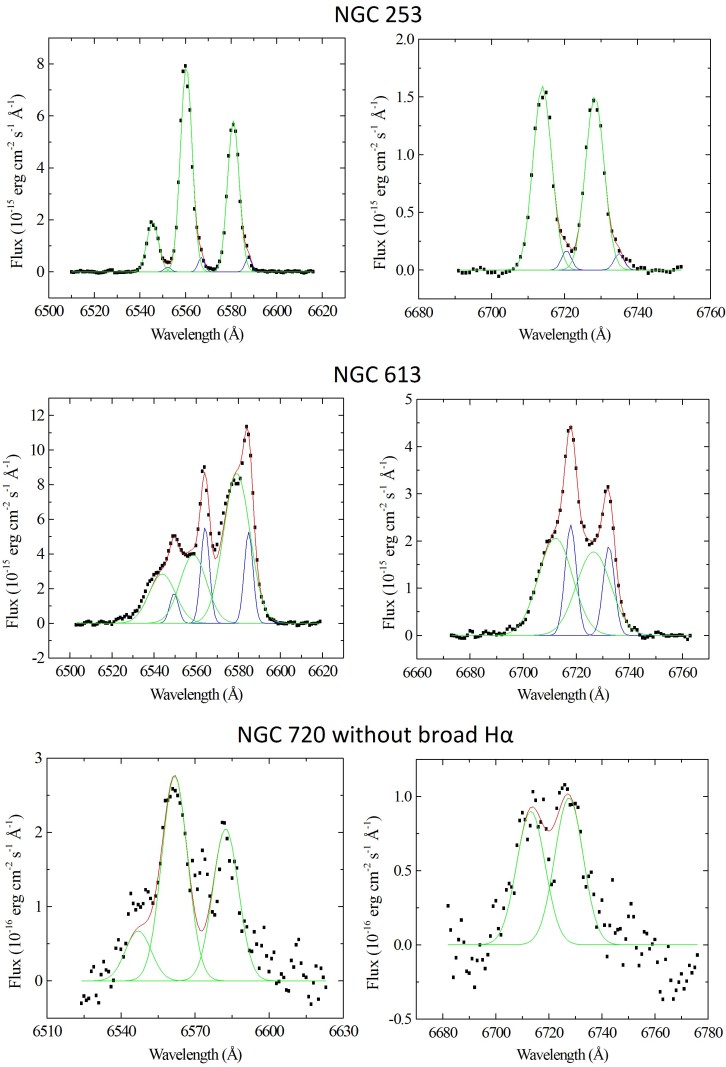}
  \caption{Gaussian fits of the blended emission lines in the data cubes of NGC 253, NGC 613, and NGC 720 (without a broad H$\alpha$ component). The Gaussian components are shown in green, blue, and magenta, and the final fits are shown in red.\label{fig01B}}
\end{center}
\end{figure*}

\begin{figure*}
\begin{center}
  \includegraphics[scale=0.6]{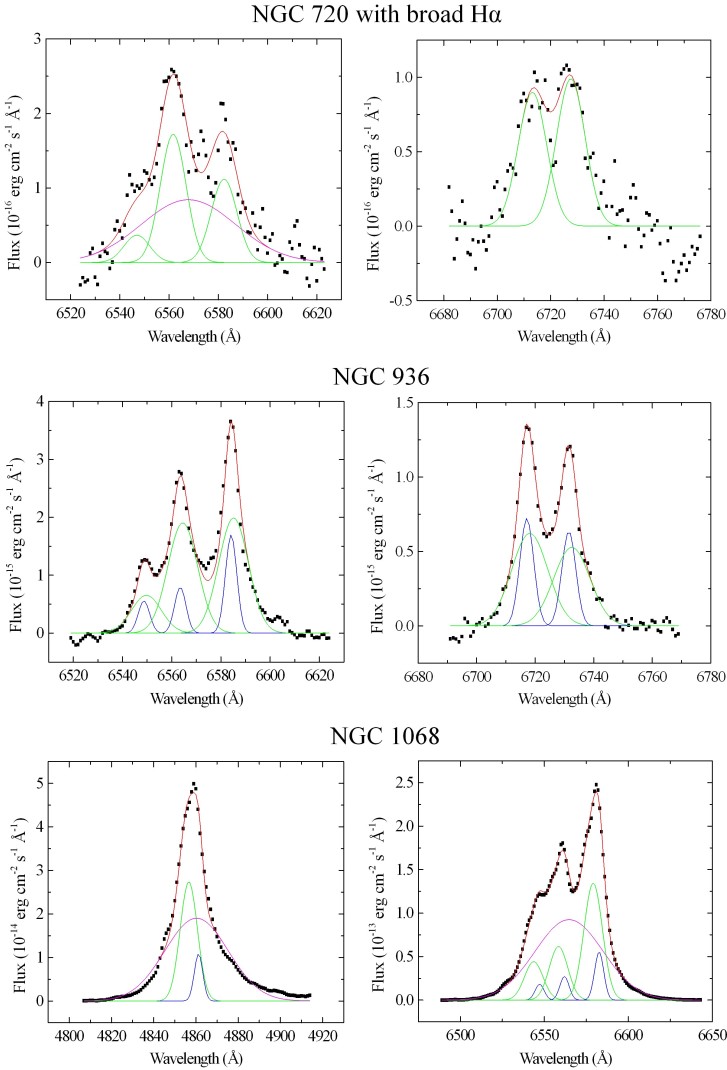}
  \caption{The same as in Fig.~\ref{fig01B}, but for the data cubes of NGC 720 (with a broad H$\alpha$ component), NGC 936, and NGC 1068.\label{fig02B}}
\end{center}
\end{figure*}

\begin{figure*}
\begin{center}
  \includegraphics[scale=0.6]{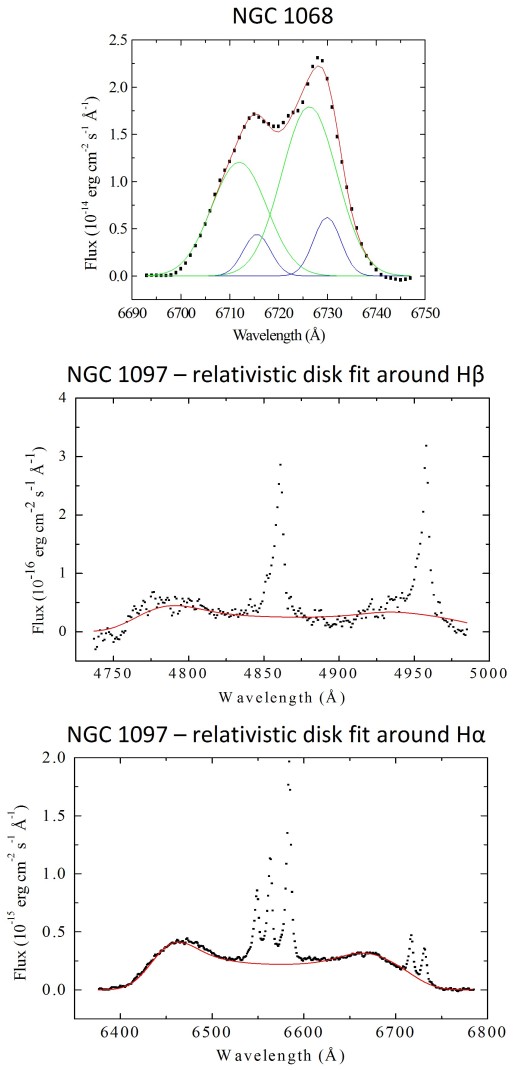}
  \caption{Top: the same as in Fig.~\ref{fig01B}, but for the [S \textsc{ii}]($\lambda$6716+$\lambda$6731) emission lines of the data cube of NGC 1068. Middle: relativistic disc fit for the spectral region around the H$\beta$ emission line of the nuclear spectrum of NGC 1097, obtained with the formalism of Chen \& Halpern (1989). The inclination of the disc, relative to the line of sight, is $i \sim 35\degr$. Bottom: the same as in the middle, but for the spectral region around the H$\alpha$ emission line.\label{fig03B}}
\end{center}
\end{figure*}

\begin{figure*}
\begin{center}
  \includegraphics[scale=0.6]{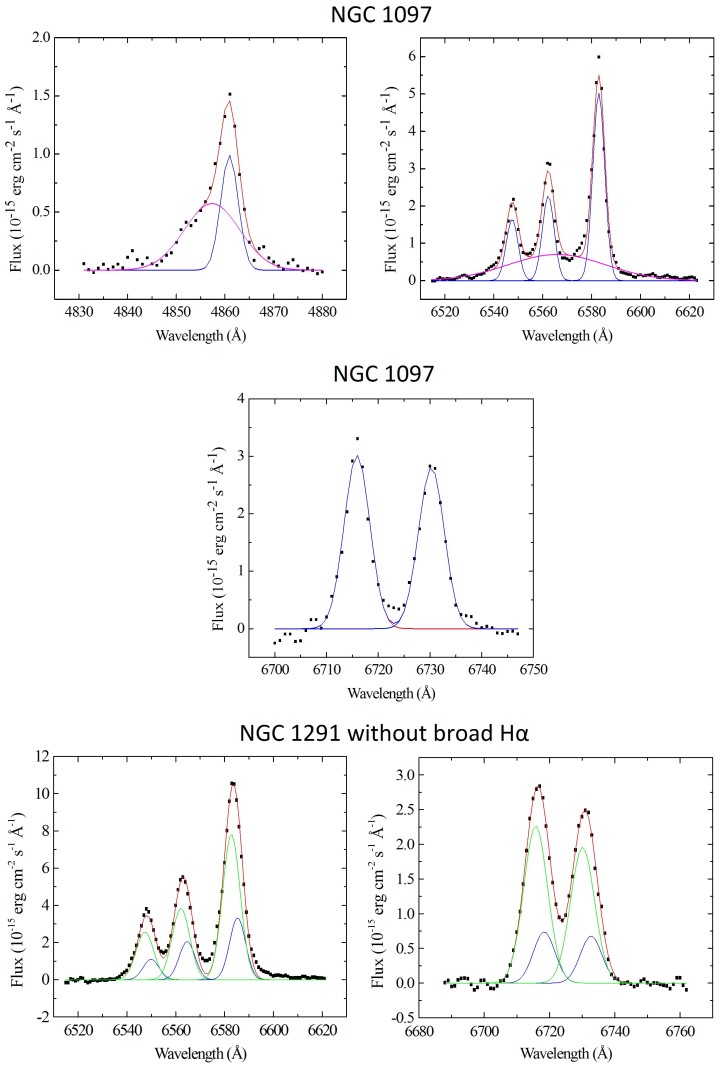}
  \caption{The same as in Fig.~\ref{fig01B}, but for the data cubes of NGC 1097 (without the emission from the relativistic disk) and NGC 1291 (without a broad H$\alpha$ component).\label{fig04B}}
\end{center}
\end{figure*}

\begin{figure*}
\begin{center}
  \includegraphics[scale=0.6]{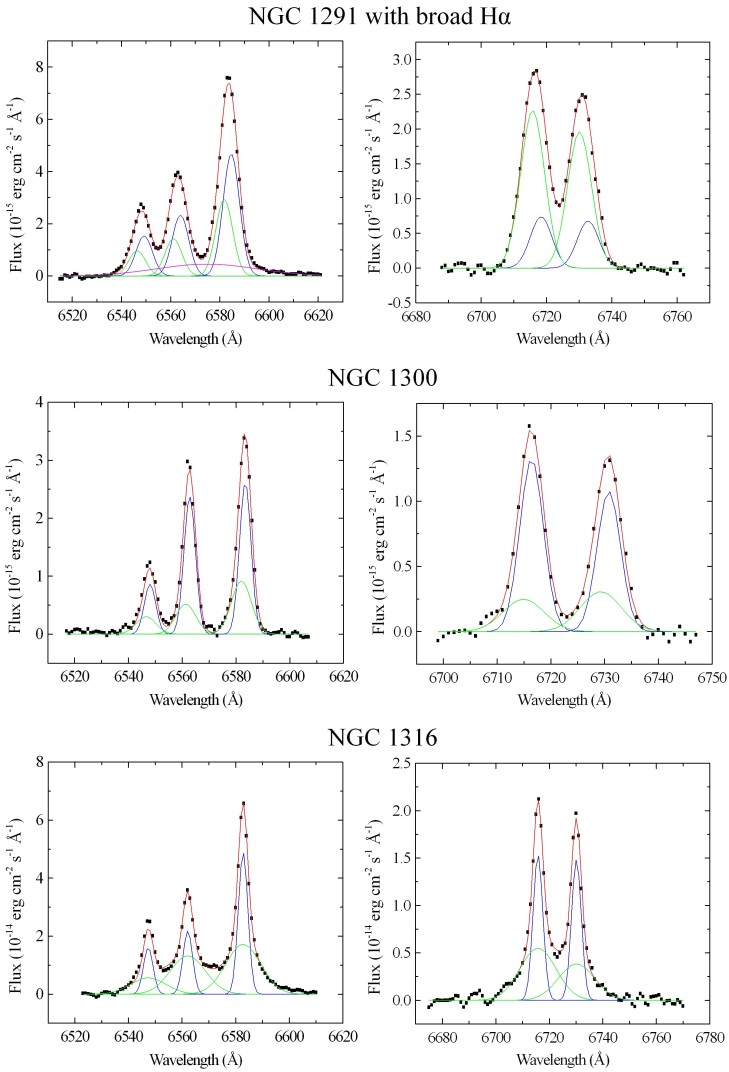}
  \caption{The same as in Fig.~\ref{fig01B}, but for the data cubes of NGC 1291 (with a broad H$\alpha$ component), NGC 1300, and NGC 1316.\label{fig05B}}
\end{center}
\end{figure*}

\begin{figure*}
\begin{center}
  \includegraphics[scale=0.6]{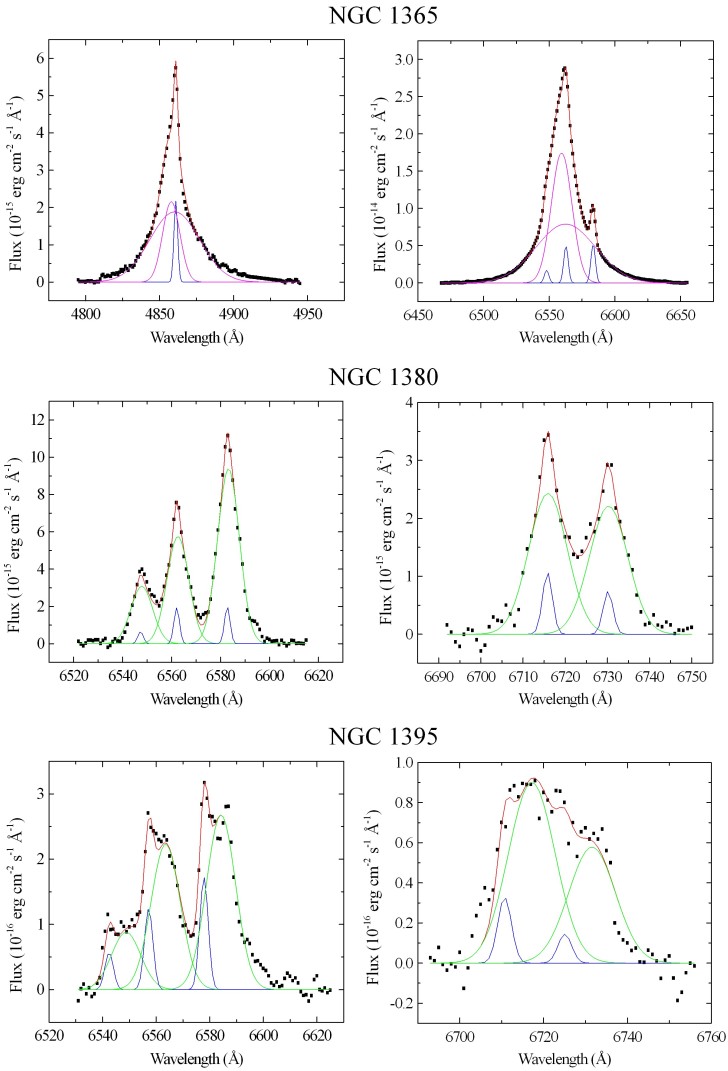}
  \caption{The same as in Fig.~\ref{fig01B}, but for the data cubes of NGC 1365, NGC 1380, and NGC 1395.\label{fig06B}}
\end{center}
\end{figure*}

\begin{figure*}
\begin{center}
  \includegraphics[scale=0.6]{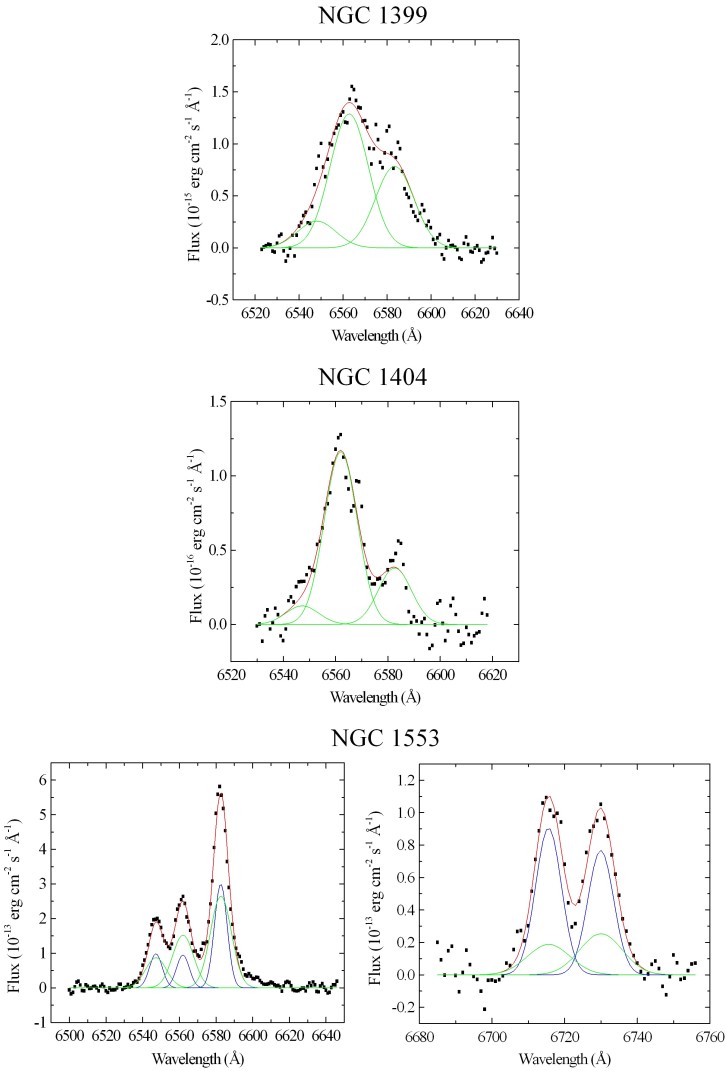}
  \caption{The same as in Fig.~\ref{fig01B}, but for the data cubes of NGC 1399, NGC 1404, and NGC 1553.\label{fig07B}}
\end{center}
\end{figure*}

\begin{figure*}
\begin{center}
  \includegraphics[scale=0.6]{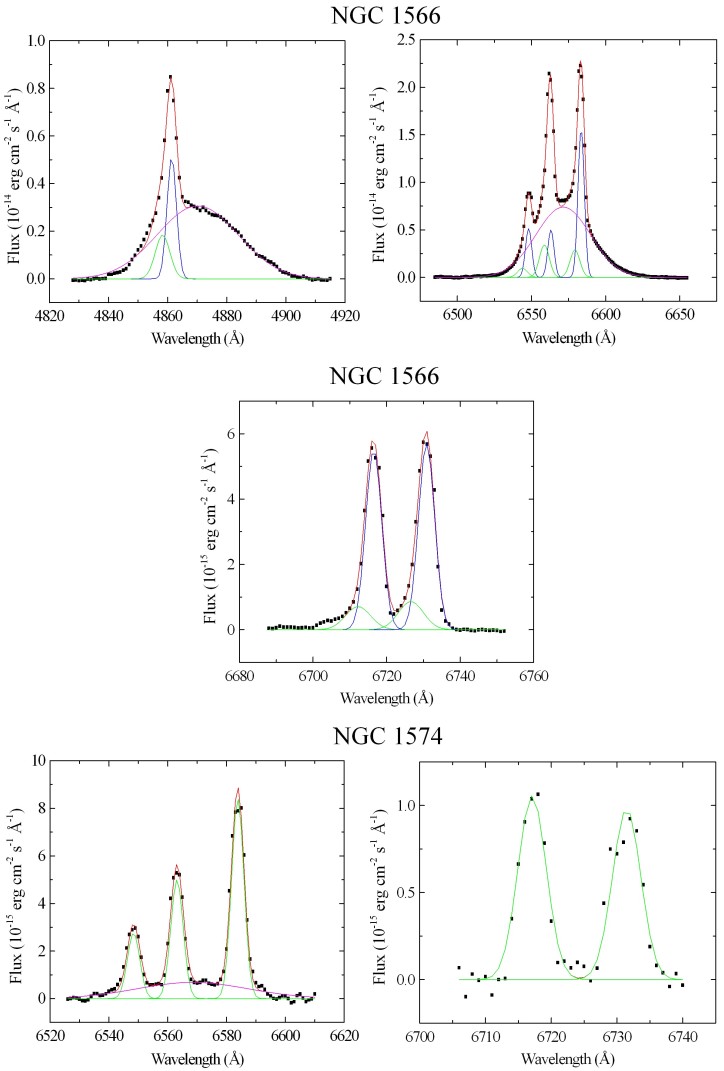}
  \caption{The same as in Fig.~\ref{fig01B}, but for the data cubes of NGC 1566 and NGC 1574.\label{fig08B}}
\end{center}
\end{figure*}

\begin{figure*}
\begin{center}
  \includegraphics[scale=0.6]{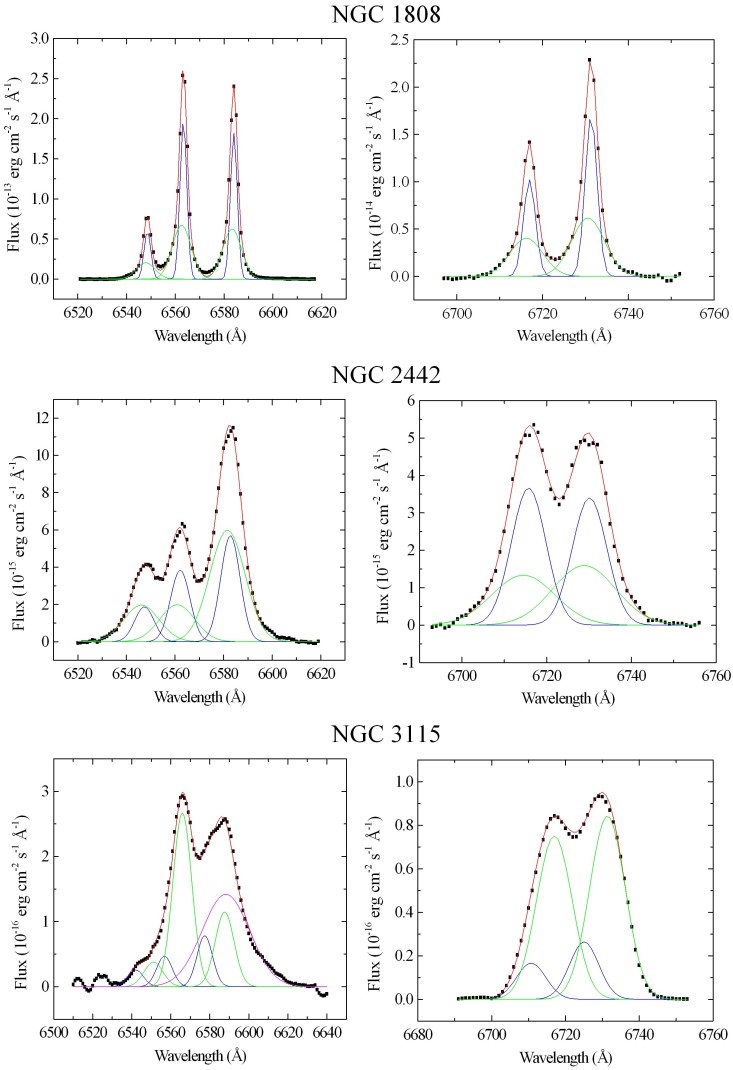}
  \caption{The same as in Fig.~\ref{fig01B}, but for the data cubes of NGC 1808, NGC 2442, and NGC 3115.\label{fig09B}}
\end{center}
\end{figure*}

\begin{figure*}
\begin{center}
  \includegraphics[scale=0.6]{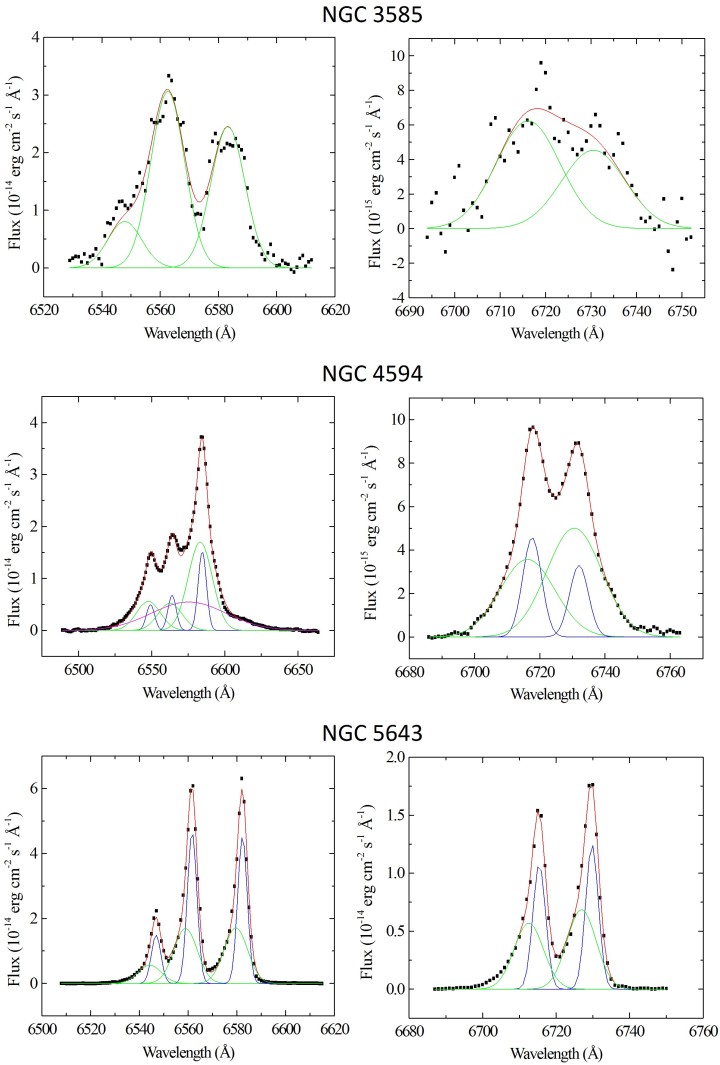}
  \caption{The same as in Fig.~\ref{fig01B}, but for the data cubes of NGC 3585, NGC 4594, and NGC 5643.\label{fig10B}}
\end{center}
\end{figure*}

\begin{figure*}
\begin{center}
  \includegraphics[scale=0.6]{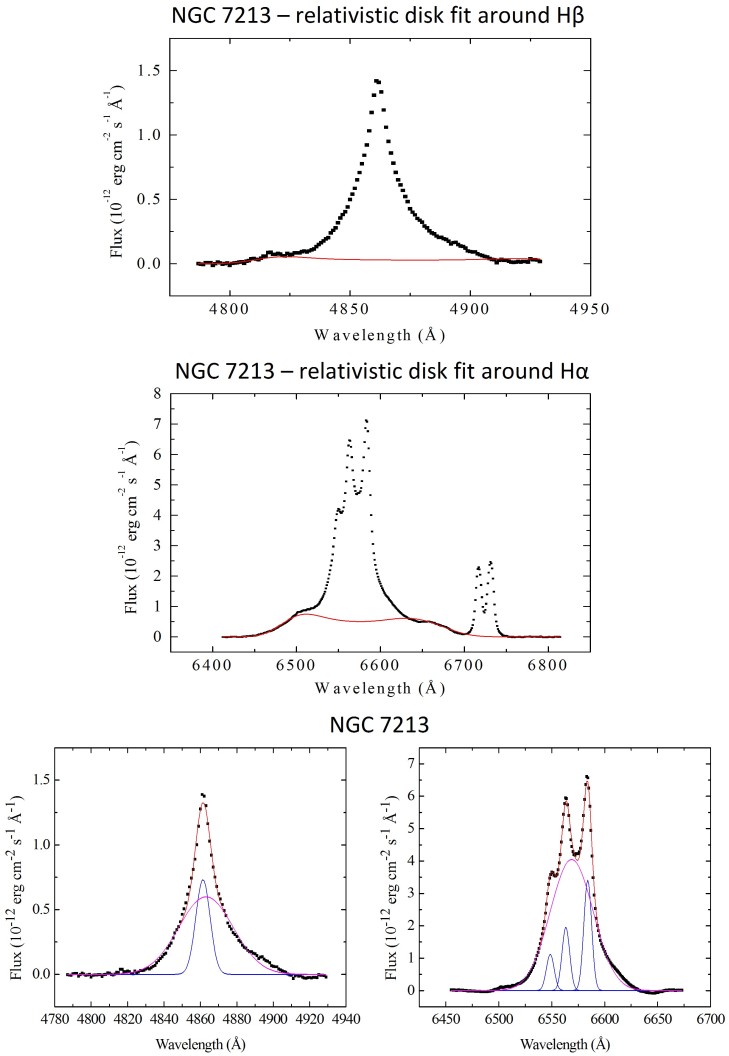}
  \caption{Top: relativistic disc fit for the spectral region around the H$\beta$ emission line of the nuclear spectrum of NGC 7213, obtained with the formalism of Chen \& Halpern (1989). The inclination of the disc, relative to the line of sight, is $i \sim 28\degr$. Middle: the same as on the top, but for the spectral region around the H$\alpha$ emission line. Bottom: the same as in Fig.~\ref{fig01B}, but for the H$\beta$ and [N \textsc{ii}] + H$\alpha$ emission lines of the data cube of NGC 7213.\label{fig11B}}
\end{center}
\end{figure*}

\begin{figure*}
\begin{center}
  \includegraphics[scale=0.4]{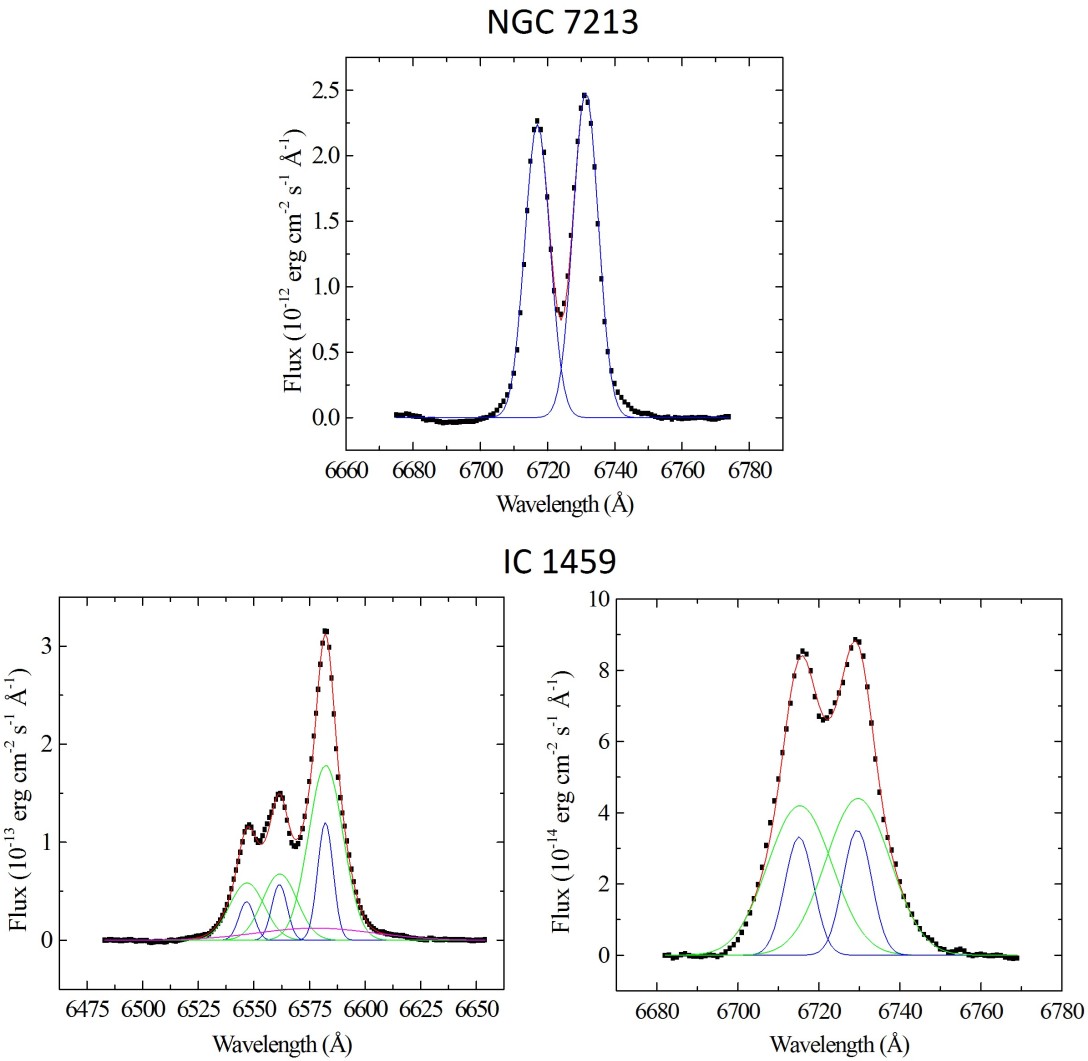}
  \caption{The same as in Fig.~\ref{fig01B}, but for the data cubes of NGC 7213 ([S \textsc{ii}]$\lambda$6716 + $\lambda$6731 emission lines) and IC 1459.\label{fig12B}}
\end{center}
\end{figure*}

\bsp	
\label{lastpage}
\end{document}